\documentclass[12pt,a4paper,notitlepage]{article}

\usepackage{graphicx}
\usepackage{amssymb}
\usepackage{amsmath}

\usepackage{amsthm}
\usepackage{amsfonts}
\usepackage{times}

\usepackage[format=hang]{subcaption}
\usepackage{color}
\usepackage[T1]{fontenc}
\usepackage{hyperref}
\usepackage[square]{natbib}
\usepackage{upgreek}

\usepackage{pgfplots}
\usepackage{todonotes}
\usepackage{bm}

\usepackage{float}

\usepackage{subcaption} 
\usepackage{siunitx}
\usepackage{tikz}
\usepackage{tikz-3dplot}
\usetikzlibrary{calc,backgrounds}

\defcitealias{KlockeZeisEtAl2014}{Klocke, Zeis et~al.}
\defcitealias{KlockeKlinkEtAl2014}{Klocke, Klink et~al.}

\pgfrealjobname{paper} 
\pgfplotsset{compat=1.14}

\definecolor{rwth1}{RGB}{0,84,159}      
\definecolor{rwth2}{RGB}{142,186,229}   
\definecolor{rwth3}{RGB}{0,97,101}      
\definecolor{rwth4}{RGB}{0,152,161}     
\definecolor{rwth5}{RGB}{87,171,39}     
\definecolor{rwth6}{RGB}{189,205,0}     
\definecolor{rwth7}{RGB}{255,237,0}     
\definecolor{rwth8}{RGB}{246,168,0}     
\definecolor{rwth9}{RGB}{227,0,102}     
\definecolor{rwth10}{RGB}{204,7,30}     
\definecolor{rwth11}{RGB}{161,16,53}    
\definecolor{rwth12}{RGB}{97,33,88}     
\definecolor{rwth13}{RGB}{122,111,172}  

\definecolor{sfb1}{RGB}{0,84,165}      
\definecolor{sfb2}{RGB}{201,0,35}      
\definecolor{sfb3}{RGB}{231,95,1}      
\definecolor{sfb4}{RGB}{127,127,127}   
\definecolor{sfb5}{RGB}{217,217,217}   

\tikzstyle{dashpattern0} = [dash pattern = ]
\tikzstyle{dashpattern1} = [dash pattern = on 4.25pt off 0.75pt]
\tikzstyle{dashpattern2} = [dash pattern = on 1.5pt off 0.5pt]
\tikzstyle{dashpattern3} = [dash pattern = on 0.75pt off 0.4pt]
\tikzstyle{dashpattern4} = [dash pattern = on 3pt off 1pt on 1pt off 1pt]
\tikzstyle{dashpattern5} = [dash pattern = on 3.75pt off 0.5pt on 0.75pt off 0.5pt on 0.75pt off 0.5pt]
\tikzstyle{dashpattern6} = [dash pattern = on 3.25pt off 0.5pt on 0.75pt off 0.5pt on 0.75pt off 0.5pt on 0.75pt off 0.5pt]
\tikzstyle{dashpattern7} = [dash pattern = on 3.25pt off 0.5pt on 0.75pt off 0.5pt on 0.75pt off 0.5pt on 0.75pt off 0.5pt on 0.75pt off 0.5pt]
\tikzstyle{dashpattern8} = [line cap=round, dash pattern = on 3.25pt off 2.75pt]
\tikzstyle{dashpattern9} = [line cap=round, dash pattern = on 0.01pt off 2pt]
\tikzstyle{dashpattern10}= [line cap=round, dash pattern = on 3.25pt off 2pt on 0.01pt off 2pt]
\tikzstyle{dashpattern11}= [line cap=round, dash pattern = on 3.5pt off 1.75pt on 0.01pt off 1.75pt on 0.01pt off 1.75pt]
\tikzstyle{dashpattern12}= [line cap=round, dash pattern = on 3.5pt off 1.75pt on 0.01pt off 1.75pt on 0.01pt off 1.75pt on 0.01pt off 1.75pt]
\tikzstyle{dashpattern13}= [line cap=round, dash pattern = on 3.5pt off 1.75pt on 0.01pt off 1.75pt on 0.01pt off 1.75pt on 0.01pt off 1.75pt on 0.01pt off 1.75pt]

\newcommand{\grad}[1]{{\rm grad} \hspace{-0.5mm} \left( #1 \right)}
\renewcommand{\div}[1]{{\rm div} \hspace{-0.5mm} \left( #1 \right)}

\newcommand{\dV}{\mathrm{d}V}

\newcommand{\dt}{\Delta t}
\newcommand{\intO}{\int_\Omega}

\newcommand{\coloneq}{\mathrel{\resizebox{\widthof{$\mathord{=}$}}{\height}{ $\!\!\resizebox{1.2\width}{0.8\height}{\raisebox{0.23ex}{$\mathop{:}$}}\!\!=\!\!$ }}}

\newcommand{\delv}{\Delta v}

\newcommand{\epsnull}{\epsilon_\mathrm{0}}
\newcommand{\epsrbar}{\bar{\epsilon}_\mathrm{r}}
\newcommand{\epsr}{\epsilon_\mathrm{r}}
\newcommand{\gv}{g_v}
\newcommand{\gjtilde}{g_{\tilde{j}}}

\newcommand{\dI}{\mathrm{d}I}
\newcommand{\dIofjd}{\mathrm{d}I(\j,d)}
\newcommand{\kEbar}{\bar{k}_\textrm{\tiny E}}
\newcommand{\kE}{k_\textrm{\tiny E}}

\newcommand{\dQdot}{\mathrm{d}\dot{Q}}
\newcommand{\rhoE}{\rho_\textrm{\tiny E}}
\newcommand{\rhoEdot}{\dot{\rho}_\textrm{\tiny E}}

\newcommand{\varv}{\delta v}

\newcommand{\D}{\bm{D}}
\newcommand{\E}{\bm{E}}
\newcommand{\Edot}{\dot{\bm{E}}}
\renewcommand{\j}{\bm{j}}

\newcommand{\n}{\bm{n}}

\newcommand{\Afuntimet}{\mathcal{A}\hspace*{-0.7mm}\left(\bm{x},t\right)}

\renewcommand{\ddot}{\dot{d}}

\newcommand{\CAT}{^\textrm{\tiny CAT}}
\newcommand{\EL}{^\textrm{\tiny EL}}
\newcommand{\ME}{^\textrm{\tiny ME}}

\newcommand{\Vdis}{V_\mathrm{dis}}
\newcommand{\dVdis}{\mathrm{d}V_\mathrm{dis}}
\newcommand{\dVdisdot}{\mathrm{d}\dot{V}_\mathrm{dis}}

\newcommand{\Veffnew}{\nu_\mathrm{dis}}
\newcommand{\Vref}{V_\mathrm{ref}}
\newcommand{\dVref}{\mathrm{d}V_\mathrm{ref}}
\newcommand{\Vel}{V_\mathrm{el}}

\newcommand{\Vco}{V_\mathrm{co}}

\newcommand{\lc}{\lambda_\mathrm{cat}}
\newcommand{\Vcat}{V_\mathrm{cat}}

\AtBeginDocument{}

\date{}

\graphicspath{{./02_Figures/}}

\begin{document}

\author{\large {\parbox{\linewidth}{\centering Tim van der Velden\footnote{Corresponding author: \\ email: tim.van.der.velden@ifam.rwth-aachen.de}\hspace{1.0mm},
                                               Stephan Ritzert,
                                               Stefanie Reese,
                                               Johanna Waimann}}\\[0.5cm]
  \hspace*{-0.1cm}
  \normalsize{\em \parbox{\linewidth}{\centering
    \vspace{3mm}
    Institute of Applied Mechanics, RWTH Aachen University, Mies-van-der-Rohe-Str. 1,\\D-52074 Aachen, Germany}
  }
}

\title{\LARGE Modeling moving boundary value problems in electrochemical machining}

\maketitle

\vspace{-3mm}

\small
{\bf Abstract.} {
This work presents a new approach to efficiently model the cathode in the moving boundary value problem of electrochemical machining. Until recently, the process simulation with finite elements had the drawback of remeshing required by the changing surface geometries. This disadvantage was overcome by a novel model formulation for the anodic dissolution that utilizes effective material parameters as well as the dissolution level as an internal variable and, thereby, does not require remeshing. Now, we extend this concept to model arbitrarily shaped and moving cathodes. Two methodologies are investigated to describe the time varying position of the cathode. In the first approach, we change the electric conductivity of elements within the cathode and, in a second approach, we apply Dirichlet boundary conditions on the nodes of corresponding elements. For both methods, elements on the cathode's surface are treated with effective material parameters. This procedure allows for the efficient simulation of industrially relevant, complex geometries without mesh adaptation. The model's performance is validated by means of analytical, numerical and experimental results from the literature. The short computation times make the approach interesting for industrial applications.

}


\normalsize

\begin{table}[htbp]
  \centering
  \caption{Relevant symbols}
  \begin{tabular}{lll}
      \hline
      Symbol                       & Unit                                          & Definition                     \\
      \hline \hline
      $\epsnull$                   & $[ \si{\ampere\second\per\volt\per\meter} ]$  & Electric constant              \\
      $\epsr$                      & $[ - ]$                                       & Relative permittivity          \\
      $\lc$                        & $[ - ]$                                       & Cathode ratio                  \\
      $\Veffnew$                   & $[ \si{\cubic\meter\per\ampere\per\second} ]$ & Effectively dissolved volume   \\
      $\rhoE$                      & $[ \si{\ampere\second\per\cubic\meter} ] $    & Electric volume charge density \\
      $\mathcal{A}$                & $[ - ]$                                       & Activation function            \\
      $d$                          & $[ - ]$                                       & Dissolution level              \\
      $\D$                         & $[ \si{\ampere\second\per\square\meter} ]$    & Electric displacement field    \\
      $\E$                         & $[ \si{\volt\per\meter} ]$                    & Electric field strength        \\
      $I$                          & $[ \si{\ampere} ]$                            & Electric current               \\
      $\j$                         & $[ \si{\ampere\per\square\meter} ]$           & Electric current density       \\
      $\kE$                        & $[ \si{\ampere\per\volt\per\meter} ]$         & Electric conductivity          \\
      $\n$                         & $[ - ]$                                       & Normal vector                  \\
      $Q$                          & $[ \si{\ampere\second} ] $                    & Electric charge                \\
      $R$                          & $[ \si{\volt\per\ampere} ] $                  & Electric resistance            \\
      $t$                          & $[ \si{\second} ] $                           & Machining time                 \\
      $v$                          & $[ \si{\volt} ] $                             & Electric potential             \\
      $\Vcat$                      & $[ \si{\cubic\meter} ] $                      & Cathode volume                 \\
      $\Vdis$                      & $[ \si{\cubic\meter} ] $                      & Dissolved volume               \\
      $\Vel$                       & $[ \si{\cubic\meter} ] $                      & Element volume                 \\
      $\Vref$                      & $[ \si{\cubic\meter} ] $                      & Reference volume               \\
      $\bm{x}$                     & $[ \si{\meter} ]$                             & Position                       \\
      $(\bullet)\CAT$              &                                               & Cathode quantity               \\
      $(\bullet)\EL$               &                                               & Electrolyte quantity           \\
      $(\bullet)\ME$               &                                               & Metal quantity                 \\                     
      \hline
  \end{tabular}
  \label{tab:constants}
\end{table}

\section{Introduction}
\label{sec:introduction}

The industry's ever increasing need for the production of complex structures that consist of high strength materials propels the use of electrochemical machining (ECM) (\cite{KlockeKlinkEtAl2014}). A decisive advantage of the process is the possibility to machine solid materials with exceptional hardness at high removal rates. Moreover, being a contactless process, ECM does not lead to tool wear and does not afflict any mechanical damage to the rim zone of the workpiece. The process utilizes the effect of electrolysis to machine the workpiece by material dissolution (\cite{DeBarrOliver1968}, \cite{McGeough1974}, \cite{KlockeKoenig2007}).

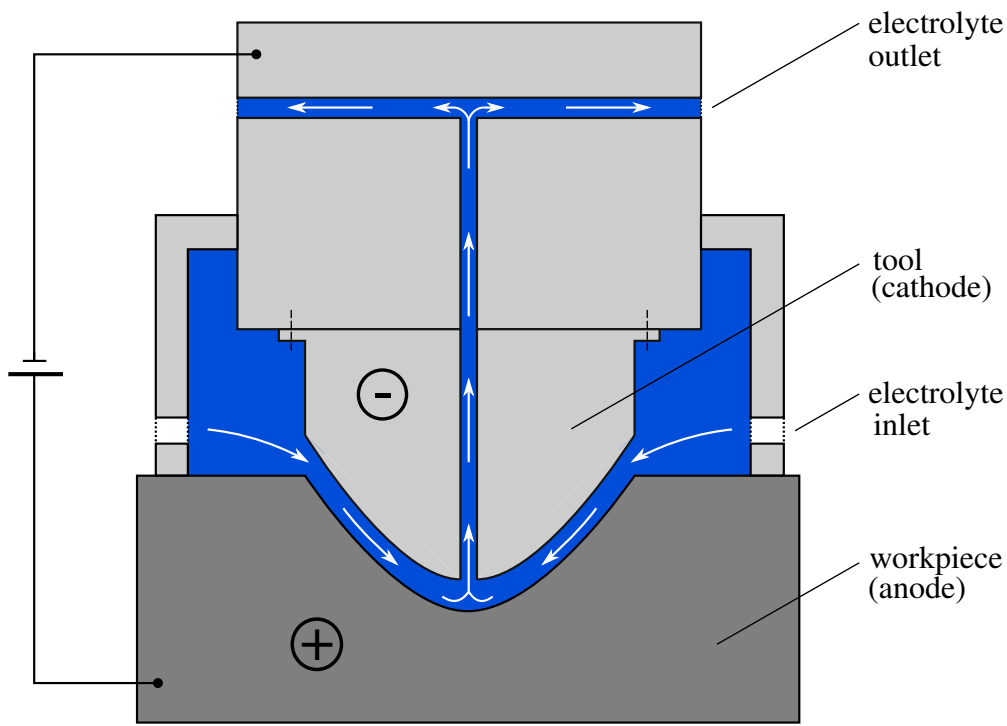
\begin{figure}[htbp]
  \centering
  \begin{tikzpicture}
           \node[inner sep=0pt] (pic) at (0,0) {\includegraphics[width=0.7\textwidth]
           {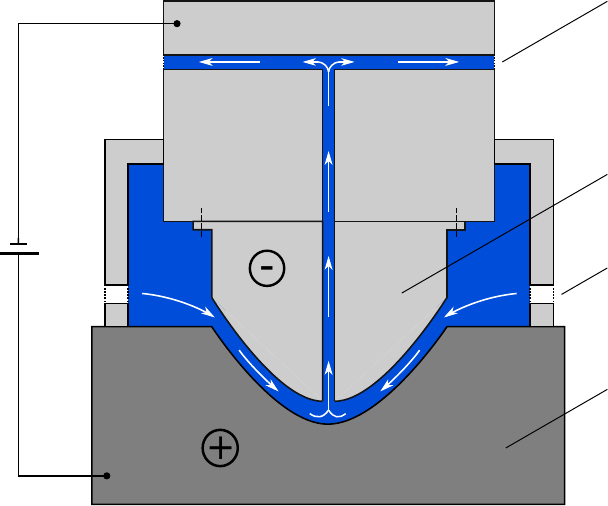}};
           \node[inner sep=0pt] (eout)  at ($(pic.east) +( 1.00, 4.60)$)  {electrolyte};
           \node[inner sep=0pt] (eout2) at ($(eout)     +(-0.41,-0.39)$)  {outlet};
           \node[inner sep=0pt] (ein)   at ($(pic.east) +( 1.00,-0.30)$)  {electrolyte};
           \node[inner sep=0pt] (ein2)  at ($(ein)      +(-0.45,-0.39)$)  {inlet};
           \node[inner sep=0pt] (tool)  at ($(pic.east) +( 0.50, 1.50)$)  {tool};
           \node[inner sep=0pt] (cat)   at ($(tool)     +( 0.44,-0.39)$)  {(cathode)};
           \node[inner sep=0pt] (wp)    at ($(pic.east) +( 1.00,-2.50)$)  {workpiece};
           \node[inner sep=0pt] (an)    at ($(wp)       +(-0.25,-0.39)$)  {(anode)};
         \end{tikzpicture} 
  \caption{Illustration of the electrochemical machining process (after \cite{KlockeKoenig2007}).}
  \label{fig:ecmsetup}
\end{figure}

Fig.~\ref{fig:ecmsetup} shows a schematic sketch for an arbitrarily shaped tool. A direct current is applied at the electrodes and yields an electric current at the workpiece's surface that causes the anodic dissolution.

The fields of application for ECM range from the blade production in turbomachinery manufacturing over the creation of shape optimized cooling boores for improved stress distributions to the precision machining of filigree structures (\cite{KlockeKoenig2007}, \cite{RajurkarSundaramEtAl2013}). This wide range of applications motivates the development of efficient simulation methods. Since it is difficult to predict the final shape of the workpiece in ECM, simulations can help to reduce costly trial and error experiments and also to speed up the design process (\cite{HindujaKunieda2013}).

Over time, different numerical methodologies were employed to model ECM: These are the finite difference method (e.g.~\cite{Tipton1964}, \cite{KoenigHuembs1977}, \cite{Kozak1998}), the boundary element method (e.g.~\cite{ChristiansenRasmussen1976}, Hansen in \cite{FasanoPrimicerio1983}, \cite{DeconinckMaggettoEtAl1985}, \cite{NarayananHindujaEtAl1986}) and also the finite element method (FEM) (e.g. \cite{AlkireBergh1978}, \cite{JainPandey1980}, \cite{HardistyMilehamEtAl1993}) that is also employed in this work. According to \cite{HindujaKunieda2013}, the FEM possesses the drawback of computationally expensive remeshing due to the changing geometries. To circumvent remeshing, e.g.~\cite{Brookes1984} tried to delete and modify elements close to the anode's surface, which also proved to be expensive.
An elegant procedure to adapt the mesh for moving interfaces was presented by \cite{DeconinckVanDammeEtAl2011}, who employ an elastic body analogy (\cite{Wuilbaut2008}) to realign the refined region of the mesh with the anodic surface.

However, even state of the art formulations like e.g.~the one of \cite{KlockeHeidemannsEtAl2018} that models precise ECM with pulsed current and oscillating cathode still need remeshing to model the anodic dissolution and the cathode movement by assigning separate domains for fluid and solid. To our best knowledge, previous models lack a continuous formulation of the dissolution process without the requirement of mesh adaption.

Therefore, we proposed a novel approach that works entirely without remeshing and models the anodic dissolution in ECM with effective material parameters and an internal variable (\cite{vanderVeldenRommesEtAl2021}). In this work, the novel approach based on effective material parameters serves to model the movement of arbitrarily shaped cathodes in ECM. Previous studies were restricted to a stationary cathode or a simplified modeling of the cathode feed. The aspiration to investigate industrially relevant and complex problems underlines the necessity of this contribution to efficiently and accurately model the cathode.

To this end, this work investigates two approaches for modeling the cathode. In method A, the electric conductivity of the elements within the cathode is modified. In method B, Dirichlet boundary conditions are applied on all nodes within the cathode. Both methodologies are based on effective material parameters for elements on the cathode's surface and, thereby, allow for a simulation entirely without remeshing.

Nowadays, following the earlier works of e.g.~\cite{AntonovaLooman2005}, \cite{Perez-AparicioTaylorEtAl2007}, numerous authors investigate electrically coupled multifield problems. \cite{HofmannWesthoffEtAl2020} conduct e.g.~electro-chemo-mechanical simulations, \cite{WuLi2018} additionally include thermal and \cite{LiuLiuEtAl2020} magnetic effects to name only a few. We, however, focus in this work on the modeling of the moving boundary value problem and, therefore, for simplicity, restrict ourselves to the electric field problem, since the electric current density is the driving force in electrochemical machining. The thermal field is thus assumed to be constant.

Moreover, the model's efficiency could further be improved by advanced finite element technology using reduced integration and hourglass stabilization, which has already been applied successfully to different multifield problems, e.g.~for electromagnetic (\cite{ReeseSvendsenEtAl2005}) and thermomechanical coupling (\cite{JuhreReese2010}) as well as gradient-extended damage (\cite{BarfuszvanderVeldenEtAl2021,BarfuszvanderVeldenEtAl2022}). Additionally, the incorporation of initially and induced anisotropic material behavior following e.g.~\cite{HolthusenBrepolsEtAl2020} and \cite{ReeseBrepolsEtAl2021} promises an interesting model extension in future works.

\textbf{Outline of the work.\quad} In Section~\ref{sec:constmod}, the electric balance equation and the constitutive laws are introduced. Thereafter, we give a brief summary of the dissolution model in Section~\ref{sec:anode} and comment on numerical aspects in Section~\ref{sec:numericalaspects}. In Section~\ref{sec:cathode}, we present the new methodologies for modeling the cathode. In Section~\ref{ssec:Ex1}, analytical reference solutions serve to validate the model's accuracy and additional studies regarding the runtime and the choice of the effective material parameters are conducted. In Sections~\ref{ssec:Ex2} and \ref{ssec:Ex3}, challenging examples with numerical and experimental references are investigated. In Section~\ref{ssec:Ex4}, the model's ability to model complex cathode shapes by simulating a blade machining process is confirmed. A conclusion is provided in Section~\ref{sec:conclusion}.

\textbf{Notational conventions.\quad} In this work, italic characters $a$, $A$ denote scalars and zeroth-order tensors and bold-face italic characters $\bm{b}$, $\bm{B}$ refer to vectors and first-order tensors. The operators $\div{\bullet}$ and $\grad{\bullet}$ denote the divergence and gradient operation of a quantity with respect to Cartesian coordinates. A dot $\cdot$ defines the single contraction of two tensors. The time derivative of a quantity is given by $\dot{(\bullet)}$. A tilde $\widetilde{(\bullet)}$ defines a prescribed quantity on the corresponding boundary. Moreover, Table~\ref{tab:constants} lists the relevant symbols and states, if applicable, the corresponding SI units.


\section{Constitutive modeling}
\label{sec:constmod}

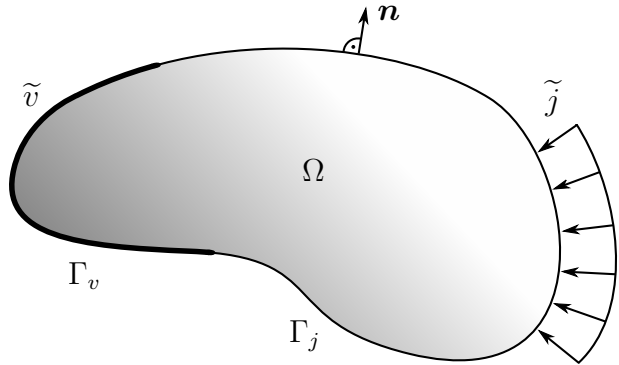
\begin{figure}[htbp]
  \centering
  \begin{tikzpicture}
           \node[inner sep=0pt] (pic) at (0,0) {\includegraphics[width=0.5\textwidth]
           {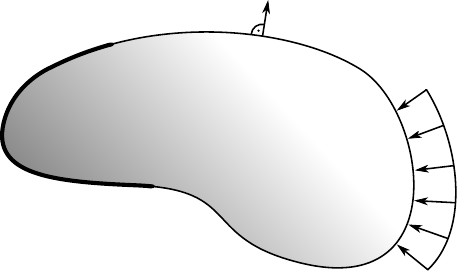}};
           \node[inner sep=0pt] (omega)  at ($(pic.east) +(-4.00, 0.20)$)  {$\Omega$};
           \node[inner sep=0pt] (gammav) at ($(pic.west) +( 1.00,-1.20)$)  {$\Gamma_v$};
           \node[inner sep=0pt] (gammaj) at ($(pic.west) +( 3.90,-2.00)$)  {$\Gamma_j$};
           \node[inner sep=0pt] (vtilde) at ($(pic.west) +( 0.30, 1.20)$)  {$\widetilde{v}$};
           \node[inner sep=0pt] (jtilde) at ($(pic.east) +(-0.85, 1.15)$)  {$\widetilde{j}$};
           \node[inner sep=0pt] (n)      at ($(pic.north)+( 1.00,-0.10)$)  {$\n$};
         \end{tikzpicture} 
  \caption{Illustration of the electric boundary value problem.}
  \label{fig:kartoffel}
\end{figure}

The present contribution focuses on the modeling of moving cathode geometries. As mentioned in Section~\ref{sec:introduction}, we consider isothermal problems and, therefore, solely present the balance of electric charge including Dirichlet and Neumann boundary conditions (see Fig.~\ref{fig:kartoffel} and cf.~\cite{Jackson1962}):
\begin{align}
  \rhoEdot + \div{\j} &= 0                 \hspace{11.1mm} \mathrm{in} \,\, \Omega    \notag                \\
  v                   &= \widetilde{v}(t)  \hspace{06.1mm} \mathrm{on} \,\, \Gamma_v  \label{eqn:strngfrmv} \\
  \j \cdot \n         &= \widetilde{j}     \hspace{11.2mm} \mathrm{on} \,\, \Gamma_j  \notag
\end{align}
The primary variable of this problem is the electric potential $v$. The movement of the cathode's surface is incorporated by considering time varying Dirichlet boundary conditions $\widetilde{v}(t)$ for the electric potential. The electric field strength $\E$, the electric displacement field $\D$ and the electric volume charge density $\rhoE$ read
\begin{equation}
  \E = - \grad{v}, \qquad   \D = \epsnull \, \epsr \, \E, \qquad   \rhoE = \div{\D}
\end{equation}
where $\epsnull$ denotes the electric constant and $\epsr$ the relative permittivity. The constitutive law of the electric current density $\j$ yields
\begin{equation}
  \j = \kE \, \E + \epsnull \, \epsr \, \Edot
\end{equation}
where $\kE$ denotes the electric conductivity and $\Edot$ the time derivative of the electric field strength $\E$. Finally, the weak form $\gv$ may be obtained as
\begin{equation}
  \gv \coloneq - \intO \left( \, \kE \, \E + 2 \, \epsnull \, \epsr \, \Edot \right) \cdot \grad{\varv} \dV + \gjtilde = 0
  \label{eqn:gv}
\end{equation}
where $\varv$ is an arbitrary test function and $\gjtilde$ a prescribed electric current density.

For the linearization of the problem and the additional consideration of electro-thermal coupling, the reader is kindly referred to \cite{vanderVeldenRommesEtAl2021}.


\section{Anodic dissolution}
\label{sec:anode}

\begin{figure}[htbp] 
  \centering 
         \begin{tikzpicture}
           \node[inner sep=0pt] (pic) at (0,0) {\includegraphics[width=0.5\textwidth]
           {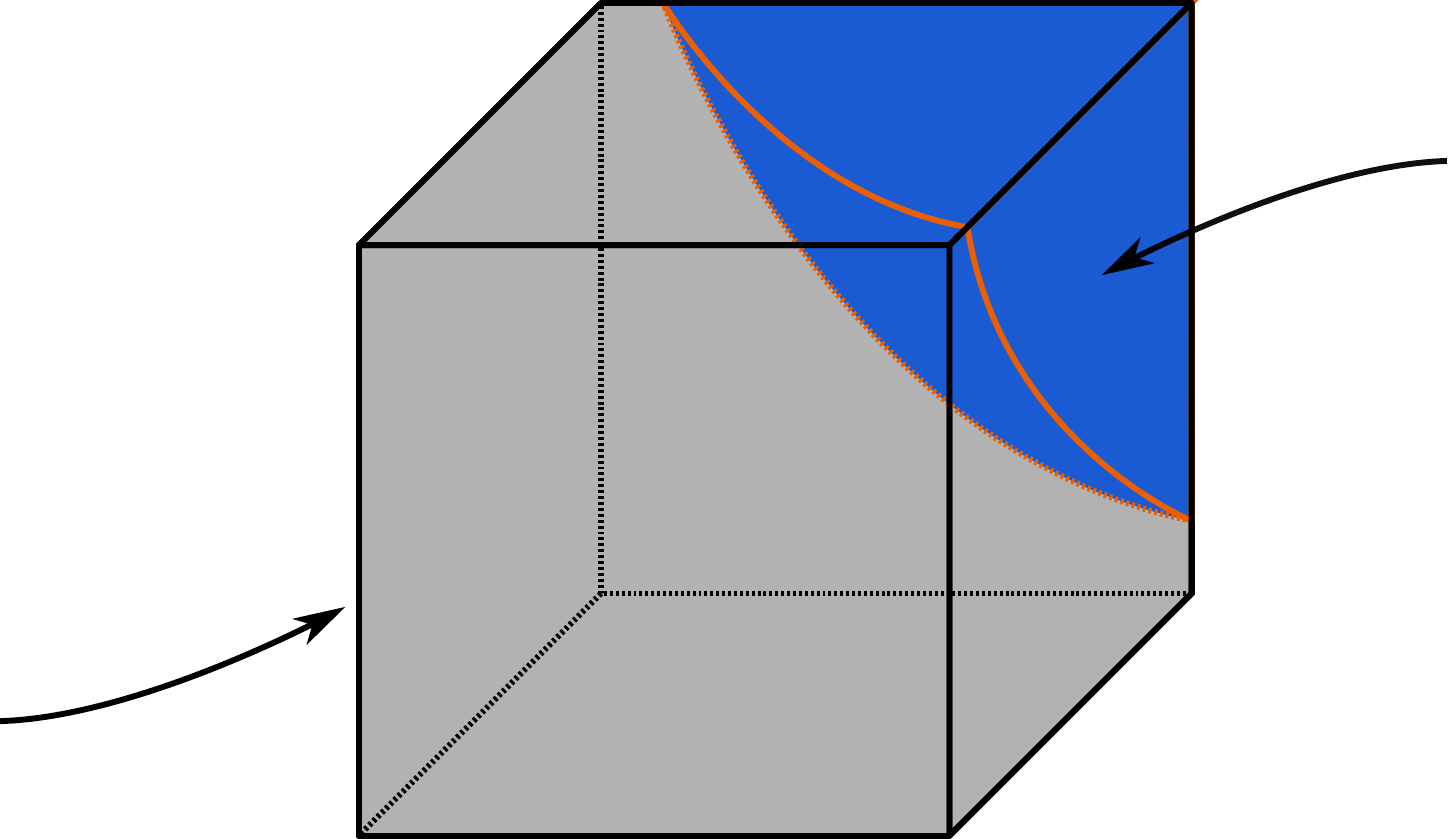}};
           \node[inner sep=0pt] (vr)    at ($(pic.south)+( -4.60, +0.60)$)  {$\dVref$};
           \node[inner sep=0pt] (vd)    at ($(pic.south)+( +4.60, +3.75)$)  {$\dVdis$};
         \end{tikzpicture} 
         \caption{Visualization of the reference volume $\dVref$ and the incremental dissolved volume $\dVdis$. The orange line surrounds the anode's surface.}
         \label{fig:dissolved_volume}
\end{figure} 

\begin{figure}[htbp] 
     \centering 
     \begin{subfigure}{.45\textwidth} 
         \centering 
         \begin{tikzpicture}
           \node[inner sep=0pt] (pic) at (0,0) {\includegraphics[width=0.8\textwidth]
           {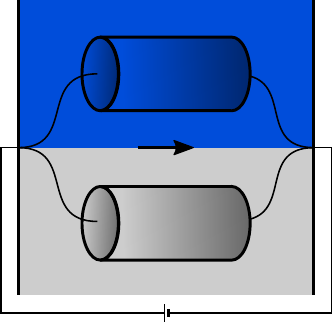}};
           \node[inner sep=0pt] (REL)   at ($(pic.west) +( 3.05, 1.52)$)  {\textcolor{white}{$R\EL$}};
           \node[inner sep=0pt] (RME)   at ($(pic.west) +( 3.05,-1.10)$)  {$R\ME$};
           \node[inner sep=0pt] (j)     at ($(pic.west) +( 3.55,-0.05)$)  {$\j$};
         \end{tikzpicture} 
         \caption{Parallel connection}
         \label{fig:Condpar}
     \end{subfigure}
     \qquad
     \begin{subfigure}{.45\textwidth} 
         \centering 
         \begin{tikzpicture}
           \node[inner sep=0pt] (pic) at (0,0) {\includegraphics[width=0.8\textwidth]
           {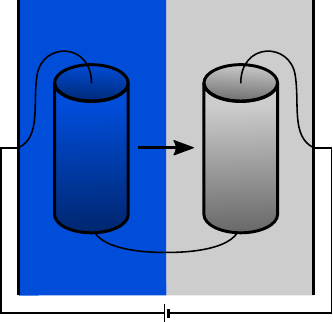}};
           \node[inner sep=0pt] (REL)   at ($(pic.west) +( 1.60, 0.30)$)  {\textcolor{white}{$R\EL$}};
           \node[inner sep=0pt] (RME)   at ($(pic.west) +( 4.13, 0.30)$)  {$R\ME$};
           \node[inner sep=0pt] (j)     at ($(pic.west) +( 3.28,-0.23)$)  {$\j$};
         \end{tikzpicture} 
         \caption{Series connection}
         \label{fig:Condser}
     \end{subfigure}

     \caption{Parallel (a) and series connection (b): Depending on the rule of mixture, different electrical circuits for a horizontal electric current density $\j$ passing between the electrodes through the electrical resistances $R\ME$ and $R\EL$ are considered.} 
     \label{fig:Cond}
\end{figure}
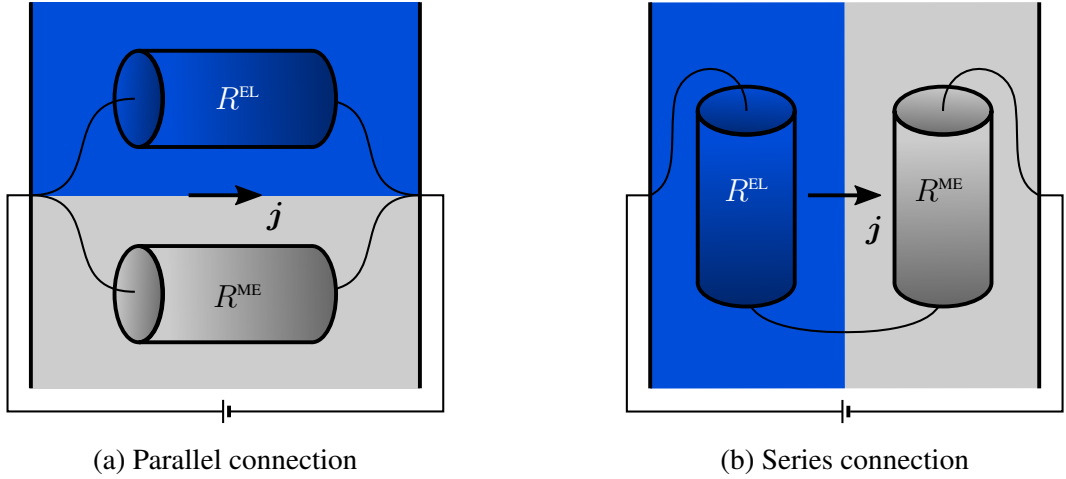

This work utilizes the model of \cite{vanderVeldenRommesEtAl2021} to describe the anodic dissolution. Analogously to damage modeling (e.g.~\cite{Kachanov1958}), an internal variable in combination with effective material parameters serves to describe the transition from an undissolved material state to a fully dissolved one. In the first situation, there is only metal present. In the second situation, the metallic material has been fully dissolved such that only electrolyte remains. The dissolution level $d \in [0,1]$ states the ratio $d = \dVdis / \dVref$ of the incremental dissolved volume $\dVdis$ to the volume of the corresponding reference volume $\dVref$ (see Fig.~\ref{fig:dissolved_volume}). Hence, the rate of the dissolution level yields the relation
\begin{equation}
  \ddot \,\, \dVref = \dVdisdot
  \label{eq:ddot}
\end{equation}
where it is assumed that $\dVref$ is constant over time. In the formulation, we employ a modified version of Faraday's law of electrolysis to correlate the dissolved volume with the passing electric charges (\cite{KlockeKoenig2007}). Using the material parameter $\Veffnew$ that represents the dissolved volume per electric charge, and the activation function $\mathcal{A}$ as well as the time derivative of the incremental electric charge $\dQdot$, the time derivative of the incrementally dissolved volume reads
\begin{equation}
  \dVdisdot = \Veffnew \, \mathcal{A} \,\, \dQdot.
\end{equation}
Moreover, we consider the definition of the activation function
\begin{equation}
  \Afuntimet = 
  \begin{cases}
    1, & \textrm{contact metal-electrolyte} \\
    0, & \mathrm{else}
  \end{cases}
  \label{eq:Afun}
\end{equation}
that depends on the position $\bm{x}$ and time $t$ and indicates whether the material has contact with the electrolyte which is a requirement for the chemical reaction. Further, the connection between the time derivative of the incremental electric charge $\dQdot$ and the incremental electric current $\dI$ is given as
\begin{equation}
  \dQdot = \dI,
  \label{eqn:dQdotdI}
\end{equation}
where $\dIofjd$ depends on the electric current density $\j$ and the dissolution level $d$. With Eq.~\eqref{eqn:dQdotdI}, we accordingly rewrite Eq.~\eqref{eq:ddot} to
\begin{equation}
  \ddot \,\, \dVref = \Veffnew \,  \Afuntimet \, \dIofjd.
\end{equation}
Finally, we define the effective material parameters. Previously, the volume average of metal and electrolyte properties (Fig.~\ref{fig:Condpar}) was considered to read
\begin{equation}
  \bar{(\bullet)}^\mathrm{p} = \left( 1 - d \right) \, (\bullet)\ME \, + \, d \,\, (\bullet)\EL.
  \label{eq:mixanpar}
\end{equation}
This constitutes a parallel connection for the rule of mixture. Motivated by the analogy of electrical circuits (Fig.~\ref{fig:Cond}), this work additionally considers a series connection (Fig.~\ref{fig:Condser}) for the rule of mixture between metal and electrolyte:
\begin{equation}
  \bar{(\bullet)}^\mathrm{s} = \left[ \, \left( 1 - d \right) \, / \, (\bullet)\ME \, + \, d \, / \, (\bullet)\EL \, \right]^{-1}
  \label{eq:mixanser}
\end{equation}
In Section~\ref{ssec:Ex1}, the influence of the rule of mixture on the dissolved volume is studied.

\section{Numerical aspects}
\label{sec:numericalaspects}

For the time and spatial discretization, we employ the backward Euler method and the finite element method, respectively. For detailed information on the latter, the reader is kindly referred to \cite{Hughes1987} and \cite{ZienkiewiczTaylorEtAl2005}. The specific element residual vectors and stiffness matrices may be found in \cite{vanderVeldenRommesEtAl2021}.

The formulation is implemented into the finite element software FEAP (\cite{TaylorGovindjee2020}). The activation function $\mathcal{A}$ (Eq.~\eqref{eq:Afun}) is incorporated into the solution algorithm by an update procedure at the end of every time step. When a finite element dissolves completely, it triggers the update procedure and activates all adjacent metal elements that have a shared surface with the dissolved element.
The cut-off volume $\Vco$, which was investigated in \cite{vanderVeldenRommesEtAl2021}, denotes the theoretically dissolved volume in an integration point that is neglected when $d_{n+1} > 1$ is reset to $d_{n+1} = 1$. In the present formulation, $\Vco$ is allocated to the activated elements. Thereby, the formulation allows for the use of larger time steps and, hence, reduces computation time.


\section{Cathode modeling}
\label{sec:cathode}

The driving force of the electric current is the potential difference $\delv$ which is applied between cathode and anode. The domain of the cathode is characterized by the property $v = 0$. To describe moving cathode geometries with arbitrary shapes, the model has to account for the time varying Dirichlet boundary conditions $\widetilde{v}(t)$, i.e.~to prescribe a zero electric potential on the surface and within the tool's geometry. Here, it is particularly advantageous that the new modeling approach does not require any mesh adaptation during the simulation despite the time dependent boundary conditions.

\subsection{Method A}
\label{ssec:methodA}

\begin{figure}[htbp] 
     \centering 
     \begin{subfigure}{.45\textwidth} 
         \centering 
         \begin{tikzpicture}
           \node[inner sep=0pt] (pic) at (0,0) {\includegraphics[width=\textwidth]
           {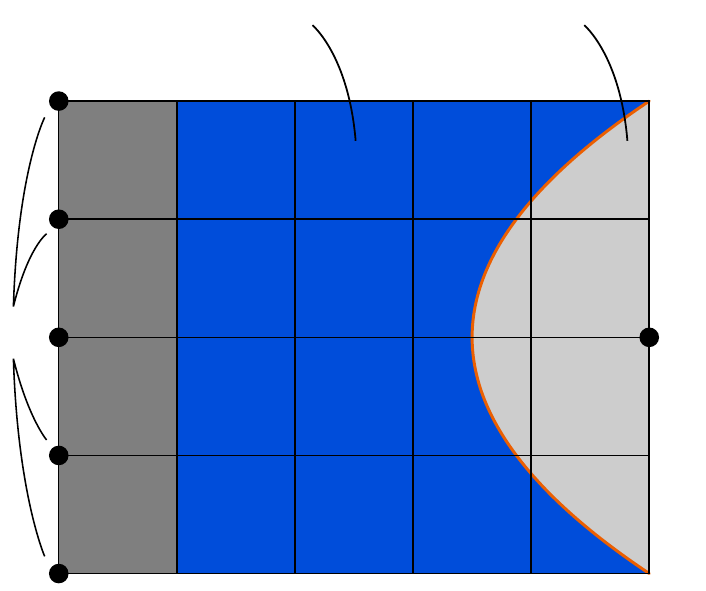}};
           \node[inner sep=0pt] (el)    at ($(pic.north)+(-1.40,-0.25)$)  {electrolyte};
           \node[inner sep=0pt] (ca)    at ($(pic.north)+( 1.60,-0.25)$)  {cathode};
           \node[inner sep=0pt] (vca)   at ($(pic.east)+(-0.10,-0.40)$)  {$\widetilde{v}_\mathrm{ca}$};
           \node[inner sep=0pt] (van)   at ($(pic.west)+( 0.15,-0.40)$)  {$\widetilde{v}_\mathrm{an}$};
           \node[inner sep=0pt] (kec)   at ($(pic.east)+(-1.15,-1.10)$)  {$\kE\CAT$};
           \node[inner sep=0pt] (keb)   at ($(pic.east)+(-2.50,-1.10)$)  {\textcolor{sfb3}{$\kEbar$}};
           \node[inner sep=0pt] (kee)   at ($(pic.east)+(-3.55,-1.10)$)  {\textcolor{white}{$\kE\EL$}};
           \node[inner sep=0pt]         at ($(pic.east)+(-4.75,-1.10)$)  {\textcolor{white}{$\kE\EL$}};
           \node[inner sep=0pt] (kea)   at ($(pic.east)+(-6.00,-1.10)$)  {$\kE\ME$};
         \end{tikzpicture} 
         \caption{Time $t_0$}
         \label{fig:MethodAt1}
     \end{subfigure}
     \qquad
     \begin{subfigure}{.45\textwidth} 
         \centering 
         \begin{tikzpicture}
           \node[inner sep=0pt] (pic) at (0,0) {\includegraphics[width=\textwidth]
           {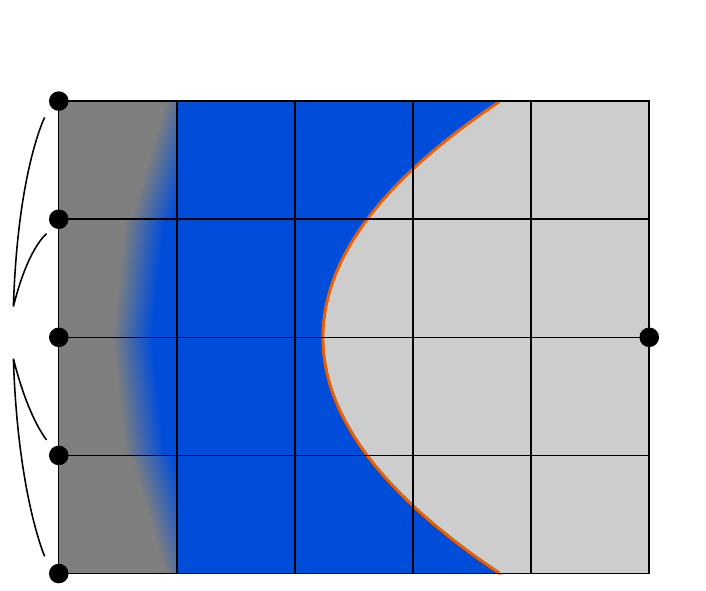}};
           \node[inner sep=0pt] (vca)   at ($(pic.east)+(-0.10,-0.40)$)  {$\widetilde{v}_\mathrm{ca}$};
           \node[inner sep=0pt] (van)   at ($(pic.west)+( 0.15,-0.40)$)  {$\widetilde{v}_\mathrm{an}$};
           \node[inner sep=0pt] (kec)   at ($(pic.east)+(-2.35,-1.10)$)  {$\kE\CAT$};
           \node[inner sep=0pt]         at ($(pic.east)+(-1.15,-1.10)$)  {$\kE\CAT$};
           \node[inner sep=0pt] (keb)   at ($(pic.east)+(-3.50,-1.10)$)  {\textcolor{sfb3}{$\kEbar$}};
           \node[inner sep=0pt] (kee)   at ($(pic.east)+(-4.755,-1.10)$) {\textcolor{white}{$\kE\EL$}};
           \node[inner sep=0pt] (kea)   at ($(pic.east)+(-6.10,-1.10)$)  {$\kEbar$};
         \end{tikzpicture} 
         \caption{Time $t_1$}
         \label{fig:MethodAt2}
     \end{subfigure}
     \caption{Illustration of method A. The cathode moves towards the anode from time $t_0$ to time $t_1$. The electric potential $v$ is fixed at one node within the cathode and the cathode feed is modeled by assigning each element within the cathode an infinite electric conductivity $\kE\CAT$. Exemplarily, the electric conductivity in a horizontal line of elements is shown at $t_0$ and $t_1$ and a black dot denotes a fixed degree of freedom.} 
     \label{fig:MethodA}
     
     \vspace*{10mm}

     \centering 
     \begin{subfigure}{.45\textwidth} 
         \centering 
         \begin{tikzpicture}
           \node[inner sep=0pt] (pic) at (0,0) {\includegraphics[width=\textwidth]
           {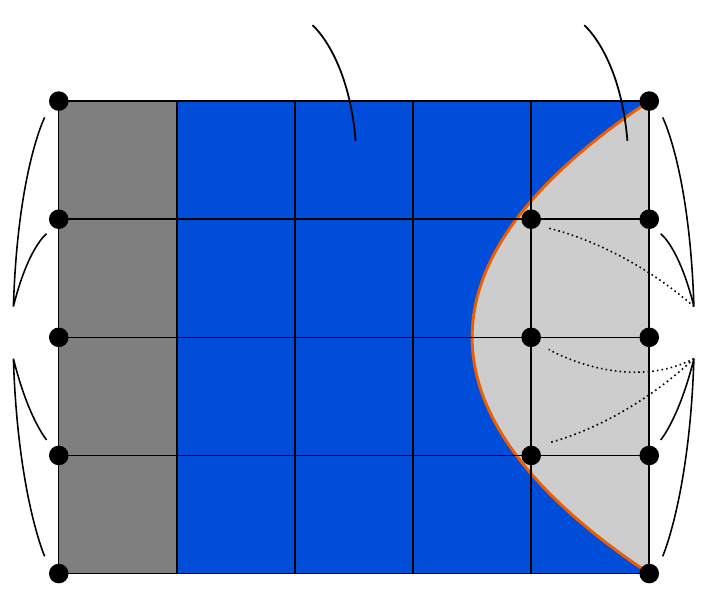}};
           \node[inner sep=0pt] (el)    at ($(pic.north)+(-1.40,-0.25)$)  {electrolyte};
           \node[inner sep=0pt] (ca)    at ($(pic.north)+( 1.60,-0.25)$)  {cathode};
           \node[inner sep=0pt] (vca)   at ($(pic.east)+(-0.10,-0.40)$)  {$\widetilde{v}_\mathrm{ca}$};
           \node[inner sep=0pt] (van)   at ($(pic.west)+( 0.15,-0.40)$)  {$\widetilde{v}_\mathrm{an}$};
           \node[inner sep=0pt] (keb)   at ($(pic.east)+(-2.50,-1.10)$)  {\textcolor{sfb3}{$\kEbar$}};
           \node[inner sep=0pt] (kee)   at ($(pic.east)+(-3.55,-1.10)$)  {\textcolor{white}{$\kE\EL$}};
           \node[inner sep=0pt]         at ($(pic.east)+(-4.75,-1.10)$)  {\textcolor{white}{$\kE\EL$}};
           \node[inner sep=0pt] (kea)   at ($(pic.east)+(-6.00,-1.10)$)  {$\kE\ME$};
         \end{tikzpicture} 
         \caption{Time $t_0$}
         \label{fig:MethodBt1}
     \end{subfigure}
     \qquad
     \begin{subfigure}{.45\textwidth} 
         \centering 
         \begin{tikzpicture}
           \node[inner sep=0pt] (pic) at (0,0) {\includegraphics[width=\textwidth]
           {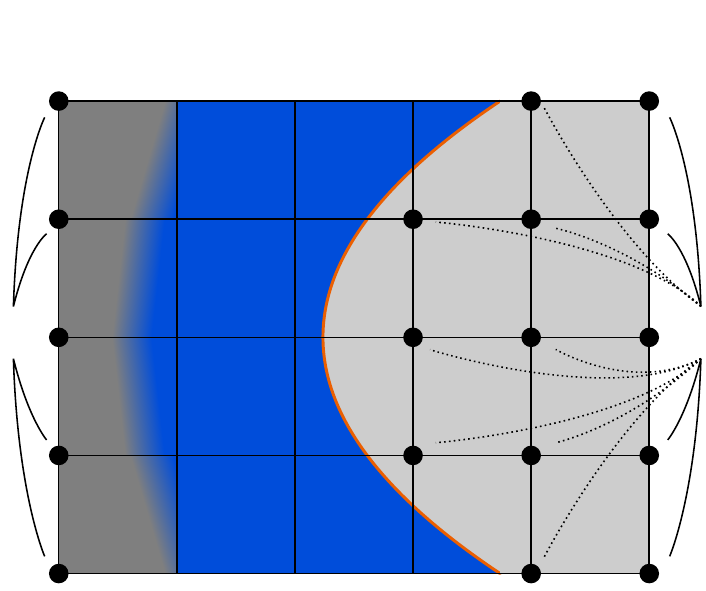}};
           \node[inner sep=0pt] (vca)   at ($(pic.east)+(-0.10,-0.40)$)  {$\widetilde{v}_\mathrm{ca}$};
           \node[inner sep=0pt] (van)   at ($(pic.west)+( 0.15,-0.40)$)  {$\widetilde{v}_\mathrm{an}$};
           \node[inner sep=0pt] (keb)   at ($(pic.east)+(-3.50,-1.10)$)  {\textcolor{sfb3}{$\kEbar$}};
           \node[inner sep=0pt] (kee)   at ($(pic.east)+(-4.755,-1.10)$) {\textcolor{white}{$\kE\EL$}};
           \node[inner sep=0pt] (kea)   at ($(pic.east)+(-6.10,-1.10)$)  {$\kEbar$};
         \end{tikzpicture} 
         \caption{Time $t_1$}
         \label{fig:MethodBt2}
     \end{subfigure}
     \caption{Illustration of method B. The cathode moves towards the anode from time $t_0$ to time $t_1$. The electric potential $v$ is fixed at all nodes within the cathode and the cathode feed is modeled by assigning additional Dirichlet boundary conditions. A black dot denotes a fixed degree of freedom.} 
     \label{fig:MethodB}
\end{figure}

Two different methods are employed to model the aforementioned condition. In the first, method A, the electric conductivities of the finite elements within the cathode are modified and the corresponding values are set numerically to infinity (Fig.~\ref{fig:MethodA}). The electric resistance of these elements is numerically zero and no change of the electric potential can be observed within the cathode. Thus, the equipotential line with $v = 0$ coincides with the cathode's surface as long as the electric potential $v$ is prescribed at one point within the cathode. The advantage of method A is the straightforward implementation, since no modification of the global system of equations is required. This approach is inspired by \cite{HardistyMilehamEtAl1993} who assign to each element a different material type depending on its classification as tool, workpiece or electrolyte.

\subsection{Method B}
\label{ssec:methodB}

In the second, method B, a zero electric potential is applied directly on all nodes within the cathode (Fig.~\ref{fig:MethodB}). To account for the cathode feed, this approach requires an adaptation and a renumbering of the global system of equations in every time step. However, this procedure is computationally still beneficial, since the number of degrees of freedom reduces for regular cathode geometries steadily throughout the simulation when the number of fixed degrees of freedom inside the cathode increases. Furthermore, method B allows for the investigation of special cathode shapes, e.g.~electrochemical wirecutting where the zero electric potential is defined only in the thin wire domain.

\subsection{Effective material parameters}

\begin{figure}[htbp] 
  \centering 
         \begin{tikzpicture}
           \node[inner sep=0pt] (pic) at (0,0) {\includegraphics[width=0.3\textwidth]
           {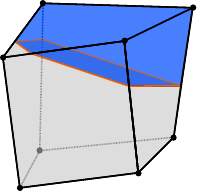}};
           \node[inner sep=0pt] (vc)    at ($(pic.south)+( -0.40, +1.90)$)  {$\Vcat$};
         \end{tikzpicture} 
         \caption{A finite element that is partially cathode and partially electrolyte. $\Vcat$ defines the finite element's volume inside the cathode and the orange line surrounds the cathode's surface.}
         \label{fig:cutelement}
\end{figure}

Both models follow the conceptual approach of the dissolution model and work for a constant finite element mesh. Naturally, the cathode's surface does not necessarily coincide with the finite elements' edges. Thus, elements exist that are partially electrolyte and partially cathode (see Fig.~\ref{fig:cutelement}). Analogously to the dissolution level $d$, we define the cathode ratio
\begin{equation}
  \lc = \frac{\Vcat}{\Vel}
\end{equation}
that states the volume of the cathode in a finite element $\Vcat$ compared to the total volume of this element $\Vel$. The polyeder's volume $\Vcat$ is computed via tetrahedralization using a Delaunay triangulation algorithm developed by \cite{Joe1991}. Moreover, following the definition of Eqs.~\eqref{eq:mixanpar} and \eqref{eq:mixanser}, we define the effective material parameters of partial cathode elements for a parallel connection as
\begin{equation}
  \bar{(\bullet)}^\mathrm{p} = \left( 1 - \lc \right) \, (\bullet)^\mathrm{\scriptscriptstyle EL} \, + \, \lc \,\, (\bullet)^\mathrm{\scriptscriptstyle CAT}
  \label{eq:mixcapar}
\end{equation}
and for a series connection respectively as
\begin{equation}
  \bar{(\bullet)}^\mathrm{s} = \left[ \, \left( 1 - \lc \right) \, / \, (\bullet)^\mathrm{\scriptscriptstyle EL} \, + \, \lc \, / \, (\bullet)^\mathrm{\scriptscriptstyle CAT} \, \right]^{-1}
  \label{eq:mixcaser}
\end{equation}
where $(\bullet)^\mathrm{\scriptscriptstyle EL}$ again denotes the material parameters of the electrolyte and $(\bullet)^\mathrm{\scriptscriptstyle CAT}$ those of the cathode.

\begin{figure}[htbp]
  \centering
  \includegraphics[width=0.9\textwidth]{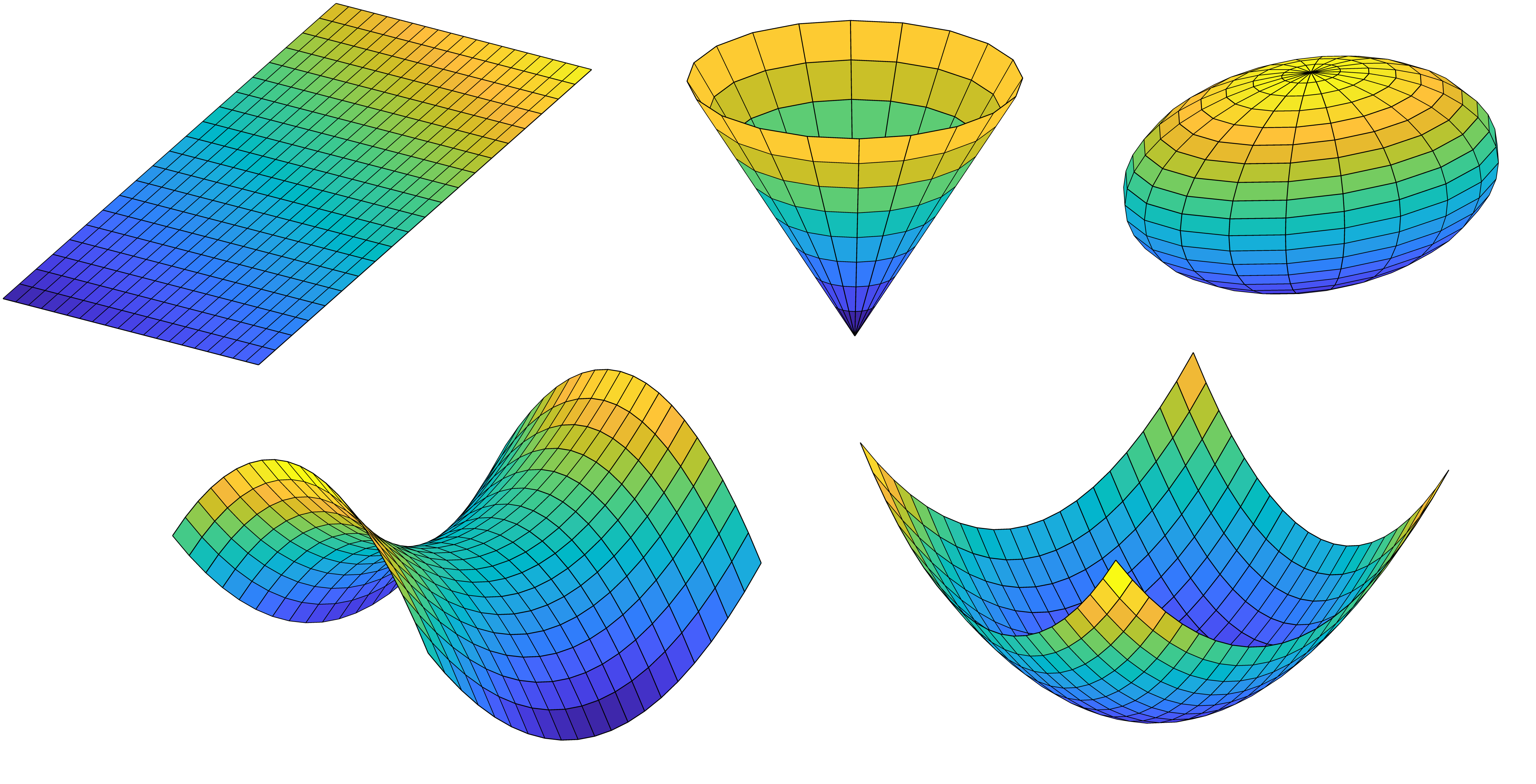}
  \caption{Basic geometries used for the cathode assembly: Plane, cone, ellipsoid and paraboloids.}
  \label{fig:geometries}
\end{figure}

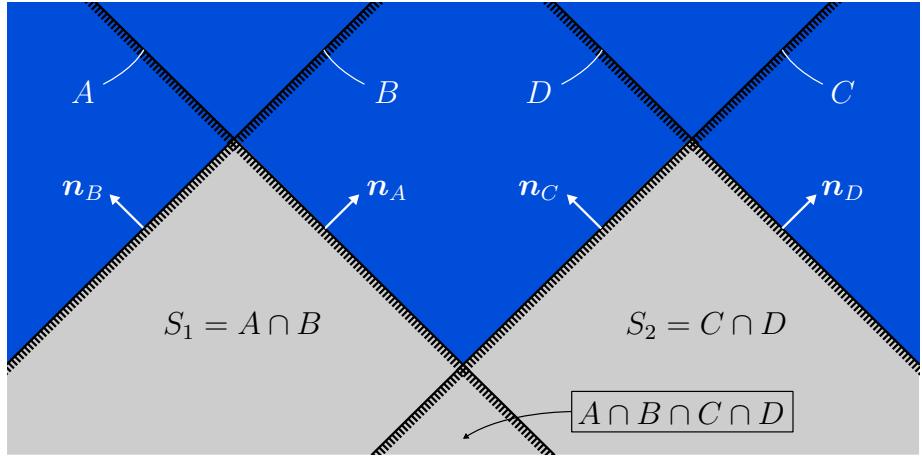
\begin{figure}[htbp]
  \centering
  \begin{tikzpicture}
           \node[inner sep=0pt] (pic) at (0,0) {\includegraphics[width=0.75\textwidth]
           {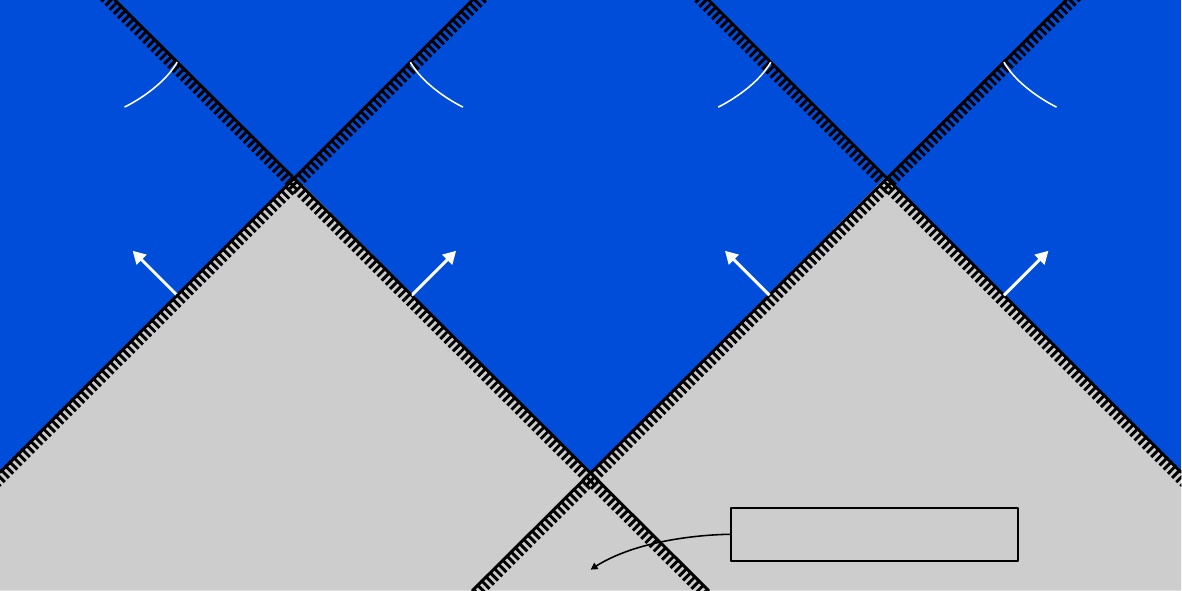}};
           \node[inner sep=0pt] (A)       at ($(pic.north)+(-5.00,-1.20)$)  {$\textcolor{white}{A}$};
           \node[inner sep=0pt] (B)       at ($(pic.north)+(-1.00,-1.20)$)  {$\textcolor{white}{B}$};
           \node[inner sep=0pt] (C)       at ($(pic.north)+( 5.00,-1.20)$)  {$\textcolor{white}{C}$};
           \node[inner sep=0pt] (D)       at ($(pic.north)+( 1.00,-1.20)$)  {$\textcolor{white}{D}$};
           \node[inner sep=0pt] (AnB)     at ($(pic.north)+(-2.90,-4.30)$)  {$S_1 = A \cap B$};
           \node[inner sep=0pt] (CnD)     at ($(pic.north)+( 3.20,-4.30)$)  {$S_2 = C \cap D$};
           \node[inner sep=0pt] (AnBCD)   at ($(pic.north)+( 2.87,-5.45)$)  {$A \cap B \cap C \cap D$};
           \node[inner sep=0pt] (nA)      at ($(pic.north)+(-1.00,-2.50)$)  {$\textcolor{white}{\n_A}$};
           \node[inner sep=0pt] (nB)      at ($(pic.north)+(-5.00,-2.50)$)  {$\textcolor{white}{\n_B}$};
           \node[inner sep=0pt] (nC)      at ($(pic.north)+( 1.00,-2.50)$)  {$\textcolor{white}{\n_C}$};
           \node[inner sep=0pt] (nD)      at ($(pic.north)+( 5.00,-2.50)$)  {$\textcolor{white}{\n_D}$};
         \end{tikzpicture} 
  \caption{Illustration of a cathode assembly by conjunction (intersection) and disjunction (union). The subsets $S_1$ and $S_2$ are obtained by the conjunction of set $A$ and $B$ and set $C$ and $D$, respectively. The disjunction of all subsets ($S_1 \cup S_2$) yields the total cathode geometry. A conjunction of set $A$, $B$, $C$ and $D$ without the definition of subsets would yield a different cathode geometry ($A \cap B \cap C \cap D \neq S_1 \cup S_2 $).}
  \label{fig:con_dis}
\end{figure}

The mathematical description of the cathode geometry follows set theory. Fig.~\ref{fig:geometries} shows different basic geometries that can be combined by conjunction in subsets. These subsets are combined by disjunction to obtain the total cathode geometry (see Fig.~\ref{fig:con_dis}). This two step procedure is necessary to fulfill the distributive property of disjunctions over conjunctions. The position vector which is inherent in each geometry models the cathode feed by incrementally updating its coordinates in every time step. The implementation features also additional options like the rotation of a geometry or the consideration of either the enclosed volume of a geometry or the volume outside of a geometry.

In a previous work, the theorem of intersecting lines served to model the cathode feed by a simplified approach. However, this procedure was restricted to processes with planar cathodes where the initial working gap width coincided with the gap width of the dynamic equilibrium state for a given feed rate. The new methodology allows for the investigation of arbitrary processes with varying process parameters as shown in Section~\ref{sec:examples}.

Fig.~\ref{fig:resulting_mbvp}, finally, provides the resulting set of equations for the presented model.

\begin{figure}[htbp]
  \centering
  \fbox{
    \begin{minipage}{0.8\textwidth}
      $\bullet$ Balance equation and boundary conditions
      \begin{align*}
        \rhoEdot + \div{\j} &= 0                 \hspace{11.1mm} \mathrm{in} \,\, \Omega    \hspace{30mm}                              \rhoE = \div{\D} \\
        v                   &= \widetilde{v}(t)  \hspace{06.1mm} \mathrm{on} \,\, \Gamma_v  \hspace{10mm} \mathrm{with} \hspace{9.5mm} \D = \epsnull \, \epsrbar \, \E \\
        \j \cdot \n         &= \widetilde{j}     \hspace{11.2mm} \mathrm{on} \,\, \Gamma_j  \hspace{28mm}                              \E = - \grad{v}
      \end{align*}
      $\bullet$ Constitutive law
      \begin{equation*}
        \j = \kEbar \, \E + \epsnull \, \epsrbar \, \Edot
      \end{equation*}
      $\bullet$ Effective material parameters anode
      \begin{equation*}
        \bar{(\bullet)}^\mathrm{p} = \left( 1 - d \right) \, (\bullet)\ME \, + \, d \,\, (\bullet)\EL   \vspace{-2mm}
      \end{equation*}
      \begin{equation*}
        \bar{(\bullet)}^\mathrm{s} = \left[ \, \left( 1 - d \right) \, / \, (\bullet)\ME \, + \, d \, / \, (\bullet)\EL \, \right]^{-1}
      \end{equation*}
      $\bullet$ Effective material parameters cathode
      \begin{equation*}
        \bar{(\bullet)}^\mathrm{p} = \left( 1 - \lc \right) \, (\bullet)^\mathrm{\scriptscriptstyle EL} \, + \, \lc \,\, (\bullet)^\mathrm{\scriptscriptstyle CAT}   \vspace{-2mm}
      \end{equation*}
      \begin{equation*}
        \bar{(\bullet)}^\mathrm{s} = \left[ \, \left( 1 - \lc \right) \, / \, (\bullet)^\mathrm{\scriptscriptstyle EL} \, + \, \lc \, / \, (\bullet)^\mathrm{\scriptscriptstyle CAT} \, \right]^{-1}
      \end{equation*}
    \end{minipage}
  }
  \caption{Resulting set of equations for the moving boundary value problem.}
  \label{fig:resulting_mbvp}
\end{figure}


\section{Numerical examples}
\label{sec:examples}

In this section, we employ the previously developed methodology to model the cathode feed in different applications. Furthermore, we compare the runtimes of method A and B. Analytical and experimental reference solutions serve to validate the methodology's performance and accuracy. Finally, we investigate an electrochemical blade manufacturing process.

The material parameters employed in the simulations stem from Table~\ref{tab:matpar}, if not explicitly stated otherwise. For simplicity, we consider only isothermal problems with a constant temperature of $50~\si{K}$.

\begin{table}[h]
  \centering
  \caption{Material parameters and physical constants}
  \vspace{-3mm}
  \begin{tabular}{lll}
      \hline
      Symbol             & Unit                                           & Value                               \\
      \hline \hline
      $\kE\CAT$          & $[ \si{\ampere\per\volt\per\meter} ]$          & $ 1.000 \times 10^{12\hphantom{-}}$ \\
      $\kE\EL$           & $[ \si{\ampere\per\volt\per\meter} ]$          & $ 1.600 \times 10^{1\hphantom{-0}}$ \\
      $\kE\ME$           & $[ \si{\ampere\per\volt\per\meter} ]$          & $ 4.625 \times 10^{6\hphantom{-0}}$ \\
      $\epsr\CAT$        & $[ - ]$                                        & $ 1.000 \times 10^{0\hphantom{-0}}$ \\
      $\epsr\EL$         & $[ - ]$                                        & $ 8.000 \times 10^{1\hphantom{-0}}$ \\
      $\epsr\ME$         & $[ - ]$                                        & $ 1.000 \times 10^{0\hphantom{-0}}$ \\
      $\epsnull$         & $[ \si{\ampere\s\per\volt\per\m} ]$            & $ 8.854 \times 10^{-12\hphantom{}}$ \\
      $\Veffnew$         & $[ \si{\cubic\meter\per\ampere\per\second} ]$  & $ 1.000 \times 10^{-11\hphantom{}}$ \\
      \hline
  \end{tabular}
  \label{tab:matpar}
\end{table}


\subsection{Planar cathode - analytical validation}
\label{ssec:Ex1}

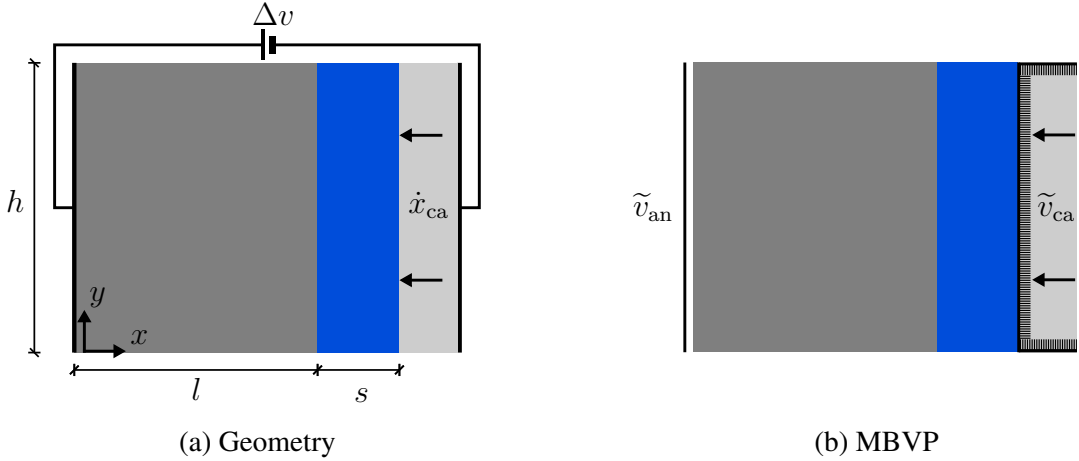
\begin{figure}[htbp] 
     \centering 
     \begin{subfigure}{.45\textwidth} 
         \centering 
         \begin{tikzpicture}
           \node[inner sep=0pt] (pic) at (0,0) {\includegraphics[width=\textwidth]
           {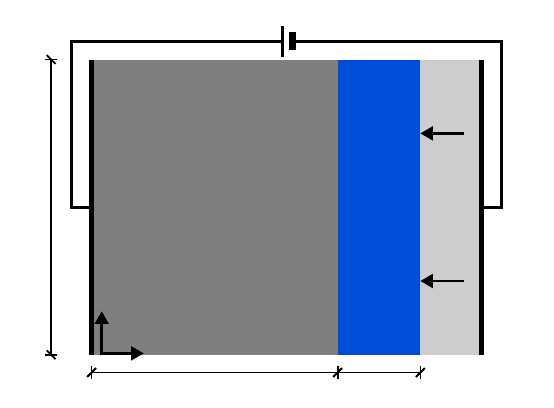}};
           \node[inner sep=0pt] (h)    at ($(pic.west) +( 0.42, 0.05)$)  {$h$};
           \node[inner sep=0pt] (l)    at ($(pic.south)+(-0.80, 0.21)$)  {$l$};
           \node[inner sep=0pt] (s)    at ($(pic.south)+( 1.37, 0.13)$)  {$s$};
           \node[inner sep=0pt] (x)    at ($(pic.south)+(-1.55, 0.90)$)  {$x$};
           \node[inner sep=0pt] (y)    at ($(x.north)  +(-0.55, 0.43)$)  {$y$};
           \node[inner sep=0pt] (xdc)  at ($(pic.east) +(-1.35, 0.00)$)  {$\dot{x}_\mathrm{ca}$};
           \node[inner sep=0pt] (dv)   at ($(pic.south)+( 0.21, 5.18)$)  {$\delv$};
         \end{tikzpicture} 
         \caption{Geometry}
         \label{fig:Ex1Geom}
     \end{subfigure}
     \qquad
     \begin{subfigure}{.45\textwidth} 
         \centering 
         \begin{tikzpicture}
           \node[inner sep=0pt] (pic) at (0,0) {\includegraphics[width=\textwidth]
           {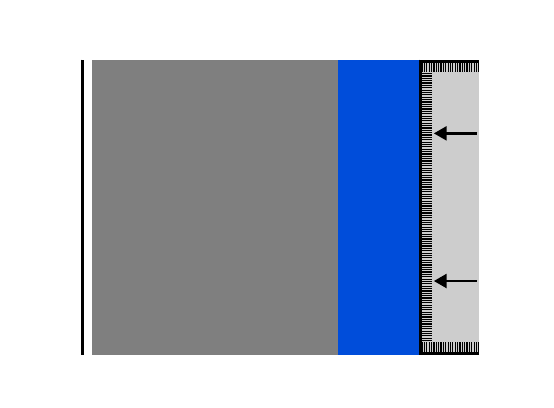}};
           \node[inner sep=0pt] (van)   at ($(pic.north)+(-2.94,-2.70)$)  {$\widetilde{v}_\mathrm{an}$};
           \node[inner sep=0pt] (vca)   at ($(van.west) +( 5.58, 0.00)$)  {$\widetilde{v}_\mathrm{ca}$};
         \end{tikzpicture} 
         \caption{MBVP}
         \label{fig:Ex1MBVP}
     \end{subfigure} 
     \caption{Geometry and moving boundary value problem for a planar cathode.} 
     \label{fig:Ex1}
\end{figure}

In the first example, we discuss a planar cathode (see Fig.~\ref{fig:Ex1}) with an analytical reference solution analogously to \cite{vanderVeldenRommesEtAl2021}. The dimensions read $l = h = 1~\si{\mm}$ with a thickness of $0.1~\si{\mm}$ and a constant feed rate of $\dot{x}_\mathrm{ca} = 0.01\si{\mm\per\s}$ is employed with a potential difference of $\delv = 20~\si{\V}$.

\begin{figure}[htbp]
  \centering
  \begin{tikzpicture}
      \begin{semilogxaxis}[
                  xlabel= $\dt$  \si{[\s]},
                  ylabel= $\left. V_\mathrm{dis}^\mathrm{FE}/V_\mathrm{dis}^\mathrm{analyt.} \right|_{t \, = \, 60~\si{s}} \si{[-]}$,
                  x dir=reverse,
                  legend pos=outer north east,legend cell align={left},
                  xmin = 0.0005,
                  xmax = 2,
                  ymin = 0.95,
                  ymax = 1.2,
                  xtick={0.001, 0.01, 0.1, 1},
                  xticklabels={$10^{-3}$, $10^{-2}$, $10^{-1}$,$1$}]
                  
          \addplot[mark=none, black, forget plot] coordinates {(10,1) (1E-4,1)};
          
          \addplot[color = sfb4, line width=1.5pt, mark=triangle*, mark options={solid}, dashpattern1]
                   table[x index = 0, y expr=\thisrowno{2}/(6.0000E-11)] 
                   {02_Figures/02_Pgf/01_Ex1/01_Structured_Meshes/02_Ansatz_DirBC/Vdis_dt.txt};
                   \addlegendentry{$10 \times 10$ parallel}
          \addplot[color = sfb4, line width=1.5pt, mark=square*, mark options={solid}, dashpattern1]
                   table[x index = 0, y expr=\thisrowno{6}/(6.0000E-11)] 
                   {02_Figures/02_Pgf/01_Ex1/01_Structured_Meshes/02_Ansatz_DirBC/Vdis_dt.txt};
                   \addlegendentry{$20 \times 20$ parallel}
          \addplot[color = sfb4, line width=1.5pt, mark=diamond*, mark options={solid}, dashpattern1]
                   table[x index = 0, y expr=\thisrowno{10}/(6.0000E-11)] 
                   {02_Figures/02_Pgf/01_Ex1/01_Structured_Meshes/02_Ansatz_DirBC/Vdis_dt.txt};
                   \addlegendentry{$40 \times 40$ parallel}
          \addplot[color = sfb4, line width=1.5pt, mark=x, mark size={1.25mm}, mark options={solid}, dashpattern1]
                   table[x index = 0, y expr=\thisrowno{14}/(6.0000E-11)] 
                   {02_Figures/02_Pgf/01_Ex1/01_Structured_Meshes/02_Ansatz_DirBC/Vdis_dt.txt};
                   \addlegendentry{$80 \times 80$ parallel}
          \addplot[color = sfb1, line width=1.5pt, mark=triangle*, dashpattern0]
                   table[x index = 0, y expr=\thisrowno{4}/(6.0000E-11)] 
                   {02_Figures/02_Pgf/01_Ex1/01_Structured_Meshes/02_Ansatz_DirBC/Vdis_dt.txt};
                   \addlegendentry{$10 \times 10$ series}
          \addplot[color = sfb1, line width=1.5pt, mark=square*, dashpattern0]
                   table[x index = 0, y expr=\thisrowno{8}/(6.0000E-11)] 
                   {02_Figures/02_Pgf/01_Ex1/01_Structured_Meshes/02_Ansatz_DirBC/Vdis_dt.txt};
                   \addlegendentry{$20 \times 20$ series}
          \addplot[color = sfb1, line width=1.5pt, mark=diamond*, dashpattern0]
                   table[x index = 0, y expr=\thisrowno{12}/(6.0000E-11)] 
                   {02_Figures/02_Pgf/01_Ex1/01_Structured_Meshes/02_Ansatz_DirBC/Vdis_dt.txt};
                   \addlegendentry{$40 \times 40$ series}
          \addplot[color = sfb1, line width=1.5pt, mark=x, mark size={1.25mm}, dashpattern0]
                   table[x index = 0, y expr=\thisrowno{16}/(6.0000E-11)] 
                   {02_Figures/02_Pgf/01_Ex1/01_Structured_Meshes/02_Ansatz_DirBC/Vdis_dt.txt};
                   \addlegendentry{$80 \times 80$ series}

      \end{semilogxaxis}
  \end{tikzpicture}
  
  \caption{Comparison of the dissolved volume at $t = 60~\si{s}$ of the finite element simulation and the analytical solution for the parallel and series connection with structured meshes for different element densities per \si{\square\mm}. The simulations with method A and B yield the same results.}
  \label{fig:Ex1_Vdis_str}
\end{figure}
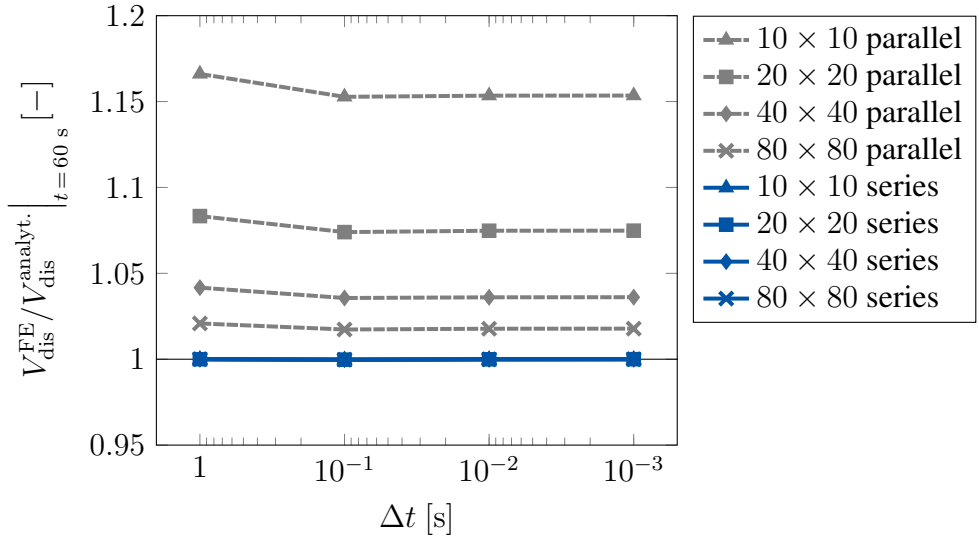

At the beginning, we utilize an initial working gap width of $s_\mathrm{init} = 0.32~\si{\mm}$ that equals the gap width of the dissolution process in the dynamic equilibrium state (cf.~\cite{KlockeKoenig2007}). Fig.~\ref{fig:Ex1_Vdis_str} shows the comparison of the parallel (Eqs.~\eqref{eq:mixanpar} and \eqref{eq:mixcapar}) the series connection (Eqs.~\eqref{eq:mixanser} and \eqref{eq:mixcaser}). The dissolved volume of the finite element simulation after $60~\si{\s}$, which is normalized with respect to the analytical solution, is given for different time increments and element densities. Structured meshes are employed.

Although the results of the parallel connection converge against the analytical solution for fine meshes, they still show an error of $+1.8~\si{\%}$ for the finest discretization.
The electric resistance of a dissolving element is underestimated by the parallel connection in this application, since the electric charges do not necessarily pass the electrolyte resistance in these elements (cf.~Fig.~\ref{fig:Cond}). Thereby, the electric current and the dissolved volume are overestimated by the model.
The series connection, however, yields the correct solution regardless of the time step size and element density. Moreover, method A and B yield identical results.

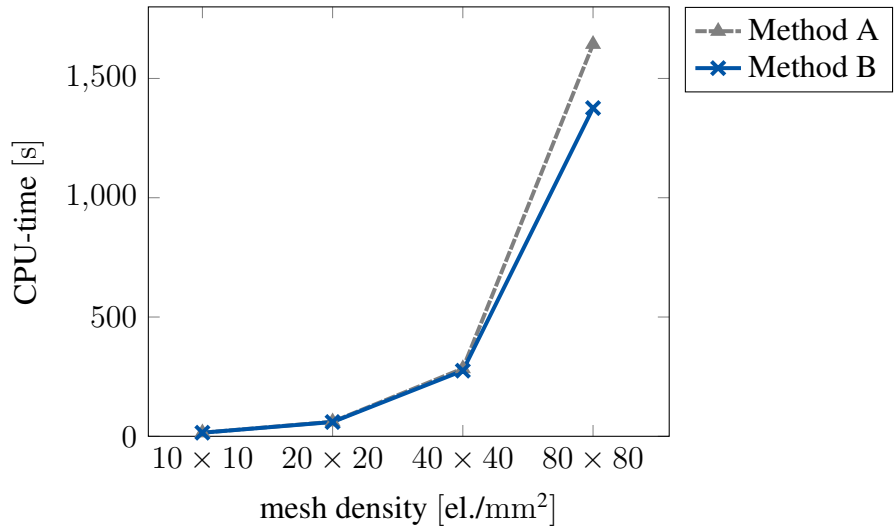
\begin{figure}[htbp]
  \centering
  \begin{tikzpicture}
      \begin{semilogxaxis}[
                  xlabel= mesh density~\text{\si{[}el./\si{\square\mm]}},
                  ylabel= CPU-time \si{[\s]},
                  legend pos=outer north east,legend cell align={left},
                  xmin = 7.5,
                  xmax = 120,
                  ymin = 0,
                  ymax = 1800,
                  xtick={10, 20, 40, 80},
                  xticklabels={$10\times10$, $20\times20$, $40\times40$,$80\times80$}]
                  
          \addplot[color = sfb4, line width=1.5pt, mark=triangle*, mark options={solid}, dashpattern1]
                   table[x index = 0, y index = 2] 
                   {02_Figures/02_Pgf/01_Ex1/01_Structured_Meshes/runtime_ed.txt};
                   \addlegendentry{Method A}                  
          \addplot[color = sfb1, line width=1.5pt, mark=x, mark size={1.25mm}, dashpattern0]
                   table[x index = 0, y index = 6] 
                   {02_Figures/02_Pgf/01_Ex1/01_Structured_Meshes/runtime_ed.txt};
                   \addlegendentry{Method B}
                  
      \end{semilogxaxis}
  \end{tikzpicture}
  \caption{Comparison of the runtime of method A and B. Structured meshes are employed with the series connection and a time increment of $\dt = 10^{-1}~\si{\second}$.}
  \label{fig:Ex1_runtime}
\end{figure}

\begin{figure}[htbp] 
  \centering 
  
  \begin{subfigure}{.32\textwidth} 
      \centering 
      \includegraphics[width=\textwidth]{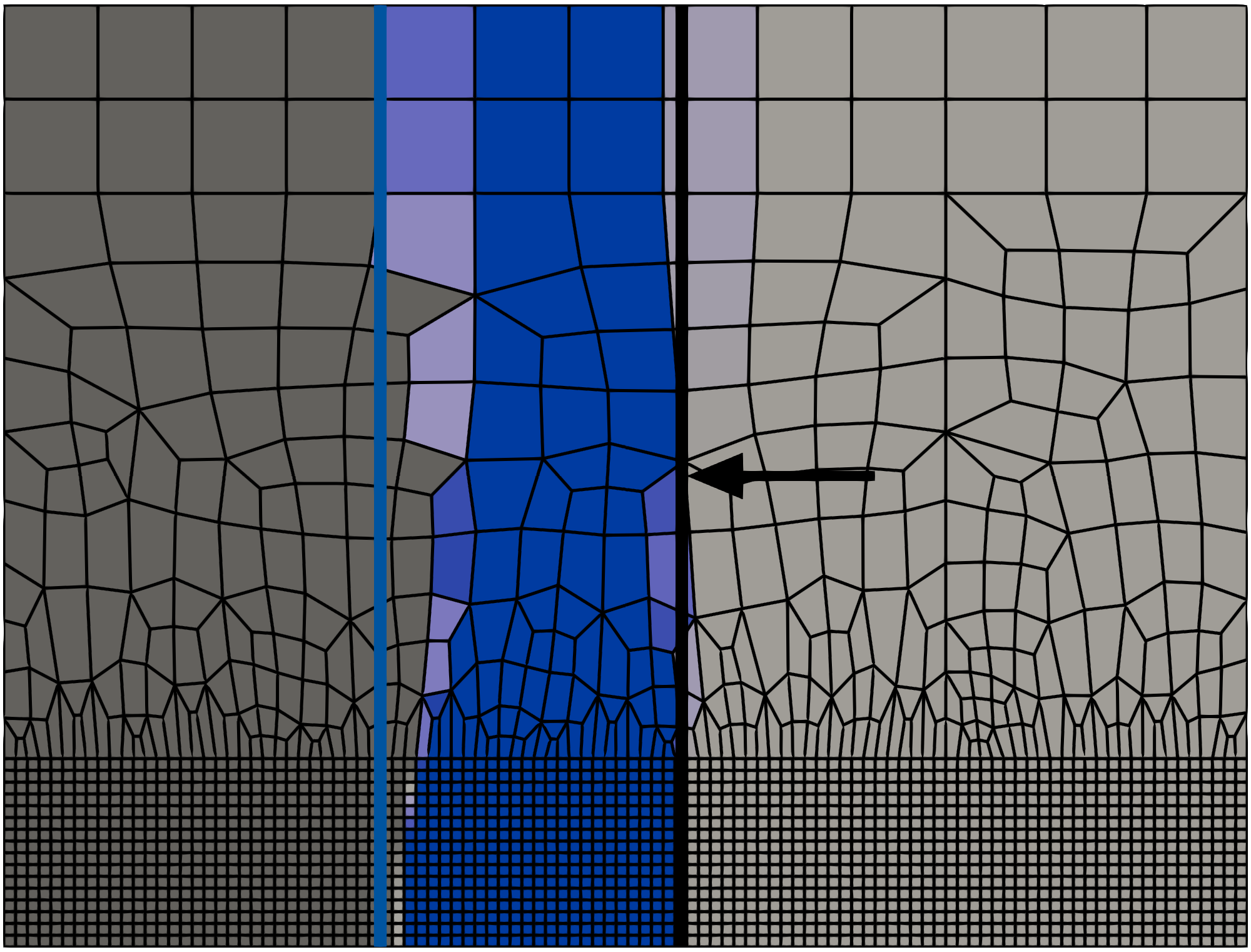}
      \caption{Mesh $10 \downarrow 80$}
      \label{fig:Ex1_dm1ver}
  \end{subfigure}
  \begin{subfigure}{.32\textwidth} 
      \centering 
      \includegraphics[width=\textwidth]{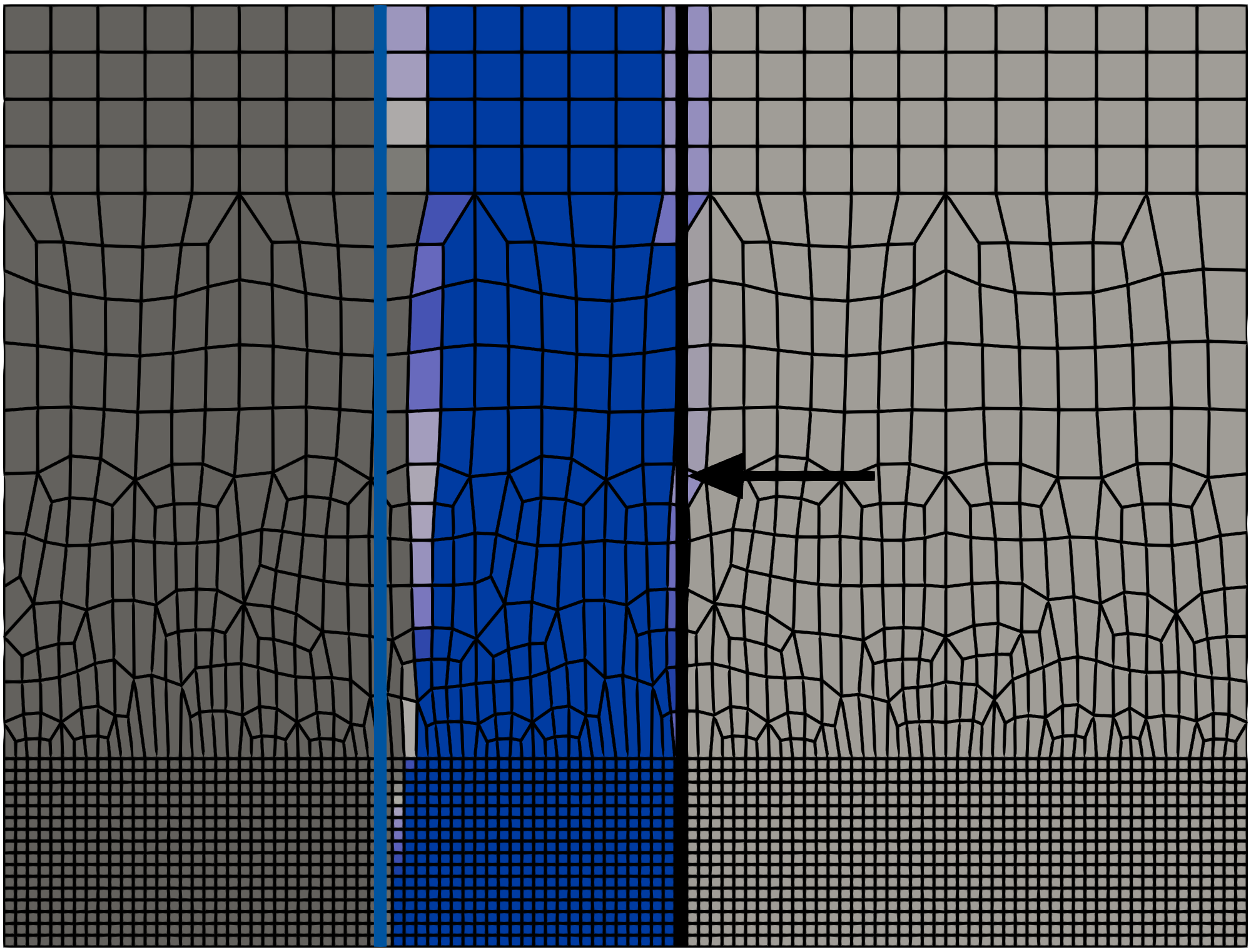}
      \caption{Mesh $20 \downarrow 80$}
      \label{fig:Ex1_dm2ver}
  \end{subfigure}
  \begin{subfigure}{.32\textwidth} 
      \centering 
      \includegraphics[width=\textwidth]{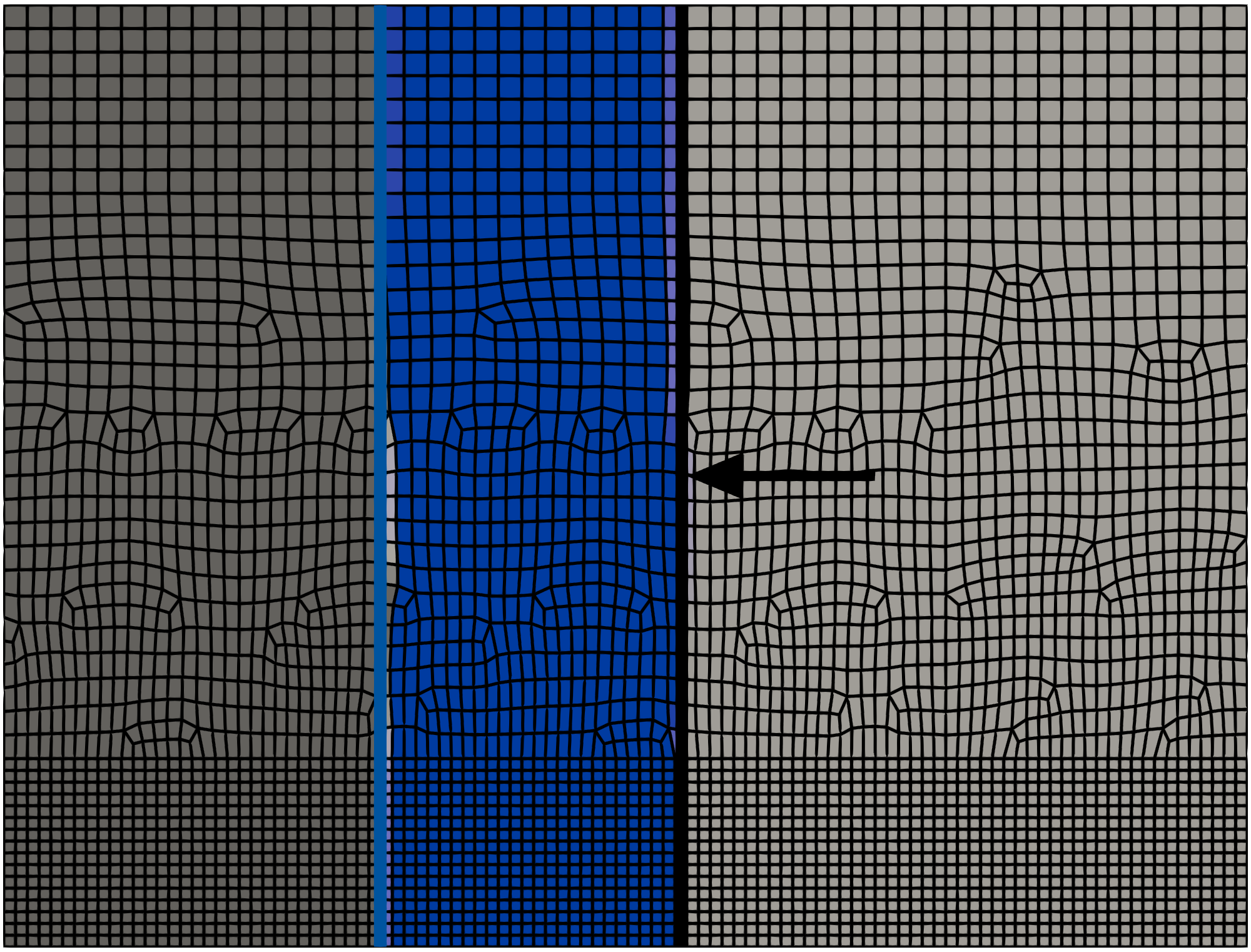}
      \caption{Mesh $40 \downarrow 80$}
      \label{fig:Ex1_dm3ver}
  \end{subfigure}
  
  \vspace{2mm}
  
  \begin{subfigure}{.32\textwidth} 
      \centering 
      \includegraphics[width=\textwidth]{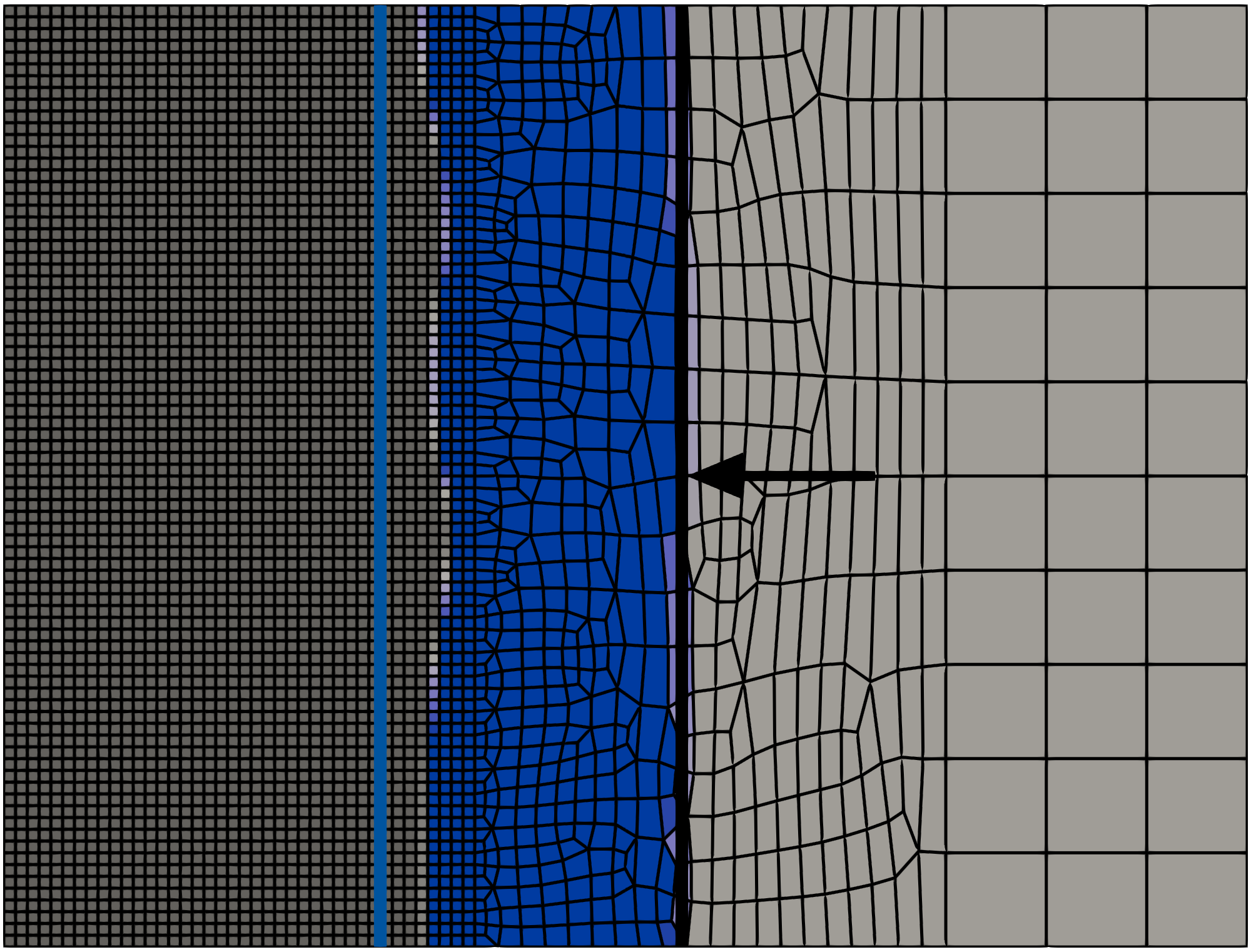}
      \caption{Mesh $80 \rightarrow 10$}
      \label{fig:Ex1_dm4hor}
  \end{subfigure}
  \begin{subfigure}{.32\textwidth} 
      \centering 
      \includegraphics[width=\textwidth]{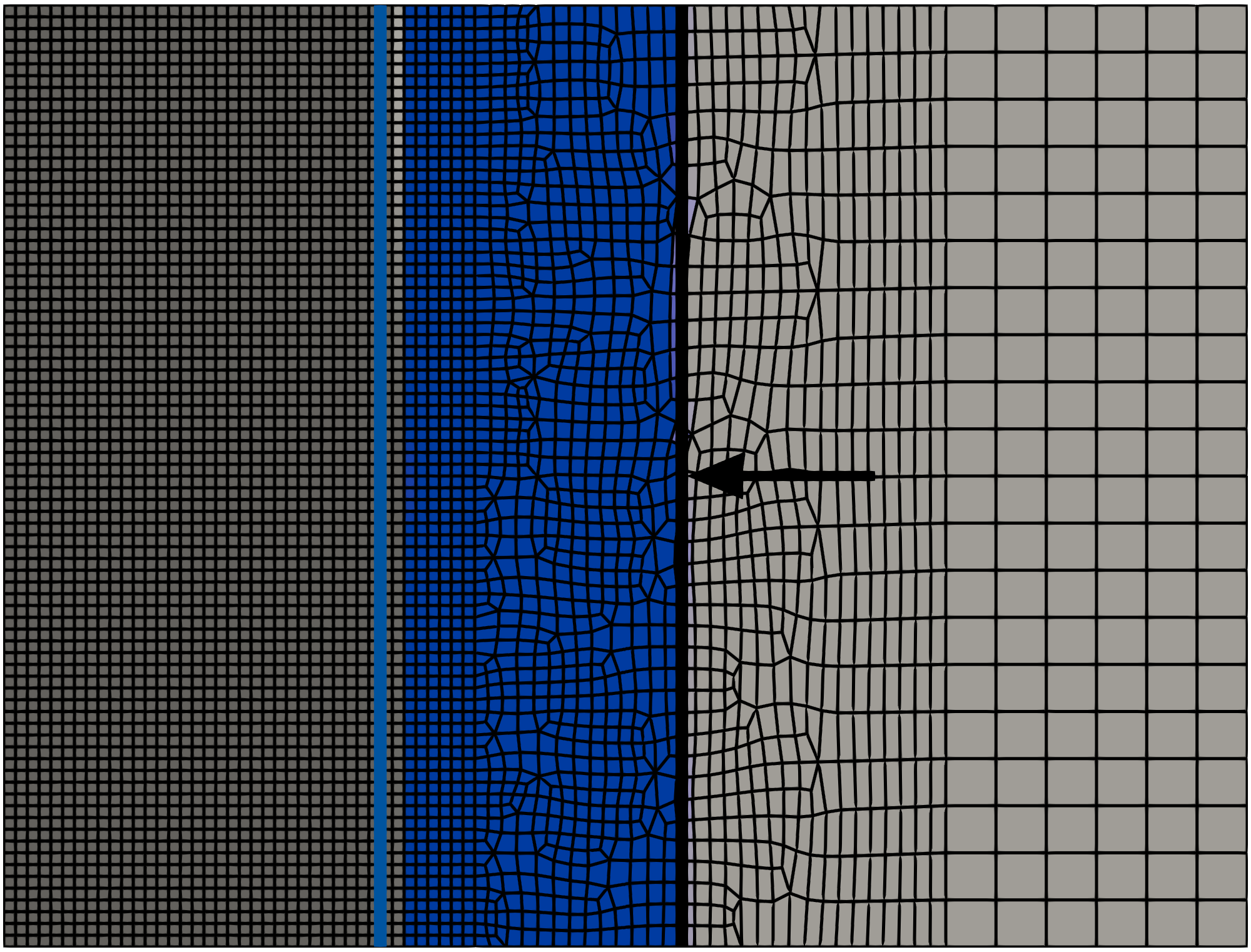}
      \caption{Mesh $80 \rightarrow 20$}
      \label{fig:Ex1_dm5hor}
  \end{subfigure}
  \begin{subfigure}{.32\textwidth} 
      \centering 
      \includegraphics[width=\textwidth]{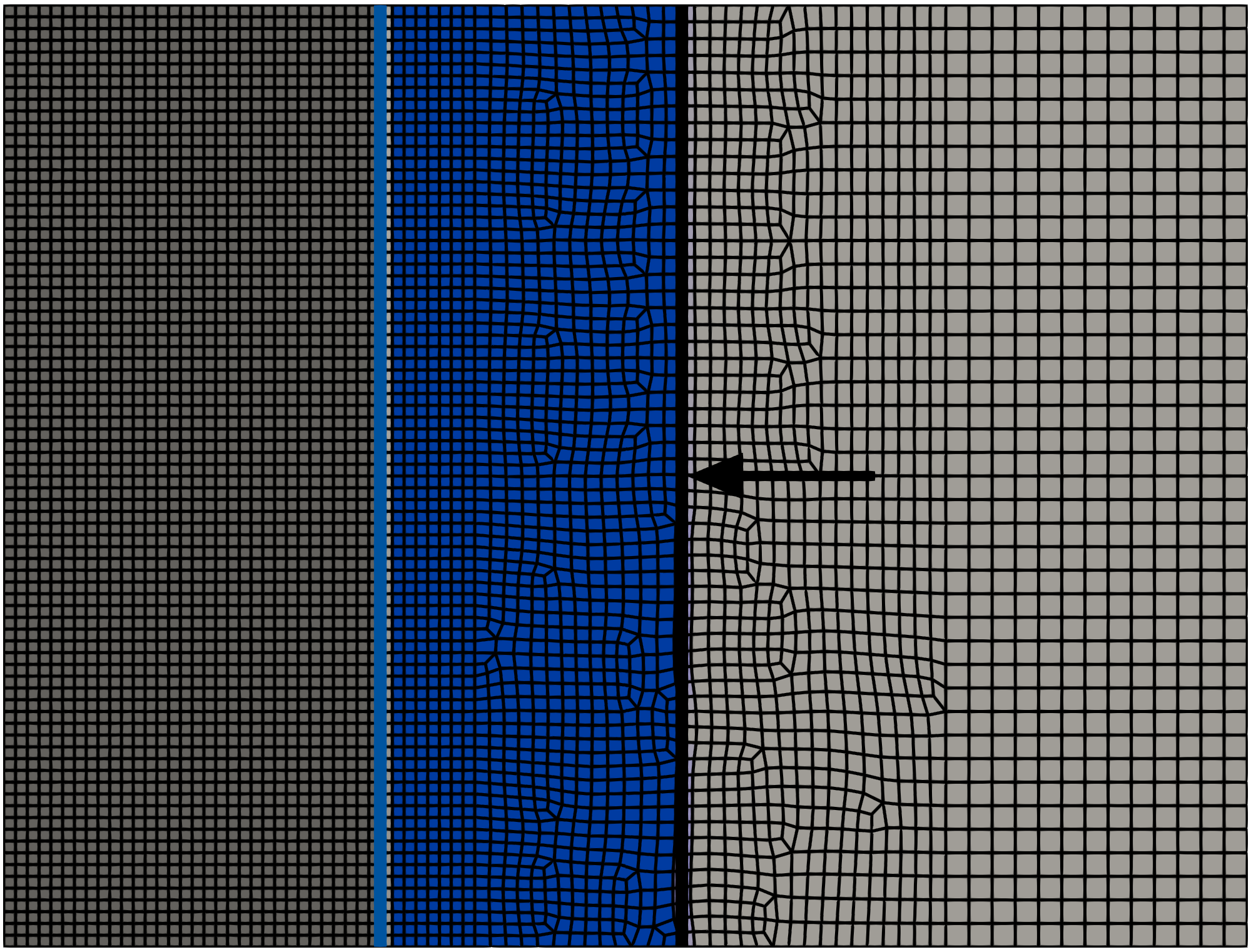}
      \caption{Mesh $80 \rightarrow 40$}
      \label{fig:Ex1_dm6hor}
  \end{subfigure}

  \vspace{3mm}
  \begin{subfigure}{.32\textwidth} 
      \centering 
      \begin{tikzpicture} 
        \node[inner sep=0pt] (pic) at (0,0) {\includegraphics[height=5mm, width=40mm]
        {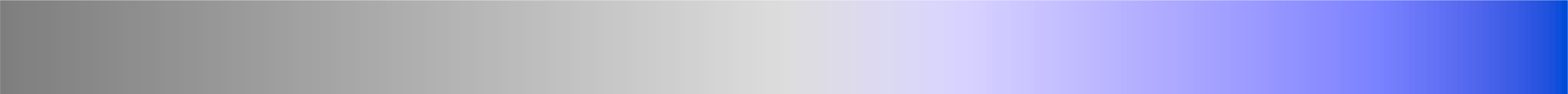}};
        \node[inner sep=0pt] (0)   at ($(pic.south)+(-2.22, 0.26)$)  {$0$};
        \node[inner sep=0pt] (1)   at ($(pic.south)+( 2.22, 0.26)$)  {$1$};
        \node[inner sep=0pt] (d)   at ($(pic.south)+( 0.00, 0.85)$)  {$d~\si{[-]}$};
      \end{tikzpicture} 
  \end{subfigure}
  \begin{subfigure}{.32\textwidth} 
      \centering 
      \begin{tikzpicture} 
        \node[inner sep=0pt] (pic) at (0,0) {\includegraphics[height=5mm, width=40mm]
        {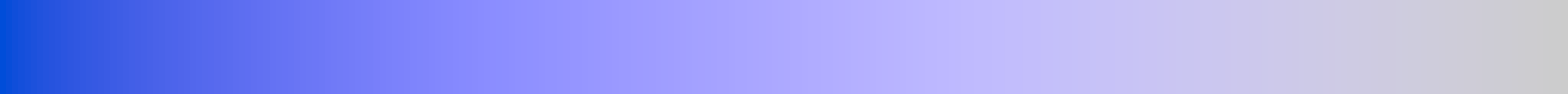}};
        \node[inner sep=0pt] (0)   at ($(pic.south)+(-2.22, 0.26)$)  {$0$};
        \node[inner sep=0pt] (1)   at ($(pic.south)+( 2.22, 0.26)$)  {$1$};
        \node[inner sep=0pt] (lc)  at ($(pic.south)+( 0.00, 0.85)$)  {$\lc~\si{[-]}$};
      \end{tikzpicture} 
  \end{subfigure}  
  
  \caption{The dissolution level $d$ and the cathode ratio $\lc$ for distorted meshes at $t = 60~\si{\second}$. The time increment reads $\dt = 10^{-3}~\si{\second}$ and the series connection as well as method B are employed. The vertical blue line provides the anode's surface according to the analytical reference solution. The vertical black line indicates the cathode surface's position and the black arrow gives the direction of the cathode feed.}
  \label{fig:Ex1_dm}     
\end{figure}
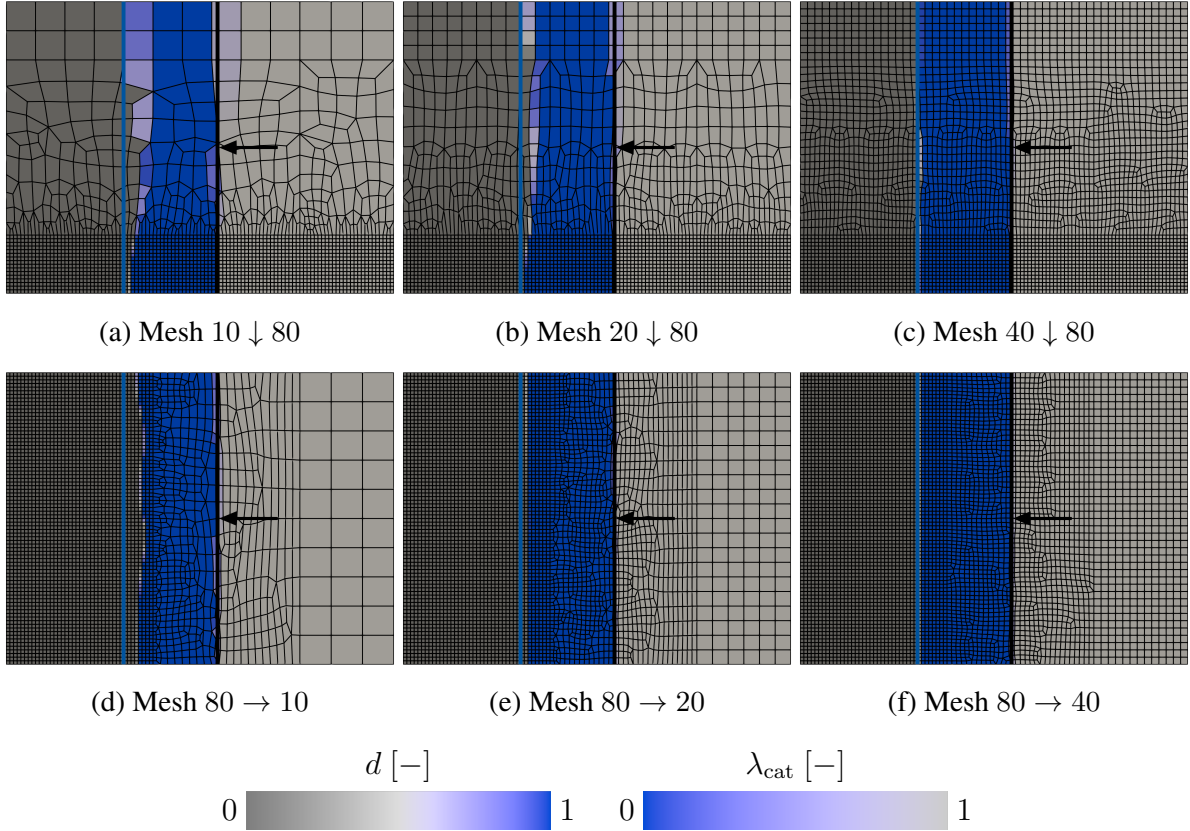

Next, the runtime of method A and B are compared in Fig.~\ref{fig:Ex1_runtime}. The coarser meshes, which have fewer degrees of freedom compared to the fine meshes, yield similar runtimes for both methods. Nevertheless for the finest mesh, method B is $16.3~\si{\%}$ faster, since the reduction of the size of the global system of equations yields a significant impact on the computational efficiency.
Afterwards, we investigate this example using vertically and horizontally distorted meshes which stem from \cite{vanderVeldenRommesEtAl2021}. Fig.~\ref{fig:Ex1_dm} shows the results for the series connection and method B with $\dt = 10^{-3}~\si{s}$. For the most severe distortions, the dissolved volume is underestimated in the simulation (mesh $10 \downarrow 80$: $-6.5~\si{\%}$, mesh $80 \rightarrow 10$: $-9.3~\si{\%}$). However, the results improve for moderate distortions (mesh $20 \downarrow 80$: $-5.1~\si{\%}$, mesh $80 \rightarrow 20$: $-3.9~\si{\%}$) and converge towards the analytical solution for minor distortions (mesh $40 \downarrow 80$: $-0.8~\si{\%}$, mesh $80 \rightarrow 40$: $-1.2~\si{\%}$). 

Figs.~\ref{fig:Ex1_Vdis_vd} and \ref{fig:Ex1_Vdis_hd} show the comparison of the parallel and series connection for different time step sizes. Both rules of mixture yield for all meshes already with the second largest time increment a stable solution that changes only marginally with a further reduction of the time increment. The parallel connection yields for the fine meshes with $\dt = 1~\si{\s}$ a solution close to the analytical result. However, this is not the model's final outcome, since convergence with respect to the time step size is not yet obtained. In the end, the dissolved volume is overestimated by the parallel connection for the different configurations and it yields also for minor distortions still an error of $+4.5~\si{\%}$ (mesh $40 \downarrow 80$) and $+4.2~\si{\%}$ (mesh $40 \rightarrow 80$).
Based on these results, the remaining examples utilize the series connection as rule of mixture.
\begin{figure}[htbp]
  \centering
  \begin{tikzpicture}
      \begin{semilogxaxis}[
                  xlabel= $\dt$  \si{[\s]},
                  ylabel= $\left. V_\mathrm{dis}^\mathrm{FE}/V_\mathrm{dis}^\mathrm{analyt.} \right|_{t \, = \, 60~\si{s}} \si{[-]}$,
                  x dir=reverse,
                  legend pos=outer north east,legend cell align={left},
                  xmin = 0.0005,
                  xmax = 2,
                  ymin = 0.85,
                  ymax = 1.15,
                  xtick={0.001, 0.01, 0.1, 1},
                  xticklabels={$10^{-3}$, $10^{-2}$, $10^{-1}$,$1$}]
                  
          \addplot[mark=none, black, forget plot] coordinates {(10,1) (1E-4,1)};
          
          \addplot[color = sfb4, line width=1.5pt, mark=square*, mark options={solid}, dashpattern1]
                   table[x index = 0, y expr=\thisrowno{2}/(6.0000E-11)] 
                   {02_Figures/02_Pgf/01_Ex1/02_Distorted_Meshes/Vdis_dt.txt};
                   \addlegendentry{$10 \downarrow 80$ parallel}
          \addplot[color = sfb4, line width=1.5pt, mark=diamond*, mark options={solid}, dashpattern1]
                   table[x index = 0, y expr=\thisrowno{6}/(6.0000E-11)] 
                   {02_Figures/02_Pgf/01_Ex1/02_Distorted_Meshes/Vdis_dt.txt};
                   \addlegendentry{$20 \downarrow 80$ parallel}
          \addplot[color = sfb4, line width=1.5pt, mark=x, mark size={1.25mm}, mark options={solid}, dashpattern1]
                   table[x index = 0, y expr=\thisrowno{10}/(6.0000E-11)] 
                   {02_Figures/02_Pgf/01_Ex1/02_Distorted_Meshes/Vdis_dt.txt};
                   \addlegendentry{$40 \downarrow 80$ parallel}
          \addplot[color = sfb1, line width=1.5pt, mark=square*, dashpattern0]
                   table[x index = 0, y expr=\thisrowno{4}/(6.0000E-11)] 
                   {02_Figures/02_Pgf/01_Ex1/02_Distorted_Meshes/Vdis_dt.txt};
                   \addlegendentry{$10 \downarrow 80$ series}
          \addplot[color = sfb1, line width=1.5pt, mark=diamond*, dashpattern0]
                   table[x index = 0, y expr=\thisrowno{8}/(6.0000E-11)] 
                   {02_Figures/02_Pgf/01_Ex1/02_Distorted_Meshes/Vdis_dt.txt};
                   \addlegendentry{$20 \downarrow 80$ series}
          \addplot[color = sfb1, line width=1.5pt, mark=x, mark size={1.25mm}, dashpattern0]
                   table[x index = 0, y expr=\thisrowno{12}/(6.0000E-11)] 
                   {02_Figures/02_Pgf/01_Ex1/02_Distorted_Meshes/Vdis_dt.txt};
                   \addlegendentry{$40 \downarrow 80$ series}
      \end{semilogxaxis}
  \end{tikzpicture}
  
  \caption{Comparison of the dissolved volume at $t = 60~\si{s}$ of the finite element simulation using method B and the analytical solution for the parallel and series connection with vertically distorted meshes.}
  \label{fig:Ex1_Vdis_vd}
\end{figure}
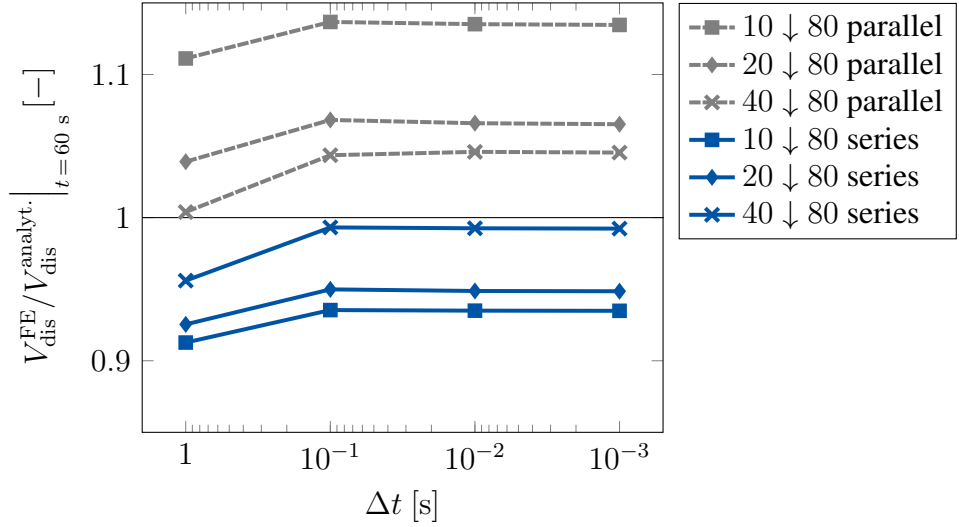
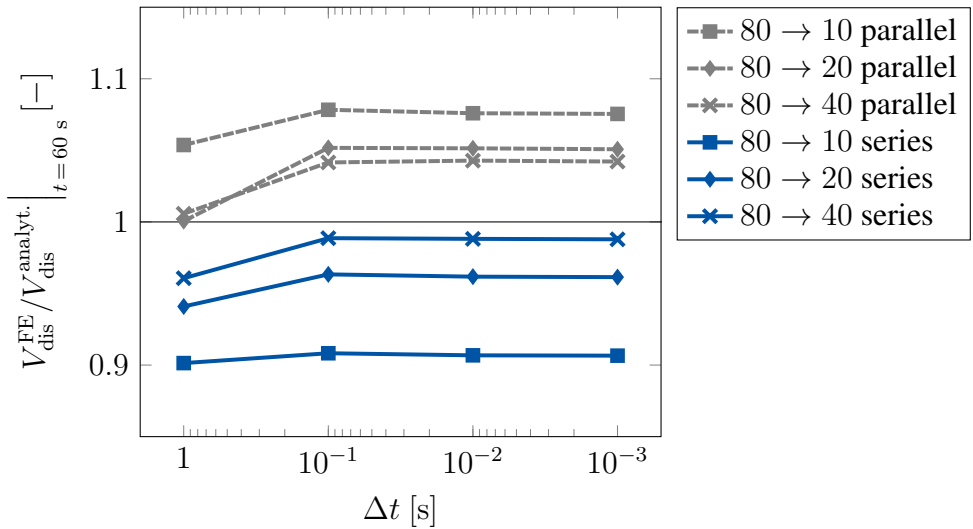
\begin{figure}[htbp]
  \centering
  \begin{tikzpicture}
      \begin{semilogxaxis}[
                  xlabel= $\dt$  \si{[\s]},
                  ylabel= $\left. V_\mathrm{dis}^\mathrm{FE}/V_\mathrm{dis}^\mathrm{analyt.} \right|_{t \, = \, 60~\si{s}} \si{[-]}$,
                  x dir=reverse,
                  legend pos=outer north east,legend cell align={left},
                  xmin = 0.0005,
                  xmax = 2,
                  ymin = 0.85,
                  ymax = 1.15,
                  xtick={0.001, 0.01, 0.1, 1},
                  xticklabels={$10^{-3}$, $10^{-2}$, $10^{-1}$,$1$}]
                  
          \addplot[mark=none, black, forget plot] coordinates {(10,1) (1E-4,1)};
          
          \addplot[color = sfb4, line width=1.5pt, mark=square*, mark options={solid}, dashpattern1]
                   table[x index = 0, y expr=\thisrowno{26}/(6.0000E-11)] 
                   {02_Figures/02_Pgf/01_Ex1/02_Distorted_Meshes/Vdis_dt.txt};
                   \addlegendentry{$80 \rightarrow 10$ parallel}
          \addplot[color = sfb4, line width=1.5pt, mark=diamond*, mark options={solid}, dashpattern1]
                   table[x index = 0, y expr=\thisrowno{30}/(6.0000E-11)] 
                   {02_Figures/02_Pgf/01_Ex1/02_Distorted_Meshes/Vdis_dt.txt};
                   \addlegendentry{$80 \rightarrow 20$ parallel}
          \addplot[color = sfb4, line width=1.5pt, mark=x, mark size={1.25mm}, mark options={solid}, dashpattern1]
                   table[x index = 0, y expr=\thisrowno{34}/(6.0000E-11)] 
                   {02_Figures/02_Pgf/01_Ex1/02_Distorted_Meshes/Vdis_dt.txt};
                   \addlegendentry{$80 \rightarrow 40$ parallel}
          \addplot[color = sfb1, line width=1.5pt, mark=square*, dashpattern0]
                   table[x index = 0, y expr=\thisrowno{28}/(6.0000E-11)] 
                   {02_Figures/02_Pgf/01_Ex1/02_Distorted_Meshes/Vdis_dt.txt};
                   \addlegendentry{$80 \rightarrow 10$ series}
          \addplot[color = sfb1, line width=1.5pt, mark=diamond*, dashpattern0]
                   table[x index = 0, y expr=\thisrowno{32}/(6.0000E-11)] 
                   {02_Figures/02_Pgf/01_Ex1/02_Distorted_Meshes/Vdis_dt.txt};
                   \addlegendentry{$80 \rightarrow 20$ series}
          \addplot[color = sfb1, line width=1.5pt, mark=x, mark size={1.25mm}, dashpattern0]
                   table[x index = 0, y expr=\thisrowno{36}/(6.0000E-11)] 
                   {02_Figures/02_Pgf/01_Ex1/02_Distorted_Meshes/Vdis_dt.txt};
                   \addlegendentry{$80 \rightarrow 40$ series}
                  
      \end{semilogxaxis}
  \end{tikzpicture}
  
  \caption{Comparison of the dissolved volume at $t = 60~\si{s}$ of the finite element simulation using method B and the analytical solution for the parallel and series connection with horizontally distorted meshes.}
  \label{fig:Ex1_Vdis_hd}
\end{figure}

\begin{figure}[htbp]
  \centering
  \includegraphics{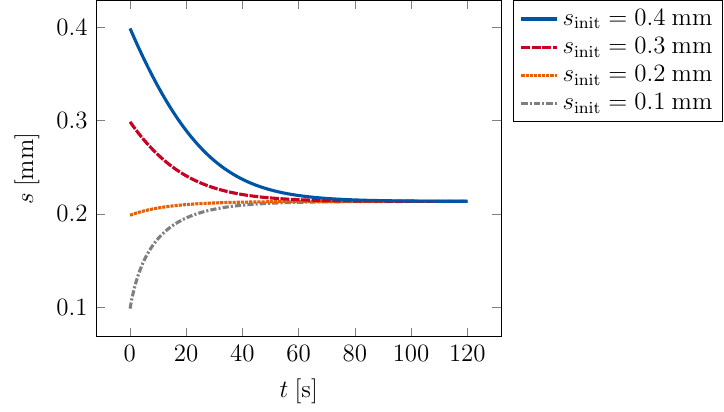}

  \caption{Investigation of the evolution of the working gap width $s$ for different initial widths $s_\mathrm{init}$. The feed rate is constant at $\dot{x}_\mathrm{ca} = 0.015~\si{\mm\per\s}$, $\dt = 0.1~\si{\s}$ and method B is employed.}
  \label{fig:wg_sinit}
\end{figure}

To demonstrate the flexibility of the new approach, Fig.~\ref{fig:wg_sinit} shows the evolution of the working gap width $s$ during the machining process with a constant feed rate $\dot{x}_\mathrm{ca} = 0.015~\si{\mm\per\s}$ and varying initial widths $s_\mathrm{init}$. All graphs concur at the analytical solution of $s = 0.213~\si{\mm}$ and confirm the correct operation of the method, since the convergence of the gap widths requires the exact modeling of the cathode.

\begin{figure}[htbp]
  \centering  
  \includegraphics{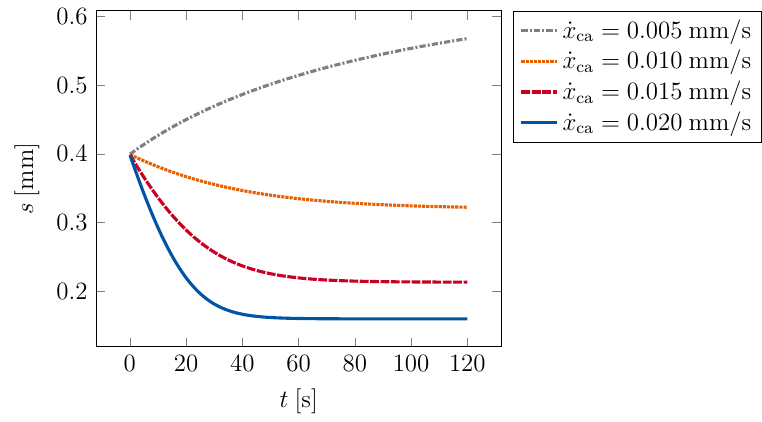}

  \caption{Investigation of the evolution of the working gap width $s$ for different feed rates $\dot{x}_\mathrm{ca}$. The initial width is  $s_\mathrm{init} = 0.4~\si{\mm}$, $\dt = 0.1~\si{\s}$ and method B is employed.}
  \label{fig:wg_xdot}
\end{figure}

Furthermore, Fig.~\ref{fig:wg_xdot} shows the evolution of the gap width $s$ for the same initial value $s_\mathrm{init}$ but different feed rates $\dot{x}_\mathrm{ca}$. For a higher feed rate, the gap width decreases (cf.~\cite{KlockeKoenig2007}) until it reaches the dynamic equilibrium state ($\dot{x}_\mathrm{ca} = 0.010 - 0.020~\si{\mm\per\s}$). For a lower feed rate, the working gap widens ($\dot{x}_\mathrm{ca} = 0.005~\si{\mm\per\s}$). This study again confirms the versatility of the new framework.


\subsection{Parabolic cathode - \cite{HardistyMileham1999}}
\label{ssec:Ex2}

\begin{figure}[htbp] 
     \centering 
     \begin{subfigure}{.45\textwidth} 
         \centering 
         \begin{tikzpicture}
           \node[inner sep=0pt] (pic) at (0,0) {\includegraphics[width=\textwidth]
           {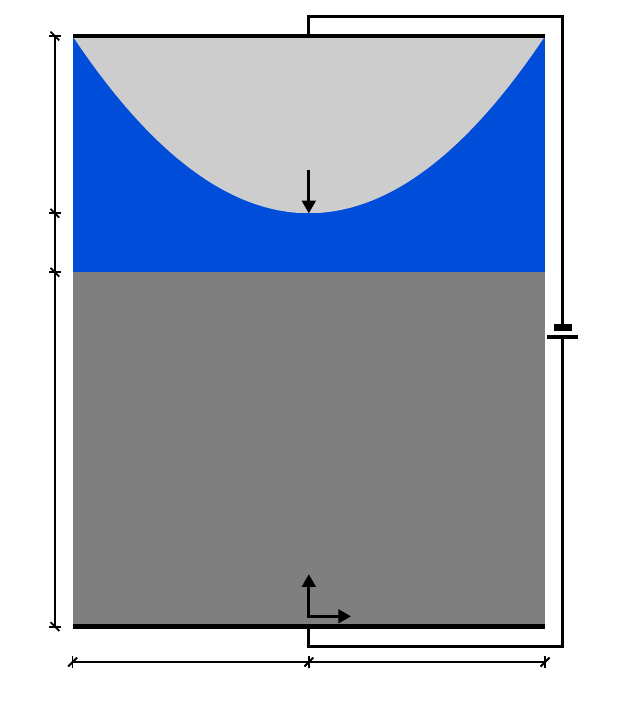}};
           \node[inner sep=0pt] (h1)   at ($(pic.west) +( 0.30,-1.00)$)  {$h_1$};
           \node[inner sep=0pt] (h2)   at ($(pic.west) +( 0.30, 2.50)$)  {$h_2$};
           \node[inner sep=0pt] (s)    at ($(pic.west) +( 0.30, 1.30)$)  {$s$};
           \node[inner sep=0pt] (l1)   at ($(pic.south)+(-1.40, 0.25)$)  {$l$};
           \node[inner sep=0pt] (l2)   at ($(pic.south)+( 1.40, 0.25)$)  {$l$};
           \node[inner sep=0pt] (x)    at ($(pic.south)+( 0.67, 1.20)$)  {$x$};
           \node[inner sep=0pt] (y)    at ($(x.north)  +(-0.52, 0.42)$)  {$y$};
           \node[inner sep=0pt] (xdc)  at ($(pic.north)+( 0.00,-1.50)$)  {$\dot{x}_\mathrm{ca}$};
           \node[inner sep=0pt] (dv)   at ($(pic.east) +(-0.05, 0.28)$)  {$\delv$};
         \end{tikzpicture} 
         \caption{Geometry}
         \label{fig:Ex2Geom}
         
     \end{subfigure}
     \qquad
     \begin{subfigure}{.45\textwidth} 
         \centering 
         \begin{tikzpicture}
           \node[inner sep=0pt] (pic) at (0,0) {\includegraphics[width=\textwidth]
           {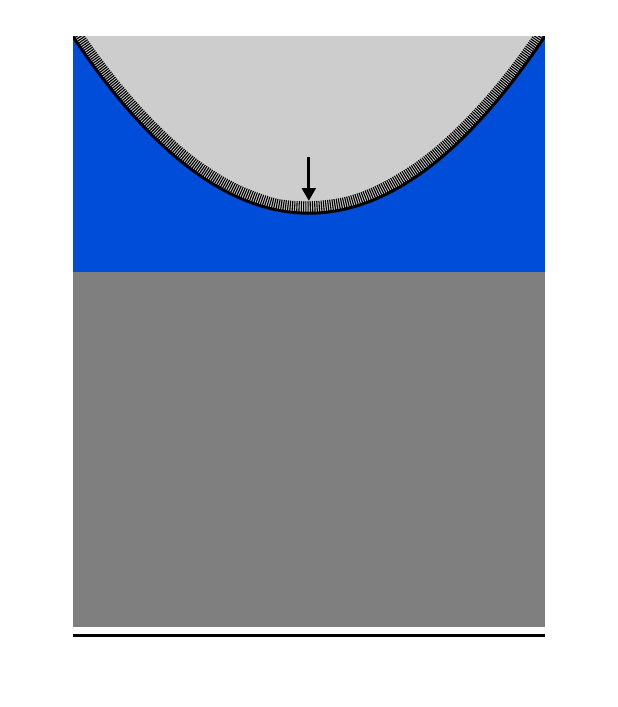}};
           \node[inner sep=0pt] (van)   at ($(pic.south)+( 0.00, 0.40)$)  {$\widetilde{v}_\mathrm{an}$};
           \node[inner sep=0pt] (vca)   at ($(pic.north)+( 0.00,-1.40)$)  {$\widetilde{v}_\mathrm{ca}$};
         \end{tikzpicture} 
         \caption{MBVP}
         \label{fig:Ex2MBVP}
     \end{subfigure} 
     \caption{Geometry and moving boundary value problem for a parabolic cathode (cf.~\cite{HardistyMileham1999}).} 
     \label{fig:Ex2}
\end{figure}
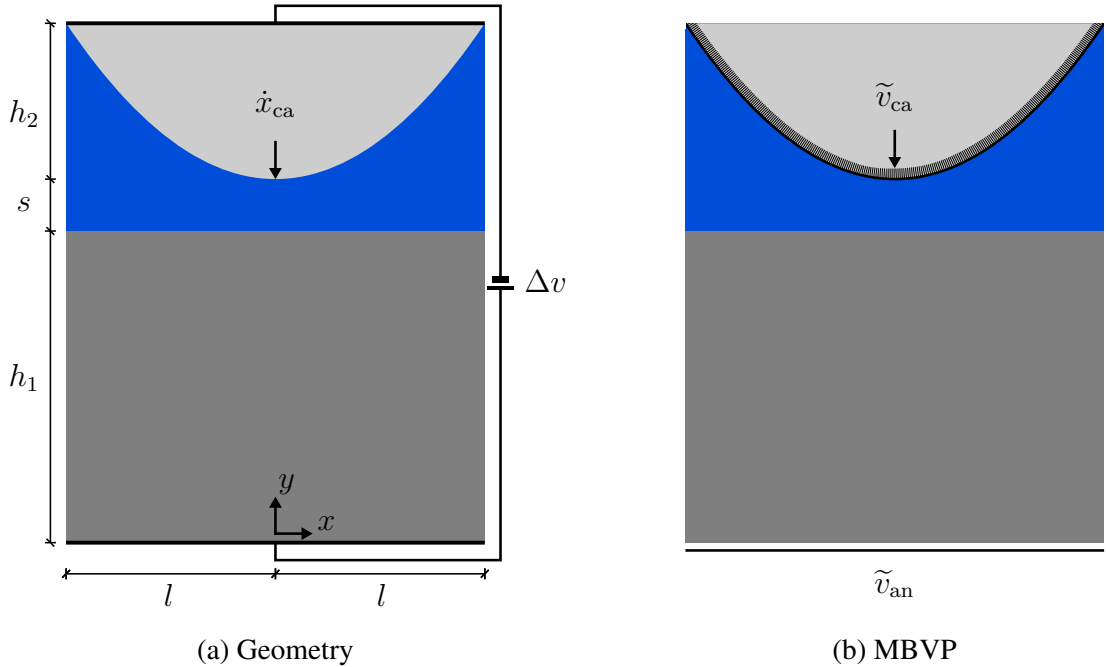

In the second example, we consider a parabolically shaped cathode (Fig.~\ref{fig:Ex2}) which has formerly been studied by \cite{HardistyMileham1999}. The material parameters read $\kE\ME = 6.670 \times 10^6~\si{\A\per\V\per\m}$, $\kE\EL = 15~\si{\A\per\V\per\m}$ and $\Veffnew = 3.696 \times 10^{-11}~\si{\cubic\m\per\A\per\s}$. The cathode moves with a constant feed rate $\dot{x}_\mathrm{ca} = 0.0145~\si{\mm\per\s}$ towards the anode with $\dt = 0.34483~\si{\s}$ yielding a displacement of $0.005~\si{\mm}$ per time step and a total displacement of $2.5~\si{\mm}$ after 500 time steps.
The geometric dimensions are $l = 2~\si{\mm}$, $h_1 = 3~\si{\mm}$, $h_2 = 1.5~\si{\mm}$, $s = 0.5~\si{\mm}$ with a thickness of $0.5~\si{\mm}$. The parabola is defined by the function $f(x) = 0.375 \, x^2 + 3.5$ and the potential difference is $\delv = 10~\si{\V}$. Due to symmetry, only the left part $x \in [-2,0]$ is simulated. We use structured meshes with finite element (el.) discretizations from $5 \times 5~\text{el./\si{\square\mm}}$ to $40 \times 40~\text{el./\si{\square\mm}}$.

\begin{figure}[htbp]
  \centering
  \includegraphics{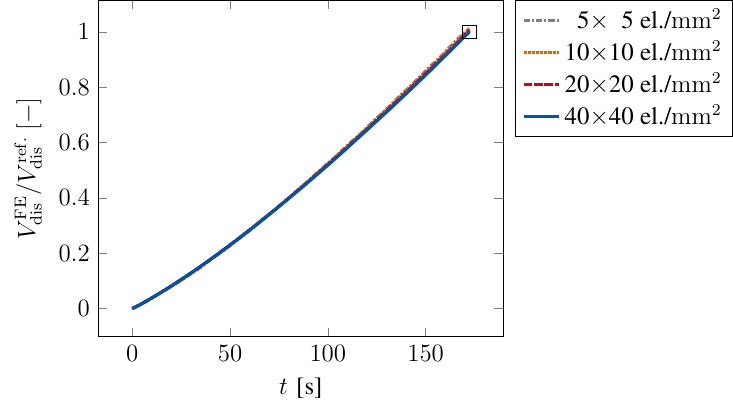}

  \vspace{-3mm}
  \caption{The dissolved volume over time computed with method B for different mesh densities. The graphs are normalized to the solution of the finest mesh with $\dt = 0.34483~\si{\second}$. The black box indicates the point of comparison in Fig.~\ref{fig:Ex2_dsln_end}.}
  \label{fig:Ex2_Vdis}
  
\end{figure}

Fig.~\ref{fig:Ex2_Vdis} shows the dissolved volume over time for the different mesh densities normalized to the solution of the finest mesh with $V_\mathrm{dis}^\mathrm{ref.} = 0.775~\si{\cubic\mm}$. The curves agree well and mesh convergence is proven.
\begin{figure}[htbp]

  
  \begin{subfigure}{.49\textwidth} 
      \centering 
      \includegraphics[width=\textwidth]{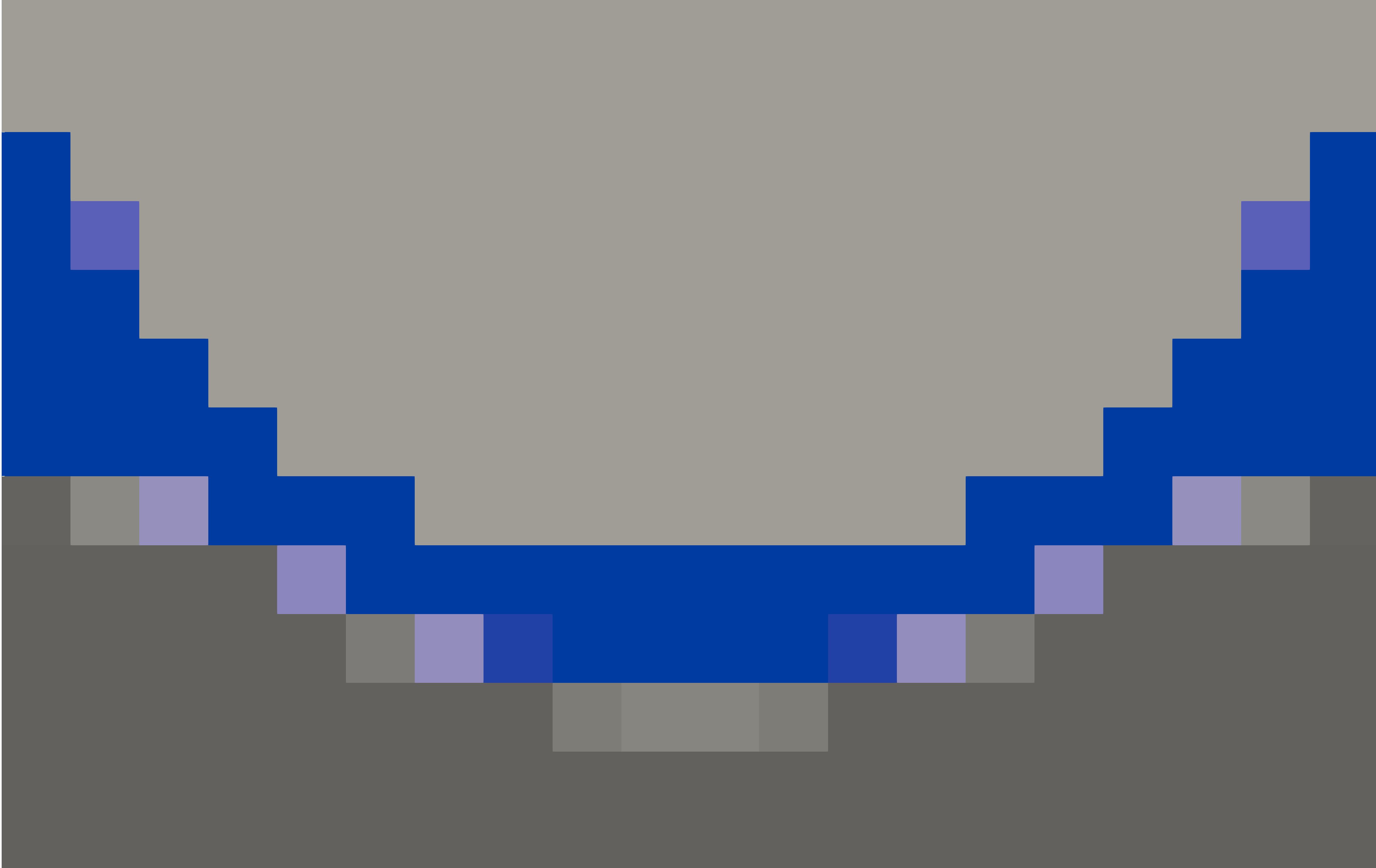}
      \caption{5$\times$5 el./\si{\square\mm}\hphantom{00}}
      \label{fig:Ex2_dsln_end_ed05}
  \end{subfigure}
  \hspace{1mm}
  \begin{subfigure}{.49\textwidth} 
      \centering 
      \includegraphics[width=\textwidth]{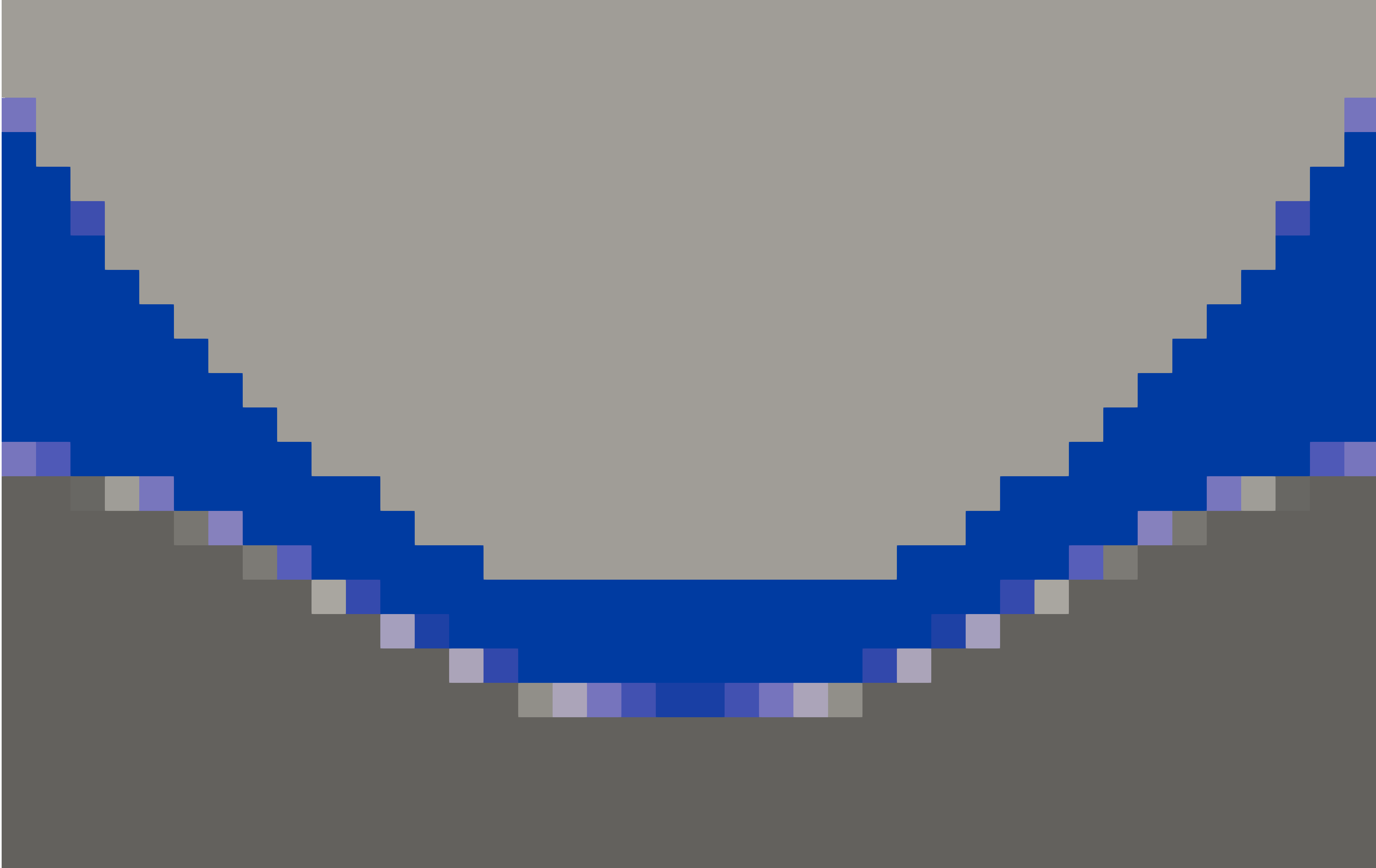}
      \caption{10$\times$10 el./\si{\square\mm}}
      \label{fig:Ex2_dsln_end_ed10}
  \end{subfigure}
  
  
  \begin{subfigure}{.49\textwidth} 
      \centering 
      \includegraphics[width=\textwidth]{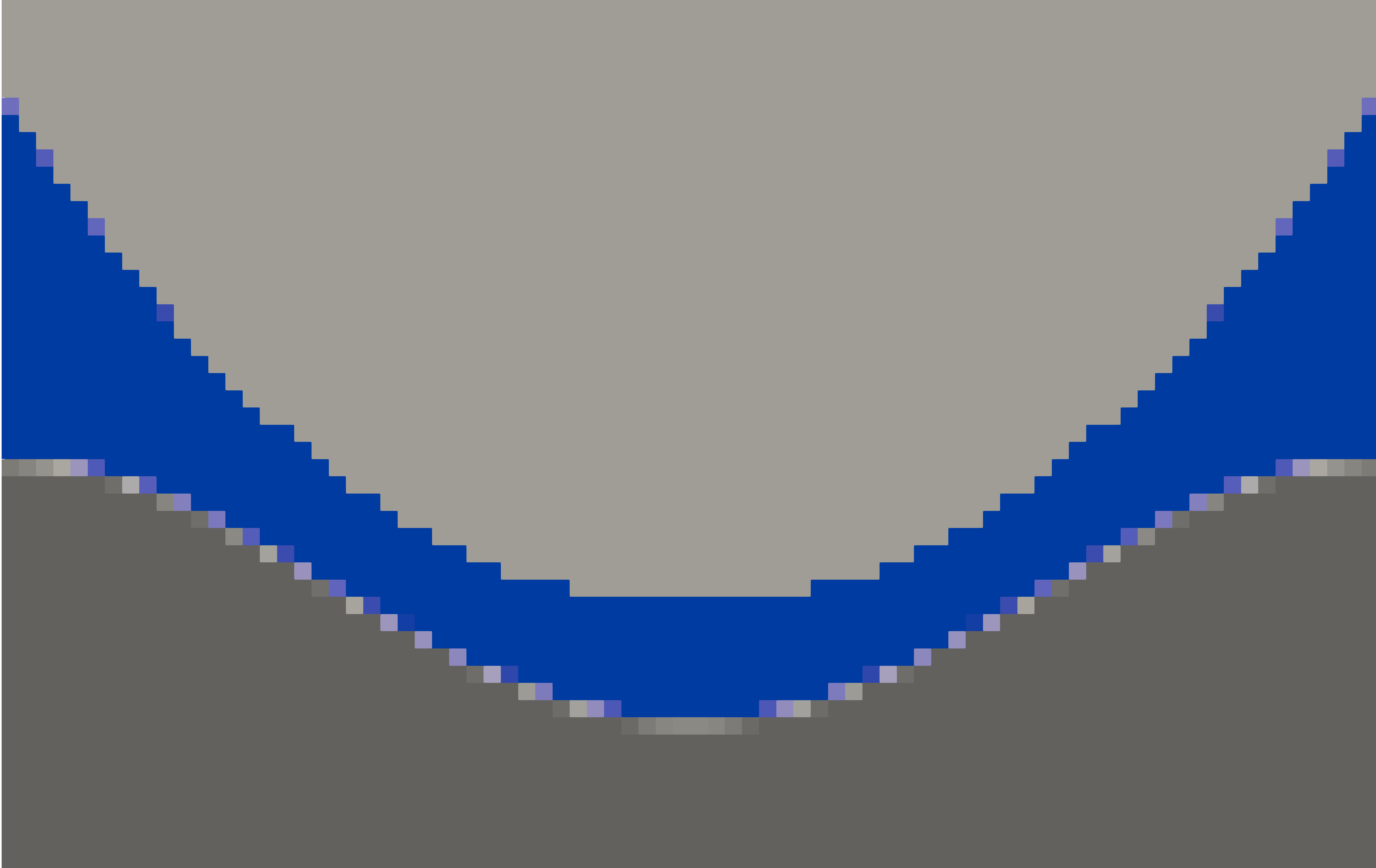}
      \caption{20$\times$20 el./\si{\square\mm}}
      \label{fig:Ex2_dsln_end_ed20}
  \end{subfigure}
  \hspace{1mm}
  \begin{subfigure}{.49\textwidth} 
      \centering 
      \begin{tikzpicture}
        \node[inner sep=0pt] (pic) at (0,0) {\includegraphics[width=\textwidth]
        {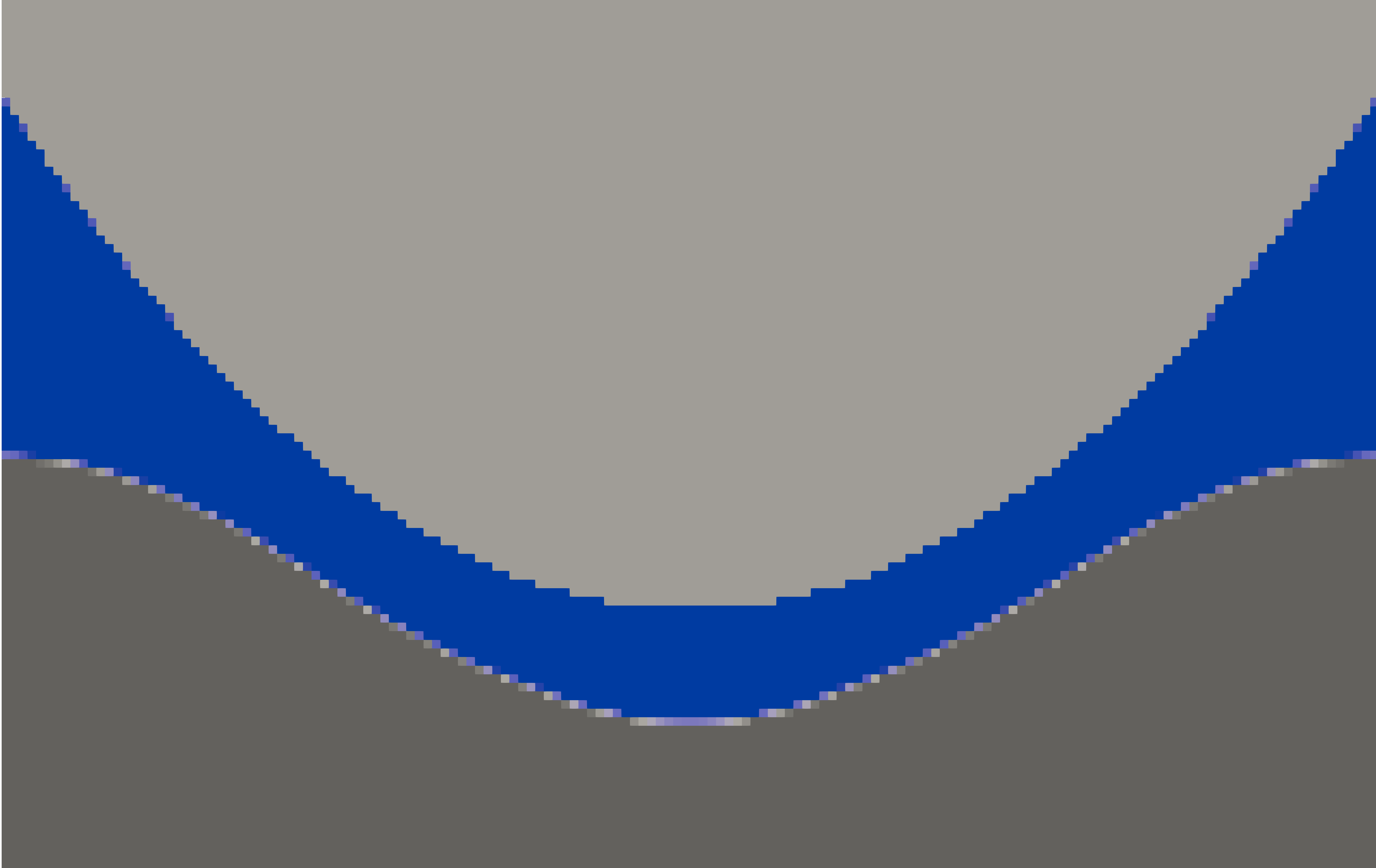}};
        \draw [latex-latex,color=white](0,-0.98) -- (-0,-1.65);
        \node[inner sep=0pt] (35mm) at (0.80,-2.05)  {\textcolor{white}{$0.345~\si{\mm}$}};
      \end{tikzpicture} 
      \caption{40$\times$40 el./\si{\square\mm}}
      \label{fig:Ex2_dsln_end_ed40}
  \end{subfigure}
  
  \vspace{1mm}
  \begin{subfigure}{.49\textwidth} 
      \centering 
      \begin{tikzpicture} 
        \node[inner sep=0pt] (pic) at (0,0) {\includegraphics[height=5mm, width=40mm]
        {03_Contour/00_Color_Maps/01_dsln_hor.png}};
        \node[inner sep=0pt] (0)   at ($(pic.south)+(-2.22, 0.26)$)  {$0$};
        \node[inner sep=0pt] (1)   at ($(pic.south)+( 2.22, 0.26)$)  {$1$};
        \node[inner sep=0pt] (d)   at ($(pic.south)+( 0.00, 0.85)$)  {$d \, [-]$};
      \end{tikzpicture} 
  \end{subfigure}
  \begin{subfigure}{.49\textwidth} 
      \centering 
      \begin{tikzpicture} 
        \node[inner sep=0pt] (pic) at (0,0) {\includegraphics[height=5mm, width=40mm]
        {03_Contour/00_Color_Maps/03_catrt_hor.png}};
        \node[inner sep=0pt] (0)   at ($(pic.south)+(-2.22, 0.26)$)  {$0$};
        \node[inner sep=0pt] (1)   at ($(pic.south)+( 2.22, 0.26)$)  {$1$};
        \node[inner sep=0pt] (lc)  at ($(pic.south)+( 0.00, 0.85)$)  {$\lc \, [-]$};
      \end{tikzpicture} 
  \end{subfigure}  
  
  \caption{Zoom to the tool tip at the end of the simulation: The dissolution level $d$ and the cathode ratio $\lc$ show mesh convergence from coarse (Fig.~\ref{fig:Ex2_dsln_end_ed05}) to fine mesh discretizations (Fig.~\ref{fig:Ex2_dsln_end_ed40}).}
  \label{fig:Ex2_dsln_end}     
\end{figure}

In Fig.~\ref{fig:Ex2_dsln_end}, we compare the shape of the anode at the end of the simulation for different mesh discretizations. With finer discretizations, the surface profile becomes smoother. Nevertheless, already the coarsest discretization allows for an accurate prediction of the final work piece shape. The working gap width at the axis of symmetry at the end of the simulation is $s = 0.345~\si{\mm}$ and agrees with the reference solution of \cite{HardistyMileham1999} of $s^\mathrm{ref.} = 0.35~\si{\mm}$.
These results confirm the model's capabilities to accurately predict the shape of the machined workpiece using coarse meshes.

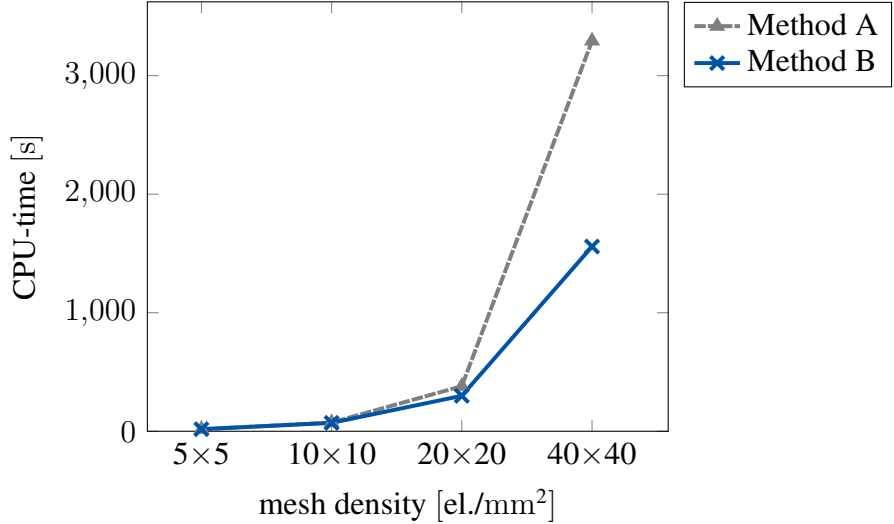
\begin{figure}[htbp]
  \centering
  \begin{tikzpicture}
      \begin{semilogxaxis}[
                  xlabel= mesh density~\text{\si{[}el./\si{\square\mm]}},
                  ylabel= CPU-time \si{[\s]},
                  legend pos=outer north east,legend cell align={left},
                  xmin = 3.75,
                  xmax = 60,
                  ymin = 0,
                  xtick={5, 10, 20, 40},
                  xticklabels={5$\times$5, 10$\times$10, 20$\times$20, 40$\times$40}]
          
          \addplot[color = sfb4, line width=1.5pt, mark=triangle*, mark options={solid}, dashpattern1]
                   table[x index = 0, y index = 2] 
                   {02_Figures/02_Pgf/02_Ex2/02_runtime/runtime_ed.txt};
                   \addlegendentry{Method A}                  
          \addplot[color = sfb1, line width=1.5pt, mark=x, mark size={1.25mm}, dashpattern0]
                   table[x index = 0, y index = 11] 
                   {02_Figures/02_Pgf/02_Ex2/02_runtime/runtime_ed.txt};
                   \addlegendentry{Method B}                  
                  
      \end{semilogxaxis}
  \end{tikzpicture}
  \caption{Comparison of the runtime of method A and B. Structured meshes are employed in the simulation with $\dt = 0.34483~\si{\second}$.}
  \label{fig:Ex2_runtime}
\end{figure}

Furthermore, the runtime of method A and B are compared in this example for different element densities in Fig.~\ref{fig:Ex2_runtime}. For coarse meshes, both methods yield similar runtimes ($5 \times 5~\text{el./\si{\square\mm}}$, $10 \times 10~\text{el./\si{\square\mm}}$). But already for the second finest mesh ($20 \times 20~\text{el./\si{\square\mm}}$), method B is with $21.4~\si{\%}$ significantly faster than method A. For the finest mesh ($40 \times 40~\text{el./\si{\square\mm}}$), the savings in computation time are $52.7~\si{\%}$.

\begin{figure}[htbp] 
  \centering 
  
  \begin{subfigure}{.32\textwidth} 
      \centering 
      \includegraphics[width=\textwidth]{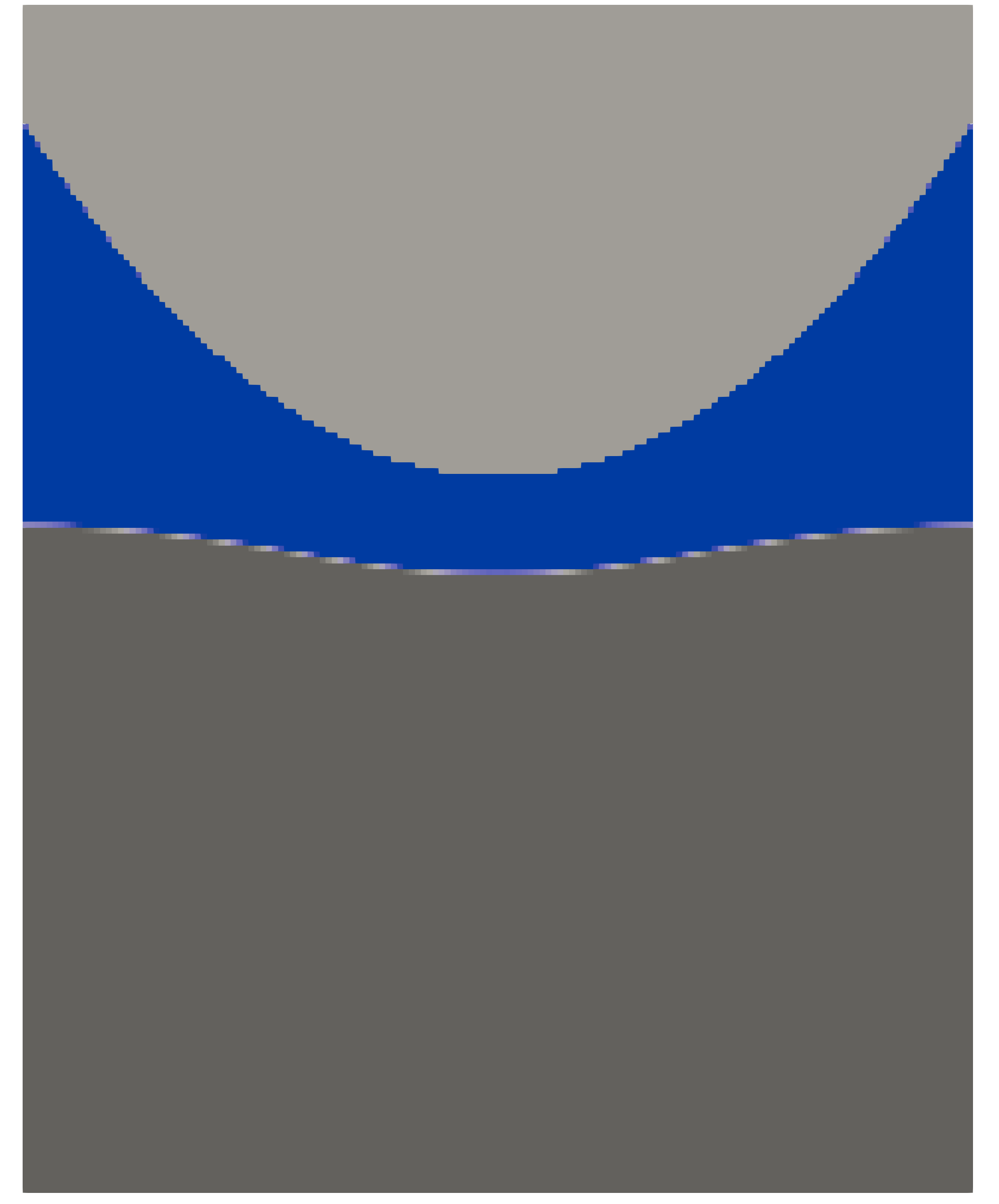}
  \end{subfigure}
  \begin{subfigure}{.32\textwidth} 
      \centering 
      \includegraphics[width=\textwidth]{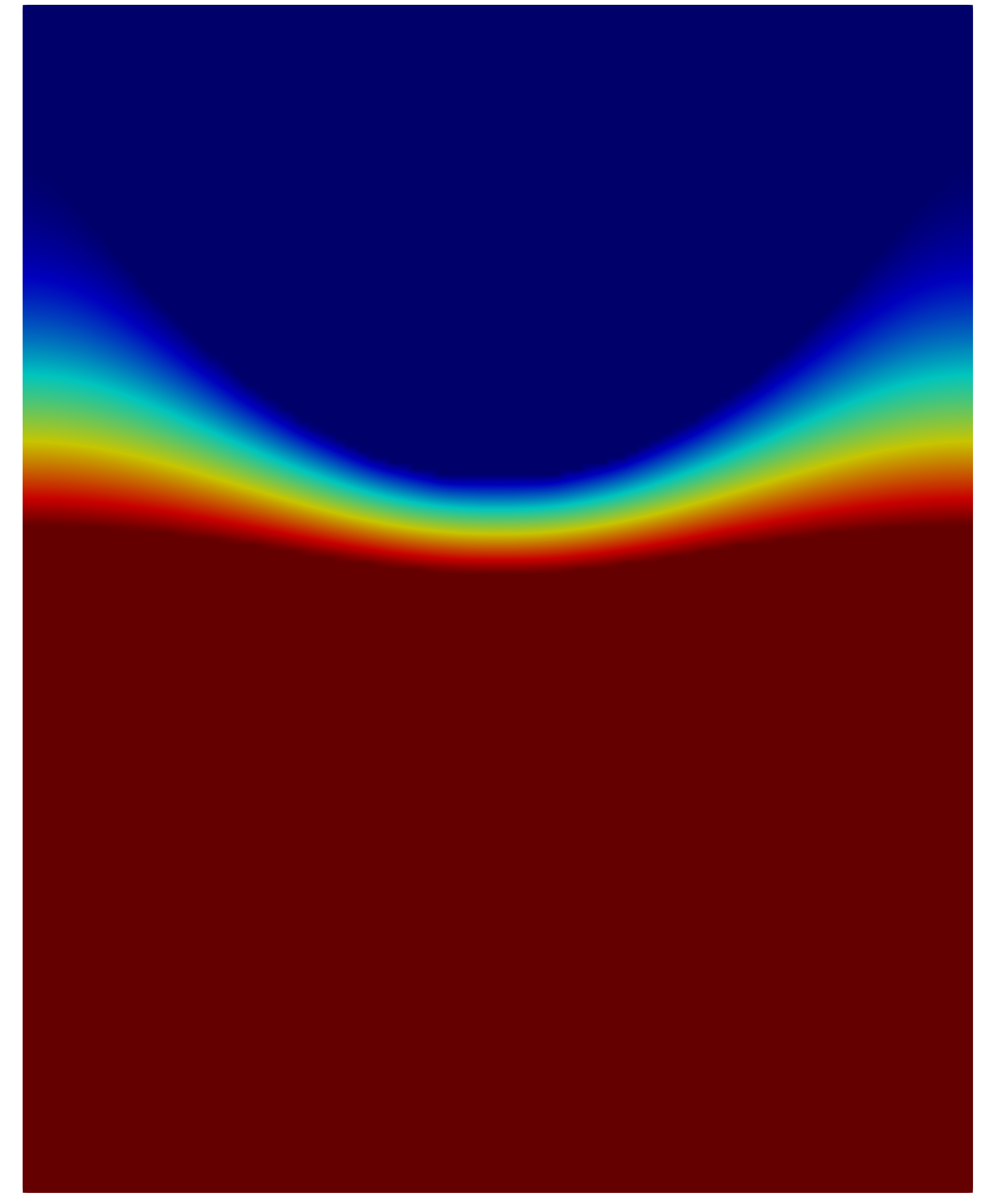}
  \end{subfigure}
  \begin{subfigure}{.32\textwidth} 
      \centering 
      \includegraphics[width=\textwidth]{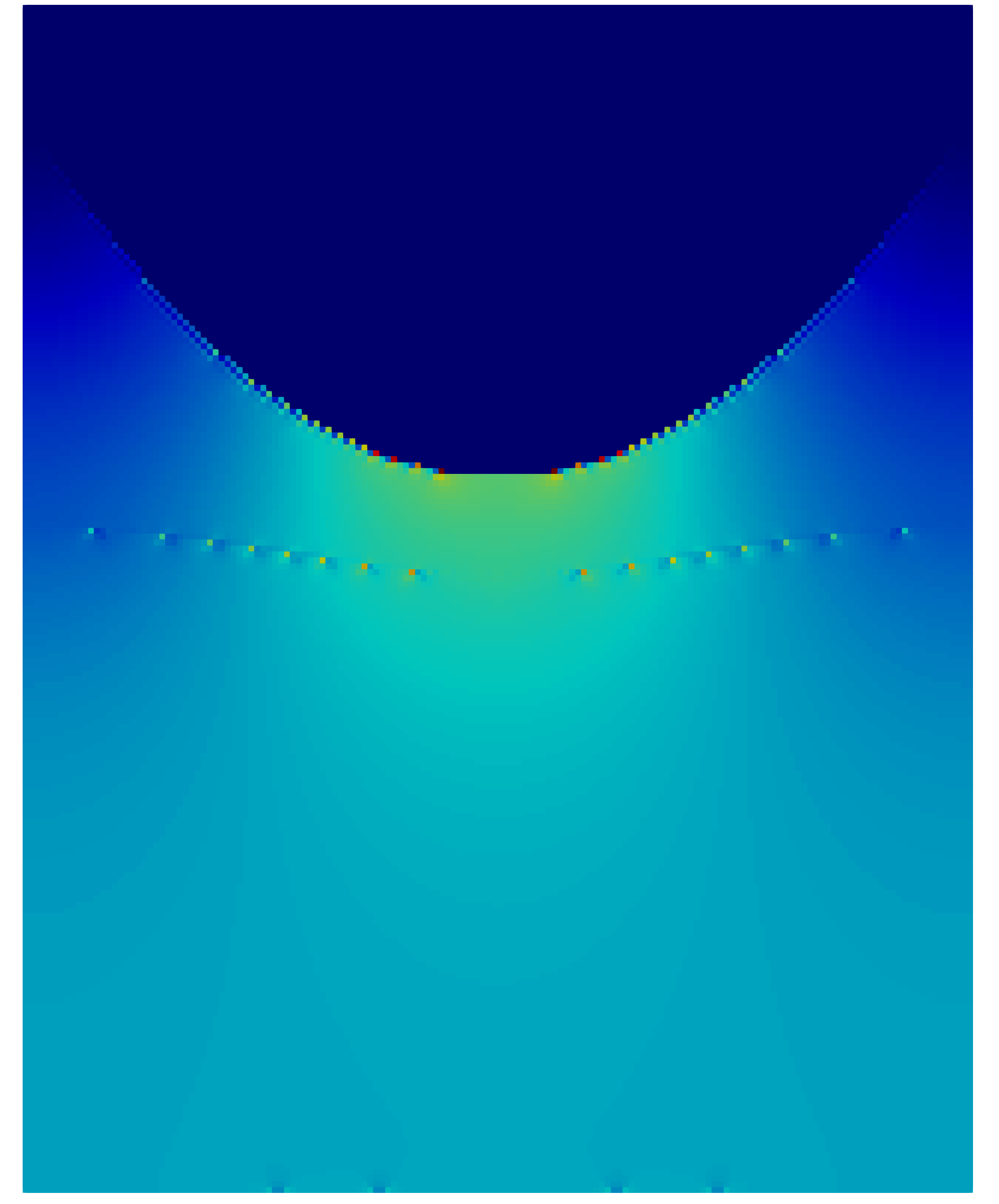}
  \end{subfigure}
  
  \vspace{1mm}  
  
  \begin{subfigure}{.32\textwidth} 
      \centering 
      \includegraphics[width=\textwidth]{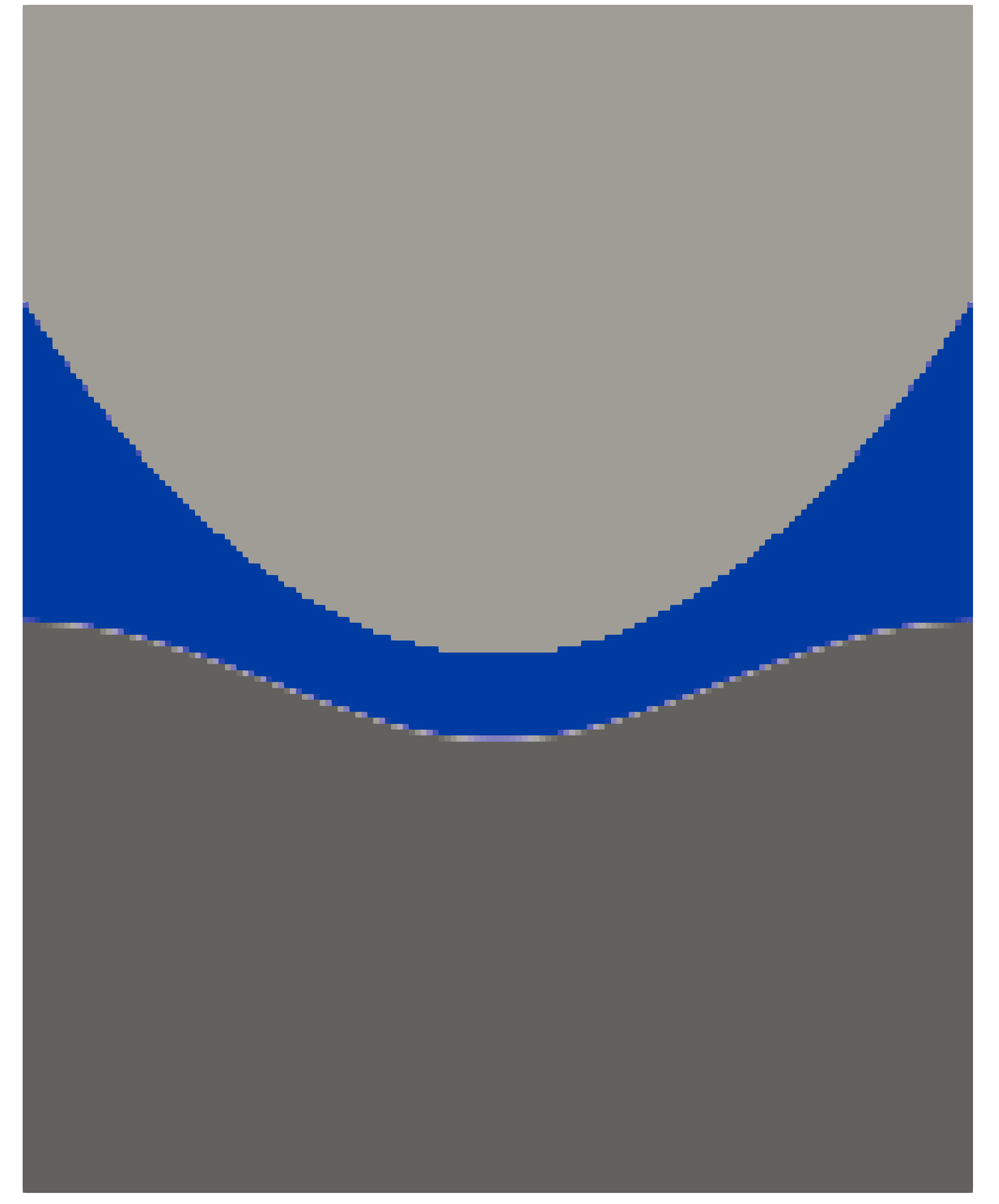}
  \end{subfigure}
  \begin{subfigure}{.32\textwidth} 
      \centering 
      \includegraphics[width=\textwidth]{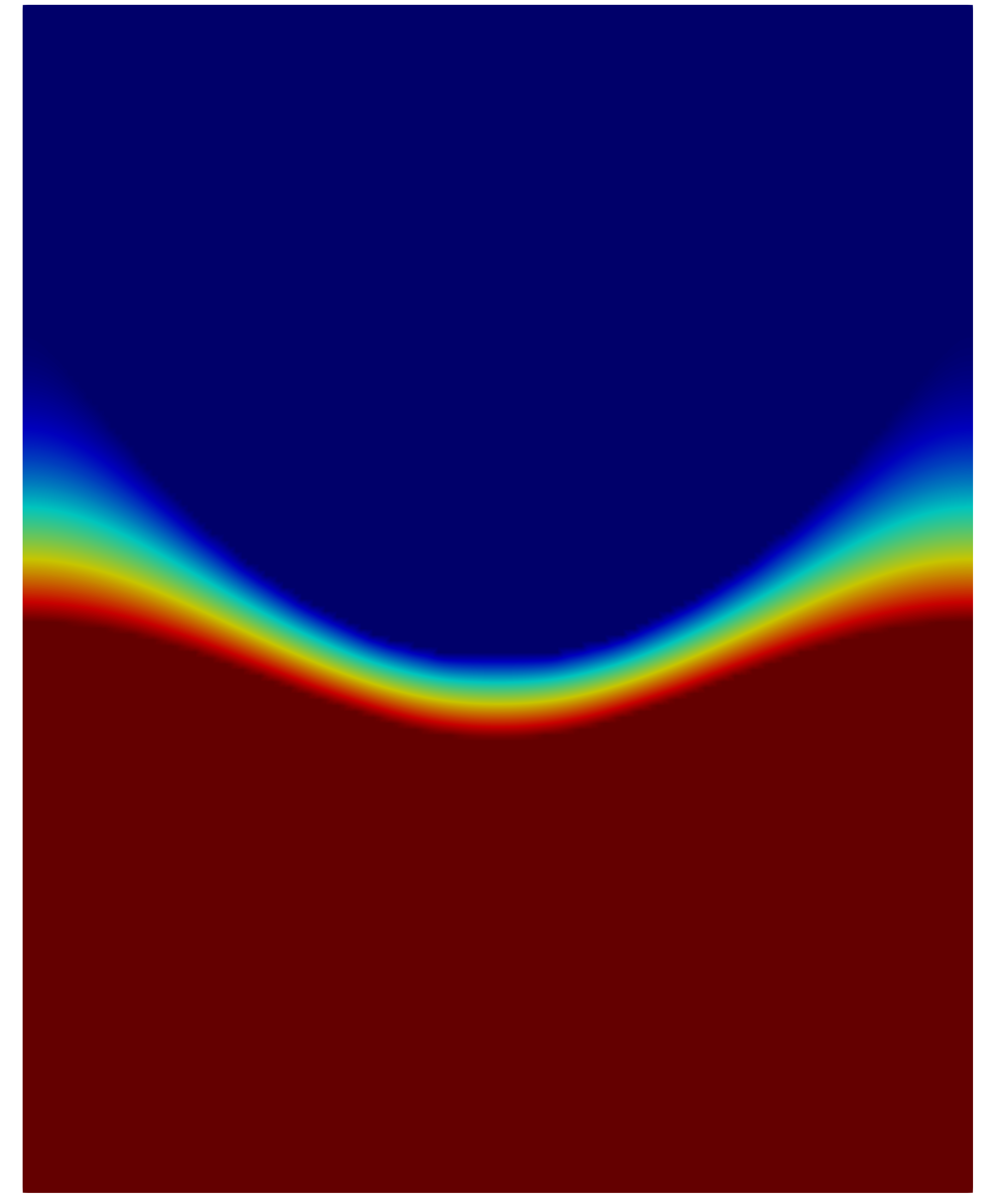}
  \end{subfigure}
  \begin{subfigure}{.32\textwidth} 
      \centering 
      \includegraphics[width=\textwidth]{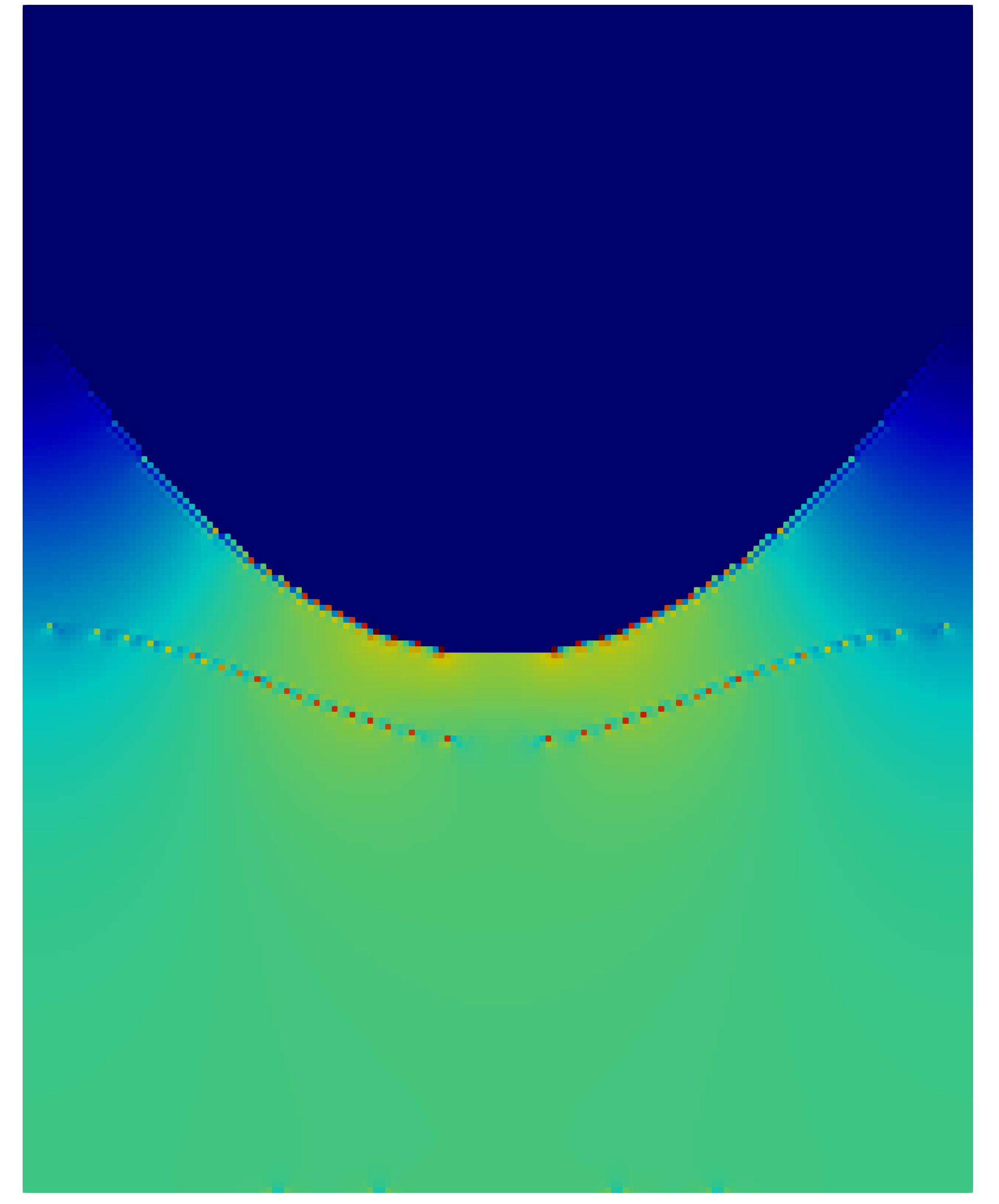}
  \end{subfigure}
  
  \vspace{1mm}

  \begin{subfigure}{.32\textwidth} 
      \centering 
      \includegraphics[width=\textwidth]{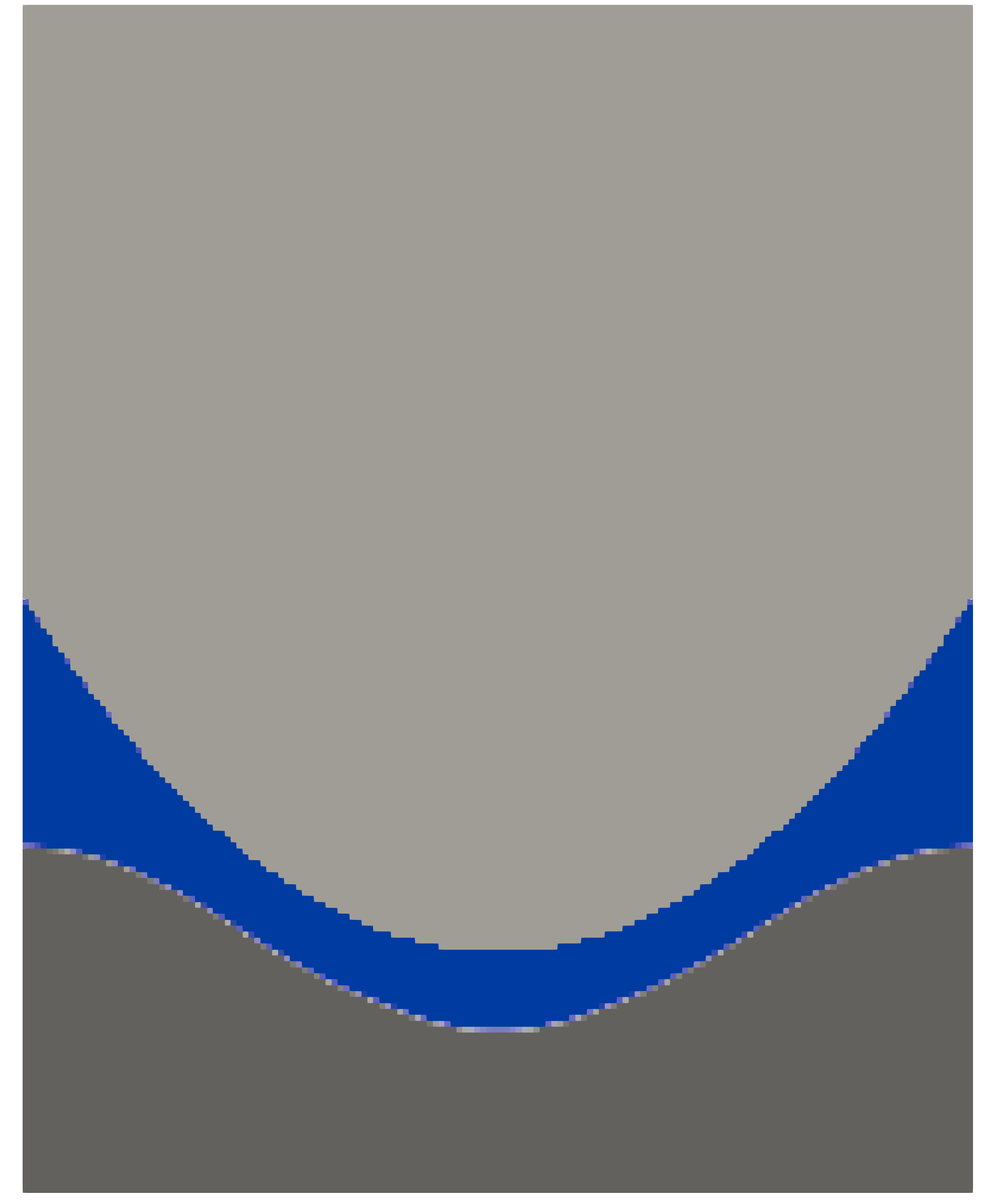}
  \end{subfigure}
  \begin{subfigure}{.32\textwidth} 
      \centering 
      \includegraphics[width=\textwidth]{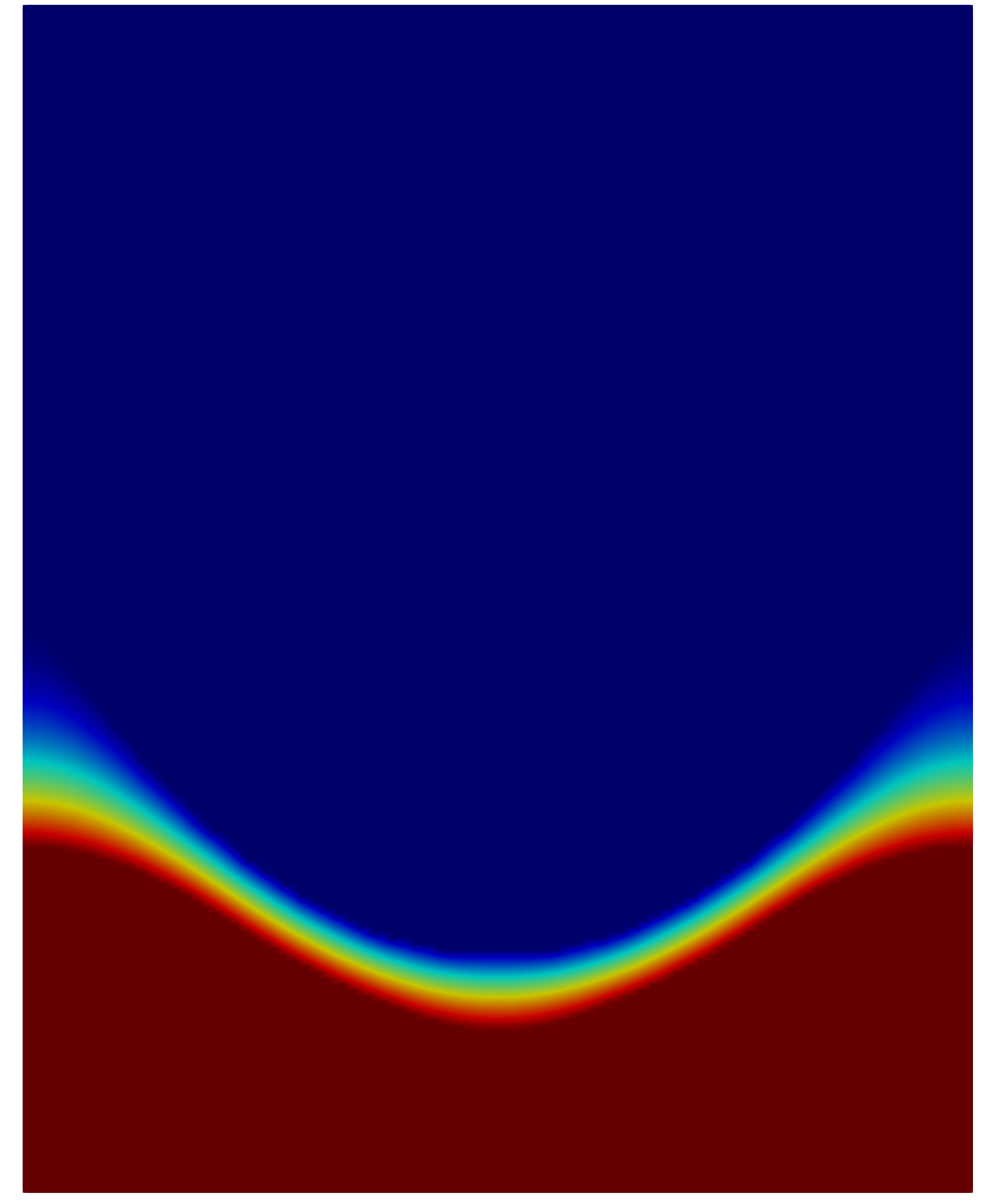}
  \end{subfigure}
  \begin{subfigure}{.32\textwidth} 
      \centering 
      \includegraphics[width=\textwidth]{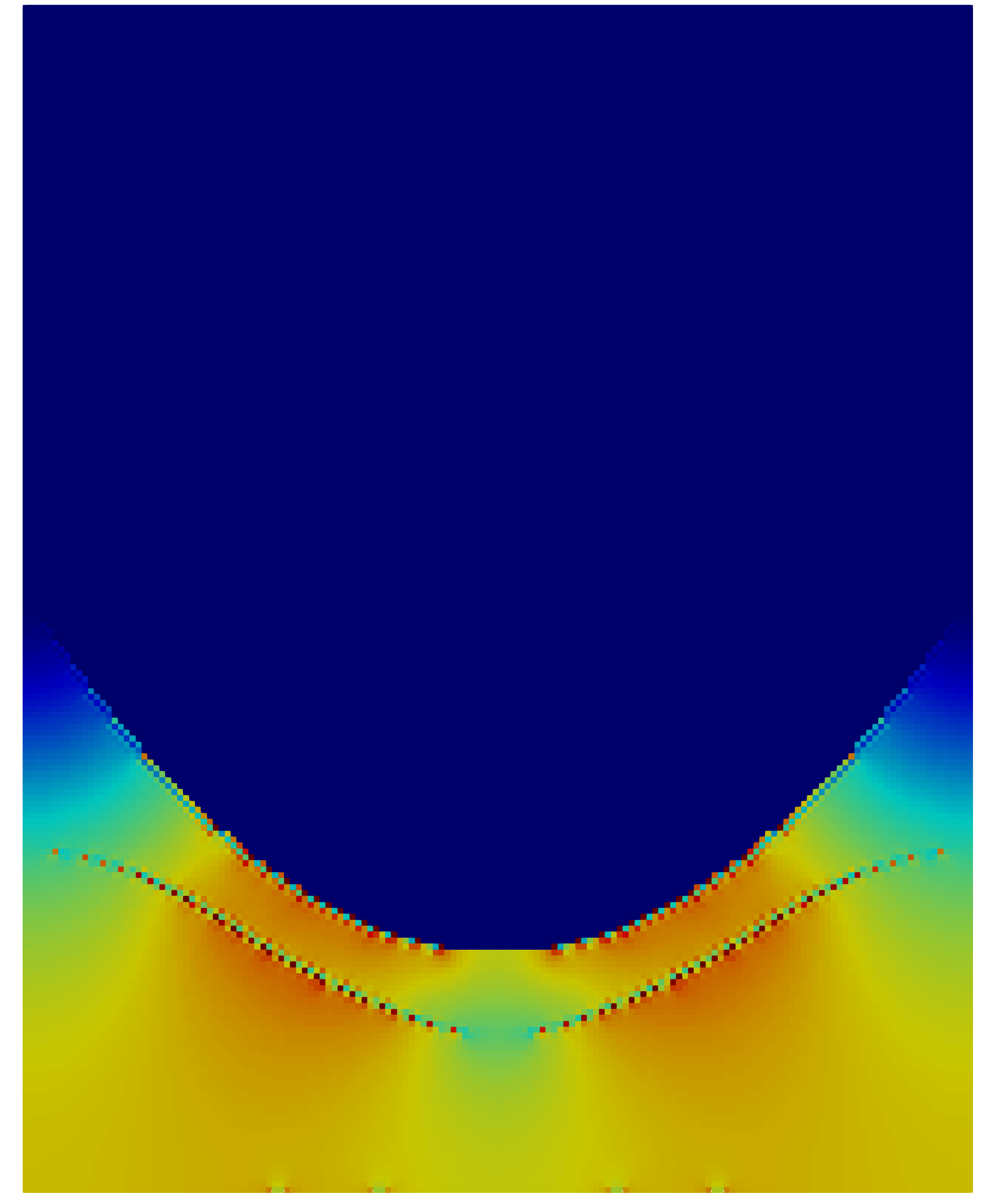}
  \end{subfigure}
  
  \vspace{4mm}
    
  \begin{subfigure}{.32\textwidth} 
      \centering 
      \begin{tikzpicture} 
        \node[inner sep=0pt] (pic) at (0,0) {\includegraphics[height=5mm, width=30mm]
        {03_Contour/00_Color_Maps/01_dsln_hor.png}};
        \node[inner sep=0pt] (0)   at ($(pic.south)+(-1.80, 0.26)$)  {$0$};
        \node[inner sep=0pt] (1)   at ($(pic.south)+( 1.80, 0.26)$)  {$1$};
        \node[inner sep=0pt] (d)   at ($(pic.south)+( 0.00, 0.85)$)  {$d~\si{[-]}$};
      \end{tikzpicture} 
  \end{subfigure}
  \begin{subfigure}{.32\textwidth} 
      \centering 
      \begin{tikzpicture} 
        \node[inner sep=0pt] (pic) at (0,0) {\includegraphics[height=5mm, width=30mm]
        {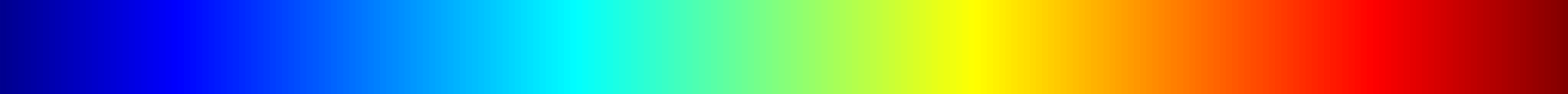}};
        \node[inner sep=0pt] (0)   at ($(pic.south)+(-1.80, 0.26)$)  {$0$};
        \node[inner sep=0pt] (10)  at ($(pic.south)+( 1.85, 0.26)$)  {$10$};
        \node[inner sep=0pt] (v)   at ($(pic.south)+( 0.00, 0.85)$)  {$v~\si{[\volt]}$};
      \end{tikzpicture} 
  \end{subfigure}
  \begin{subfigure}{.32\textwidth} 
      \centering 
      \begin{tikzpicture} 
        \node[inner sep=0pt] (pic) at (0,0) {\includegraphics[height=5mm, width=30mm]
        {03_Contour/00_Color_Maps/05_jet_hor.png}};
        \node[inner sep=0pt] (0)   at ($(pic.south)+(-1.80, 0.26)$)  {$0$};
        \node[inner sep=0pt] (8e5) at ($(pic.south)+( 2.15, 0.26)$)  {$\geq \hspace{-0.8mm} 8$e$5$};
        \node[inner sep=0pt] (j)   at ($(pic.south)+( 0.00, 0.85)$)  {$\|\j\|~\si{[\ampere\per\square\meter]}$};
      \end{tikzpicture} 
  \end{subfigure}  
  \caption{Evolution of the dissolution level $d$, the electric potential $v$ and the norm of the electric current density $\|\j\|$ for the parabolic cathode at time steps 100, 250 and 500. Due to varying dissolution levels and cathode ratios in adjacent elements, the visualization of the electric current density may yield minor non-uniformities on the surface of the anode and cathode.}
  \label{fig:Ex2_evolution}     
\end{figure}

Moreover, Fig.~\ref{fig:Ex2_evolution} shows the evolution of the dissolution level $d$, the electric potential $v$ and the electric current density $\j$ of the problem for two intermediate and the final configuration. Inside the cathode, the electric potential is $v = 0~\si{\V}$ and inside the anode $v \approx 10~\si{\V}$ with a transition zone in the electrolyte. The electric current density initially concentrates in the middle of the specimen leading to a pronounced material removal in this region. With progressing cathode feed, the electric current density's focus widens mapping a parabolic shape onto the anode.

\begin{figure}[htbp]    
  \centering
  \begin{tikzpicture}
           \node[inner sep=0pt] (pic) at (0,0) {\includegraphics[width=0.75\textwidth]
           {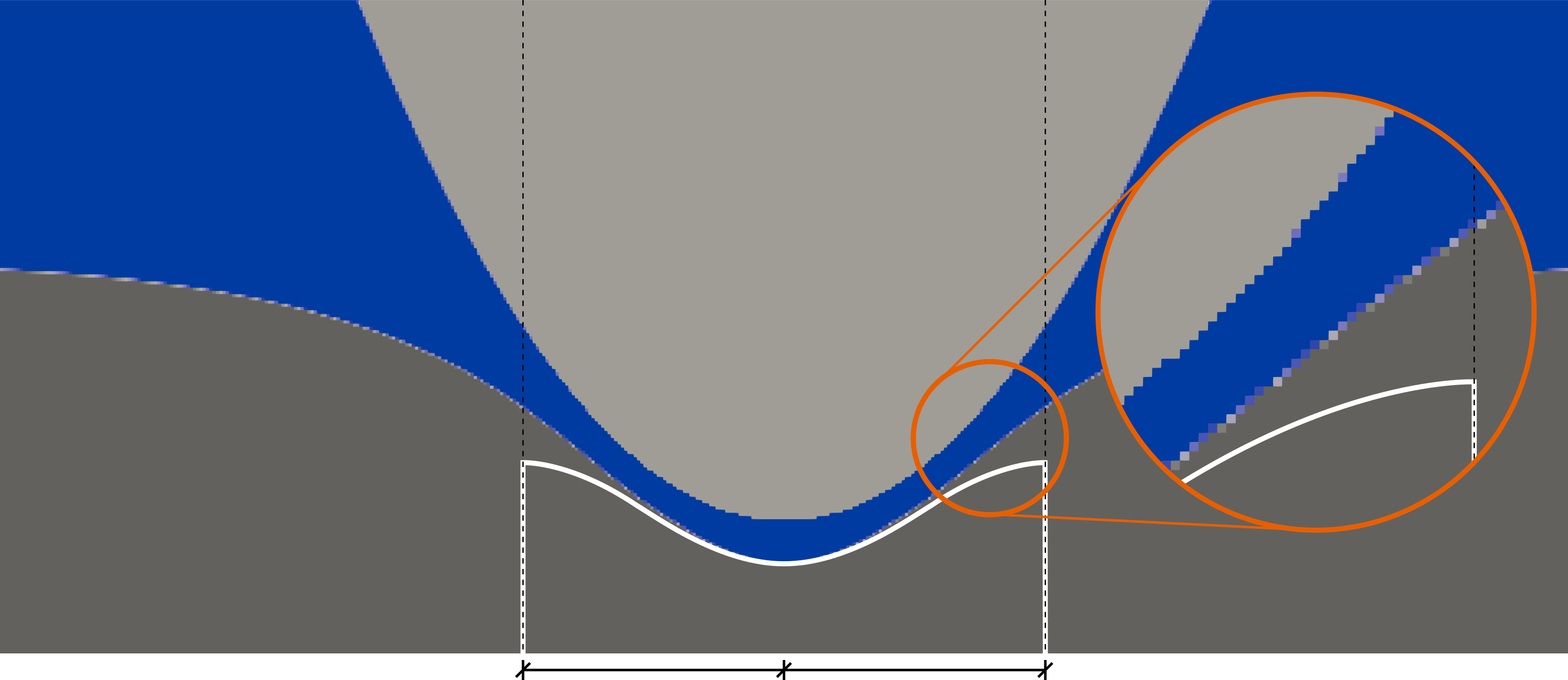}};
           \node[inner sep=0pt] (l1)   at ($(pic.south)+(-1.00,-0.25)$)  {$l$};
           \node[inner sep=0pt] (l2)   at ($(pic.south)+( 1.00,-0.25)$)  {$l$};
  \end{tikzpicture}
  \caption{Influence of specimen's length in $x$-direction on final surface shape. The white line shows the final surface of the specimen with a length of $2 \, l$. The contour plot shows the surface of the specimen with a length of $10 \, l$ in the range of $x \approx \pm \, 3 \, l$.}
  \label{fig:Ex2_lx}
\end{figure}

As discussed in \cite{HardistyMileham1999} and derived in \cite{NilsonTsuei1976}, a parabolic tool yields a parabolic workpiece. In the current example, a flattening of the workpiece's surface at the outer egdes is observed. The electric field lines are passing perpendicular to the anode's surface. Here, a zero horizontal flux is prescribed by the Neumann boundary conditions at the outer edge, which, therefore, leads to the flattening of the surface profile in this domain. Fig.~\ref{fig:Ex2_lx} shows the machined profile of the specimen with a total length of $10 \, l$. The specimen's extension releases the zero flux condition at $x = \pm \, l$ and avoids the pronounced flattening at this position.


\subsection{Wire cathode - \cite{SharmaSrivastavaEtAl2019}}
\label{ssec:Ex3}

\begin{figure}[htbp] 
     \centering 
     \begin{subfigure}{.475\textwidth} 
         \centering 
         \begin{tikzpicture}
           \node[inner sep=0pt] (pic) at (0,0) {\includegraphics[width=\textwidth]
           {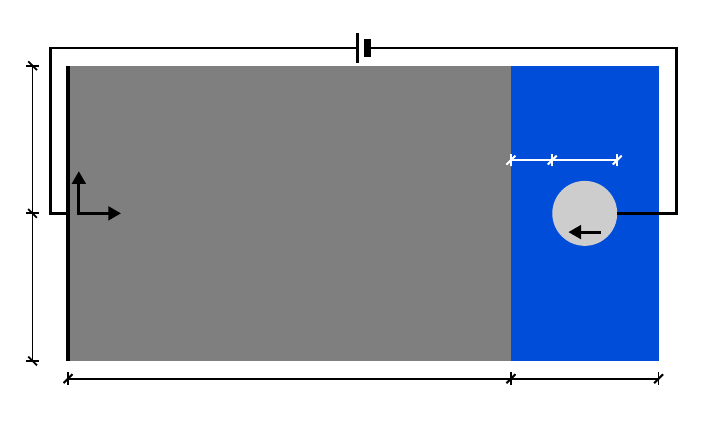}};
           \node[inner sep=0pt] (h1)   at ($(pic.west) +( 0.10,-0.80)$)  {$h$};
           \node[inner sep=0pt] (h2)   at ($(pic.west) +( 0.10, 0.80)$)  {$h$};
           \node[inner sep=0pt] (s)    at ($(pic.east) +(-1.90, 0.77)$)  {\textcolor{white}{$s$}};
           \node[inner sep=0pt] (rr)   at ($(pic.east) +(-1.30, 0.81)$)  {\textcolor{white}{$2r$}};
           \node[inner sep=0pt] (l1)   at ($(pic.south)+(-0.80, 0.13)$)  {$l_1$};
           \node[inner sep=0pt] (l2)   at ($(pic.south)+( 2.40, 0.13)$)  {$l_2$};
           \node[inner sep=0pt] (x)    at ($(pic.west) +( 1.50, 0.00)$)  {$x$};
           \node[inner sep=0pt] (y)    at ($(x.north)  +(-0.44, 0.44)$)  {$y$};
           \node[inner sep=0pt] (xdc)  at ($(pic.east) +(-1.32, 0.05)$)  {$\dot{x}_\mathrm{ca}$};
           \node[inner sep=0pt] (dv)   at ($(pic.north)+( 0.14,-0.11)$)  {$\delv$};
         \end{tikzpicture} 
         \caption{Geometry}
         \label{fig:Ex3Geom}
     \end{subfigure}
     \quad
     \begin{subfigure}{.475\textwidth} 
         \centering 
         \begin{tikzpicture}
           \node[inner sep=0pt] (pic) at (0,0) {\includegraphics[width=\textwidth]
           {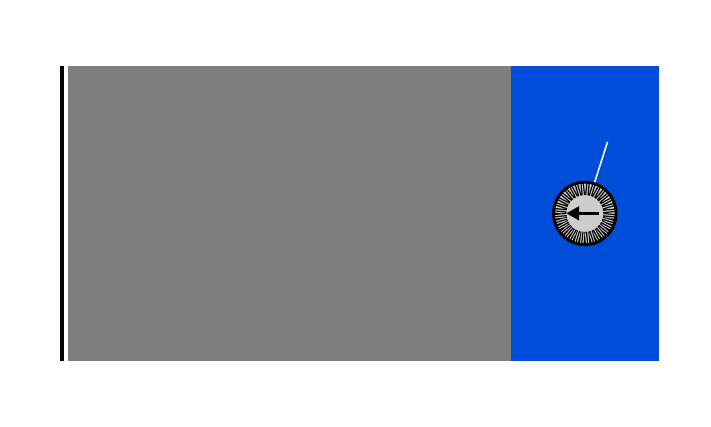}};
           \node[inner sep=0pt] (van)   at ($(pic.west) +( 0.20, 0.00)$)  {$\widetilde{v}_\mathrm{an}$};
           \node[inner sep=0pt] (vca)   at ($(pic.east) +(-1.20, 1.05)$)  {\textcolor{white}{$\widetilde{v}_\mathrm{ca}$}};
         \end{tikzpicture} 
         \caption{MBVP}
         \label{fig:Ex3MBVP}
     \end{subfigure} 
     \caption{Geometry and moving boundary value problem for a wire tool (cf.~\cite{SharmaSrivastavaEtAl2019}).} 
     \label{fig:Ex3}
\end{figure}
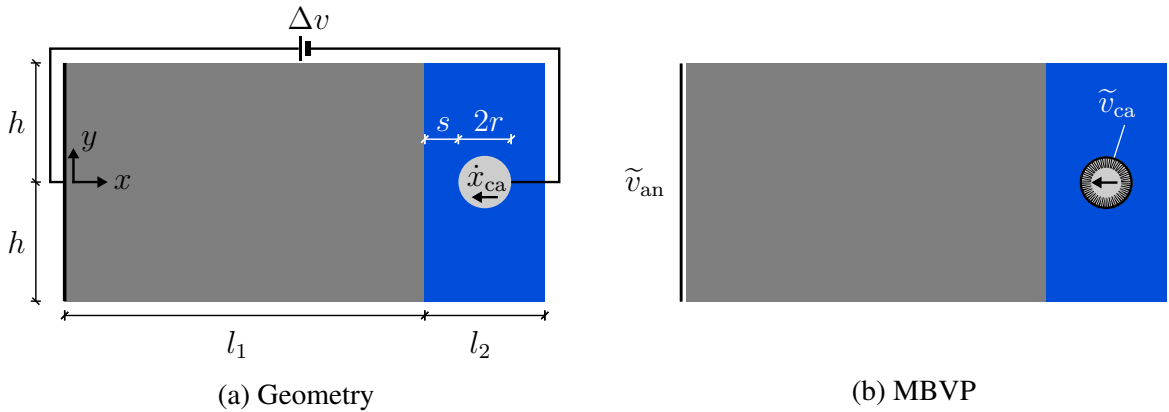

In the third example, a wire electrochemical machining application is investigated (Fig.~\ref{fig:Ex3}) and the results are compared with the experiments and simulations of \cite{SharmaSrivastavaEtAl2019}. The geometric dimensions read $l_1 = 700~\si{\um}$, $l_2 = 50~\si{\um}$, $h = 200~\si{\um}$, $s = 10~\si{\um}$, $r = 15~\si{\um}$ with a thickness of $100~\si{\um}$. The feed rate is $\dot{x}_\mathrm{ca} = 4~\si{\um\per\s}$, the potential difference $\delv = 6~\si{\V}$ and the electric conductivity $\kE\EL = 1.71~\si{\A\per\V\per\m}$. We set $\Veffnew = 1.09 \times 10^{-11}~\si{\cubic\m\per\A\per\s}$ and $\dt = 0.1~\si{\s}$. This example necessitates the use of method B to apply the cathode condition $\widetilde{v}_\mathrm{ca} = 0~\si{\V}$ only locally within the wire.

\begin{figure}[htbp]    
  \centering
  \begin{tikzpicture}
           \node[inner sep=0pt] (pic) at (0,0) {\includegraphics[width=0.75\textwidth]
           {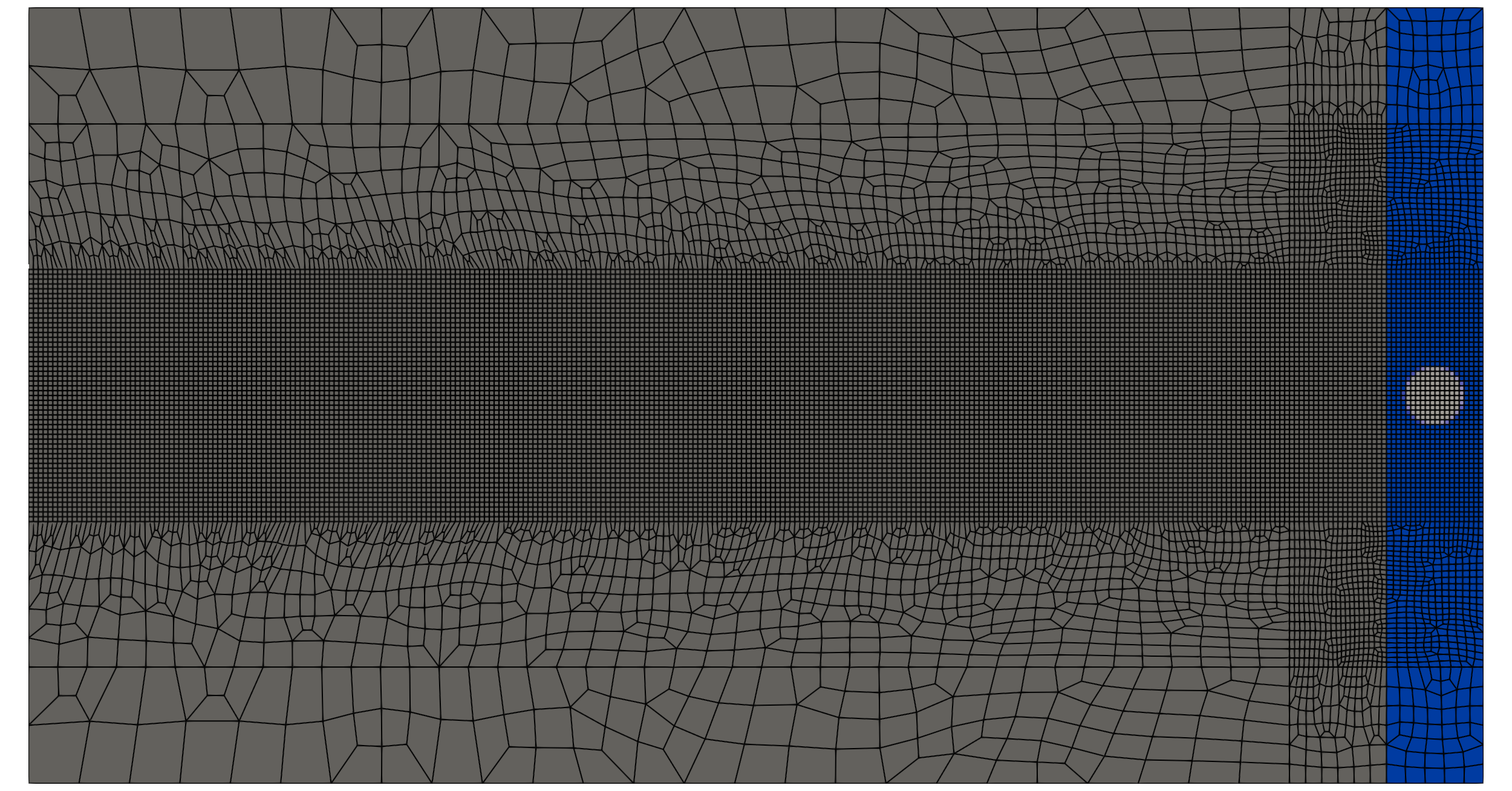}};
           \draw[color = sfb3, line width=1.5pt] (2, -1.9) rectangle (5.75, 1.9);
  \end{tikzpicture}
  \caption{Mesh for wire tool example (19995 elements). The orange box indicates the image section investigated in Fig.~\ref{fig:Ex3_evolution_d_j_zoom}.}
  \label{fig:Ex3_mesh}
\end{figure}

\begin{figure}[htbp] 
  \centering 
  \vspace*{-5mm}
  
  \begin{subfigure}{.34\textwidth} 
      \centering 
      \includegraphics[width=\textwidth]{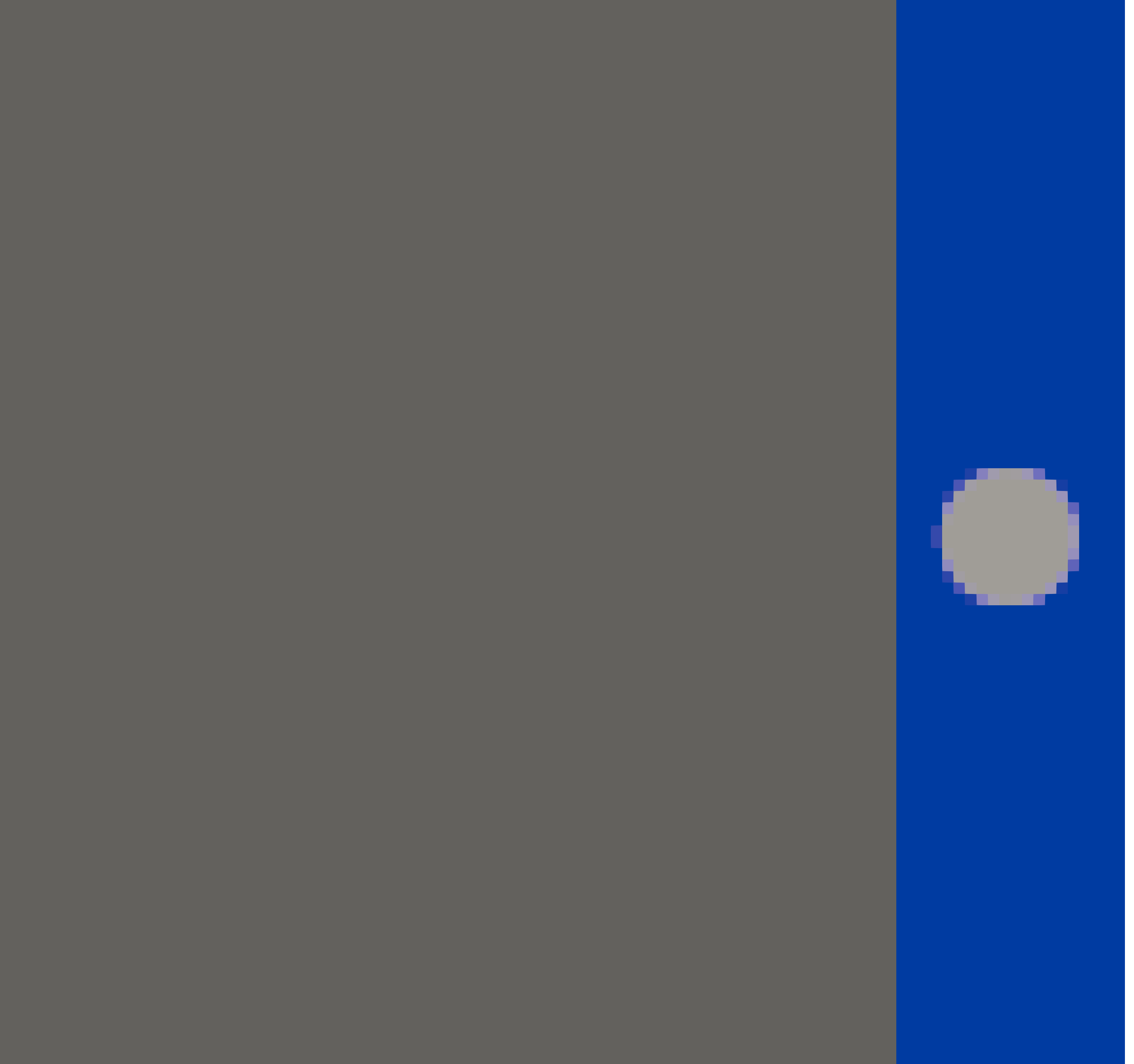}
  \end{subfigure}
  \begin{subfigure}{.34\textwidth} 
      \centering 
      \includegraphics[width=\textwidth]{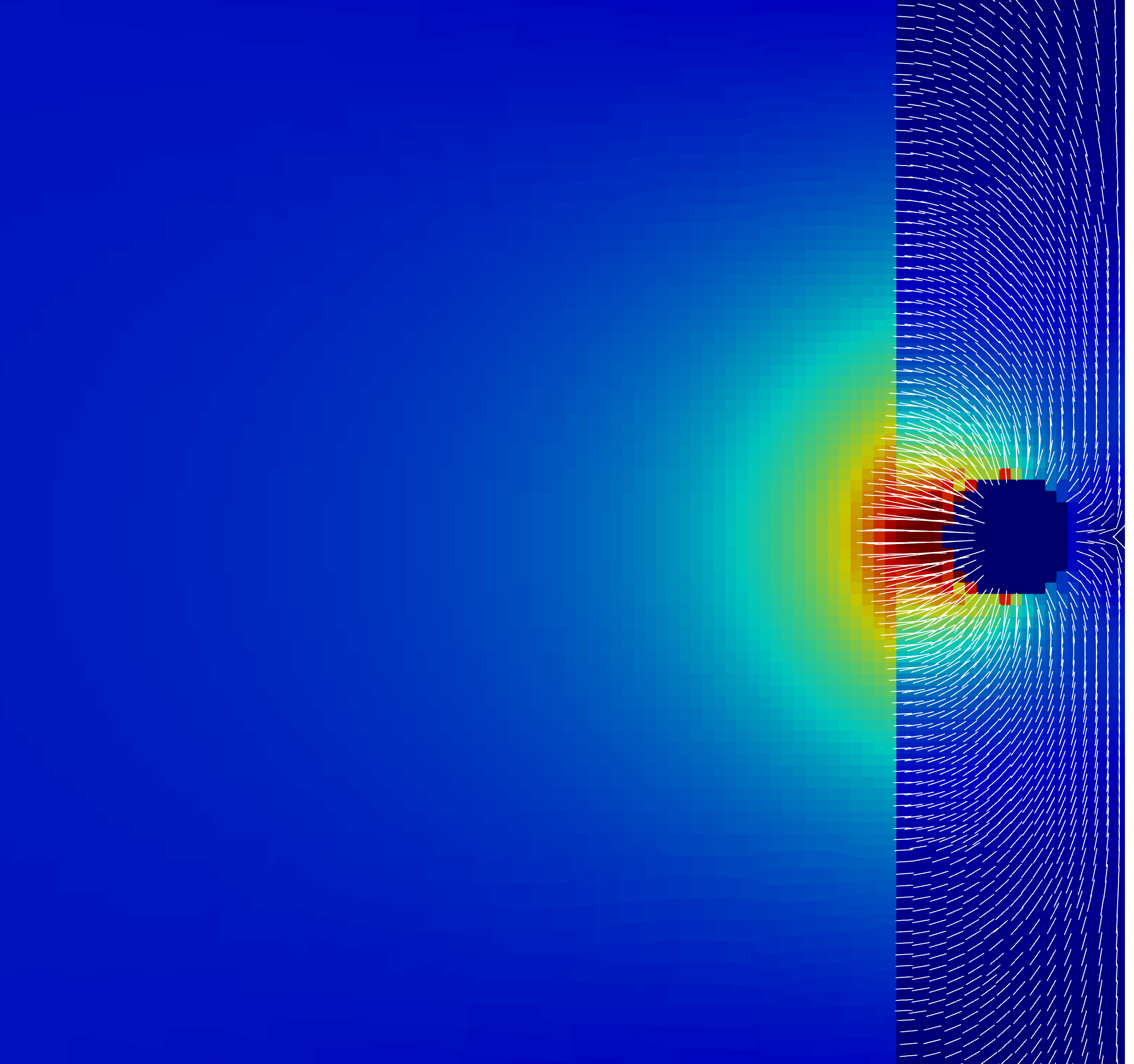}
  \end{subfigure}
  
  \vspace{1mm}
  
  \begin{subfigure}{.34\textwidth} 
      \centering 
      \includegraphics[width=\textwidth]{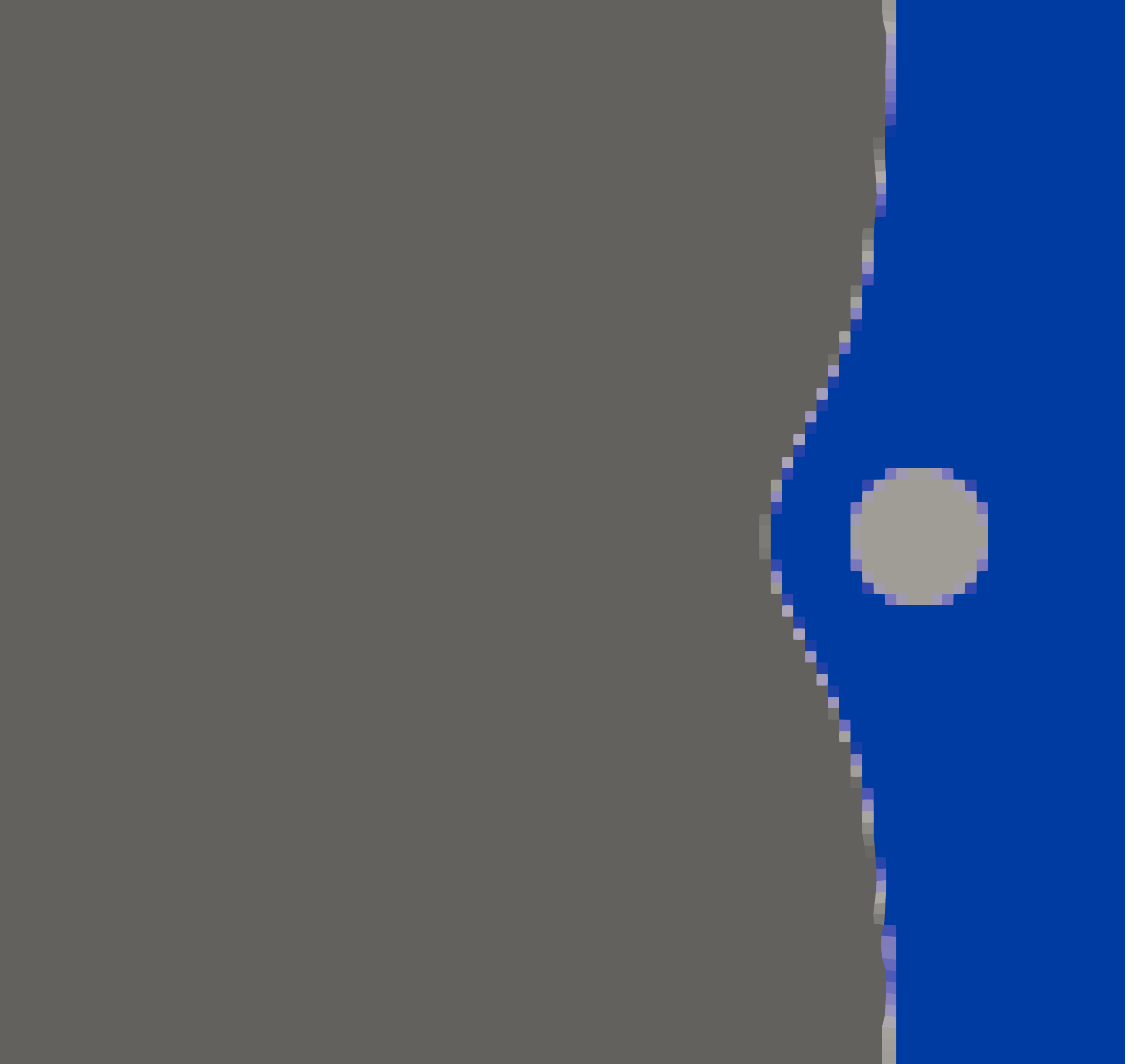}
  \end{subfigure}
  \begin{subfigure}{.34\textwidth} 
      \centering 
      \includegraphics[width=\textwidth]{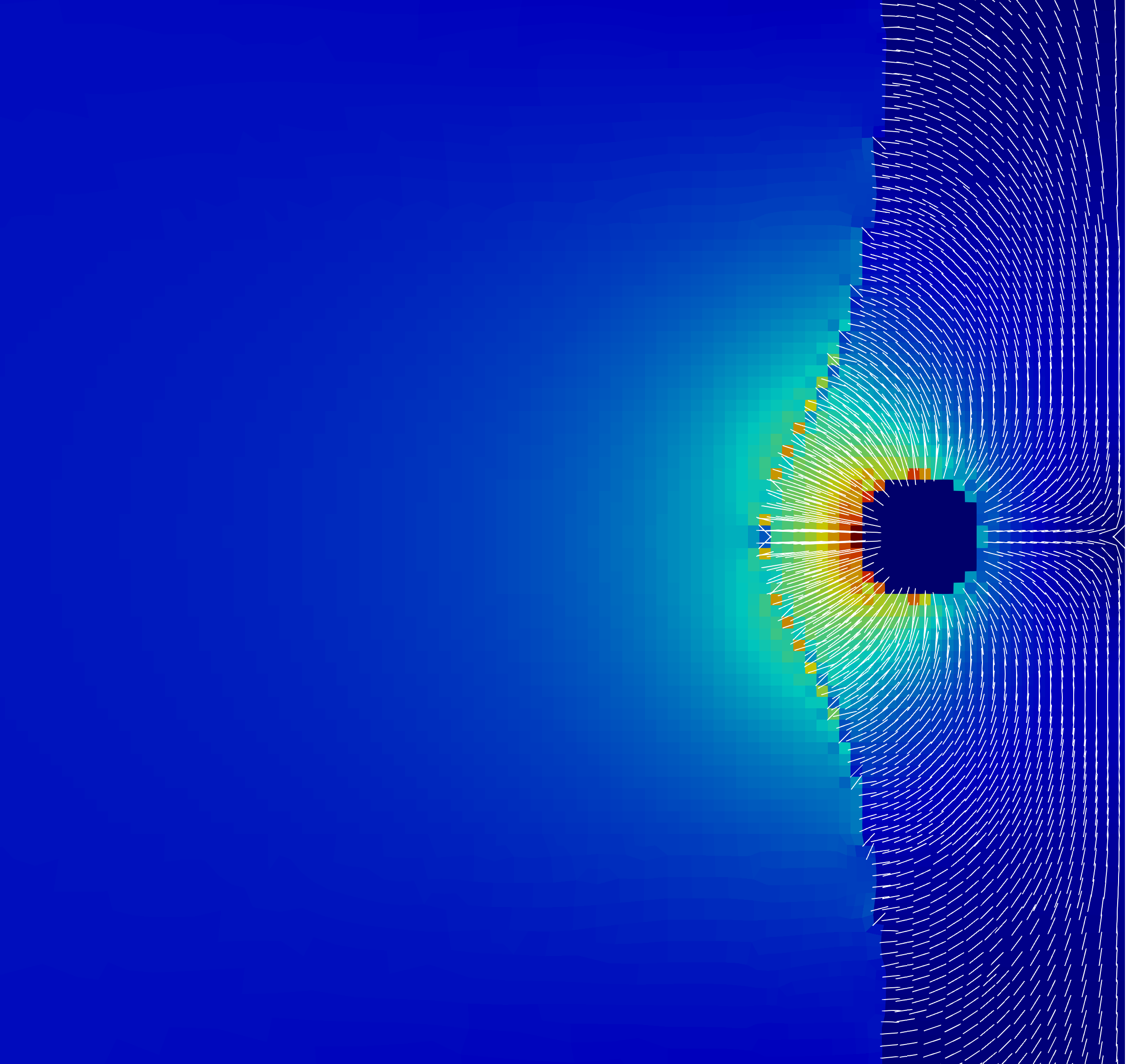}
  \end{subfigure}
  
  \vspace{1mm}  
  
  \begin{subfigure}{.34\textwidth} 
      \centering 
      \includegraphics[width=\textwidth]{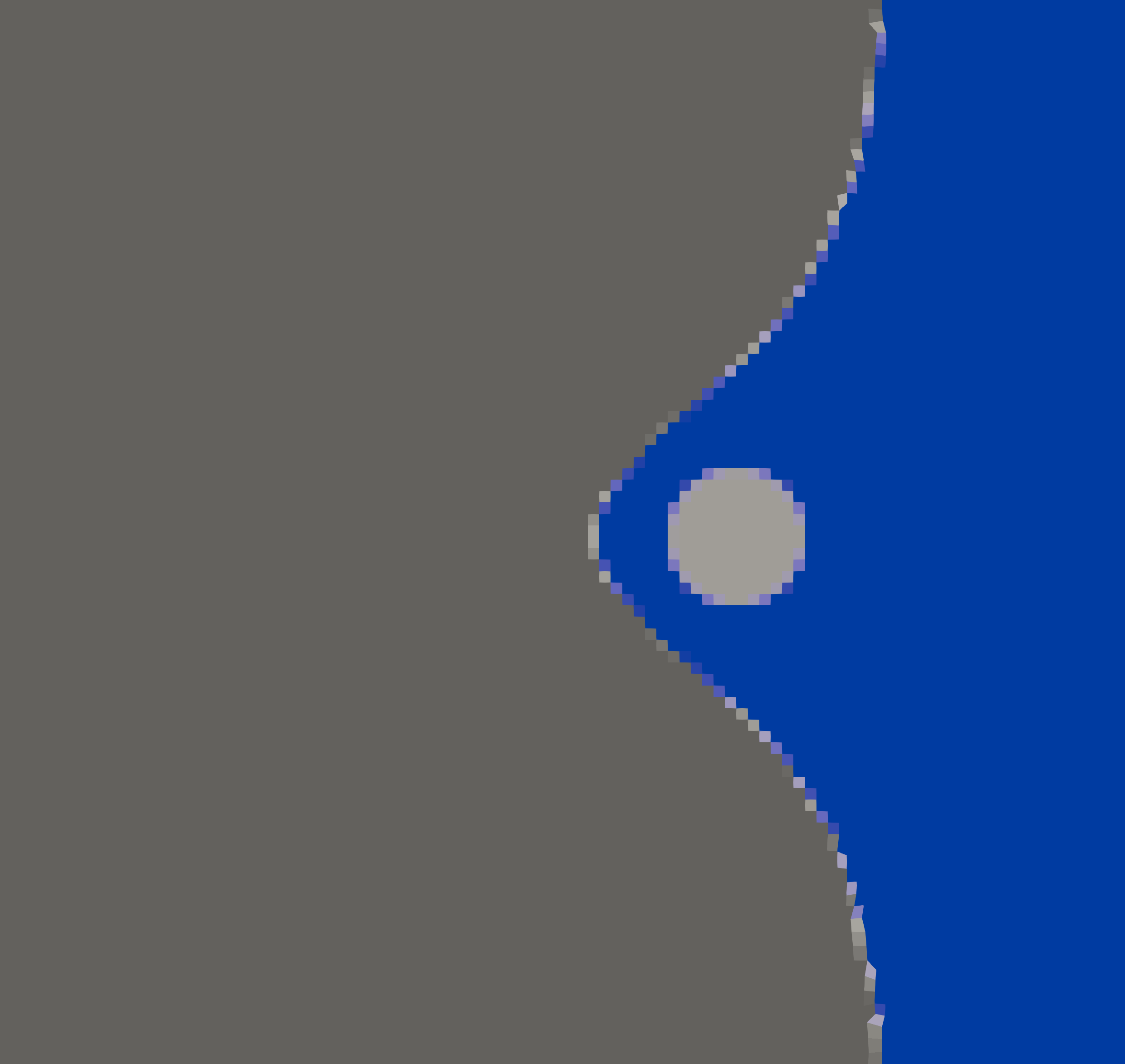}
  \end{subfigure}
  \begin{subfigure}{.34\textwidth} 
      \centering 
      \includegraphics[width=\textwidth]{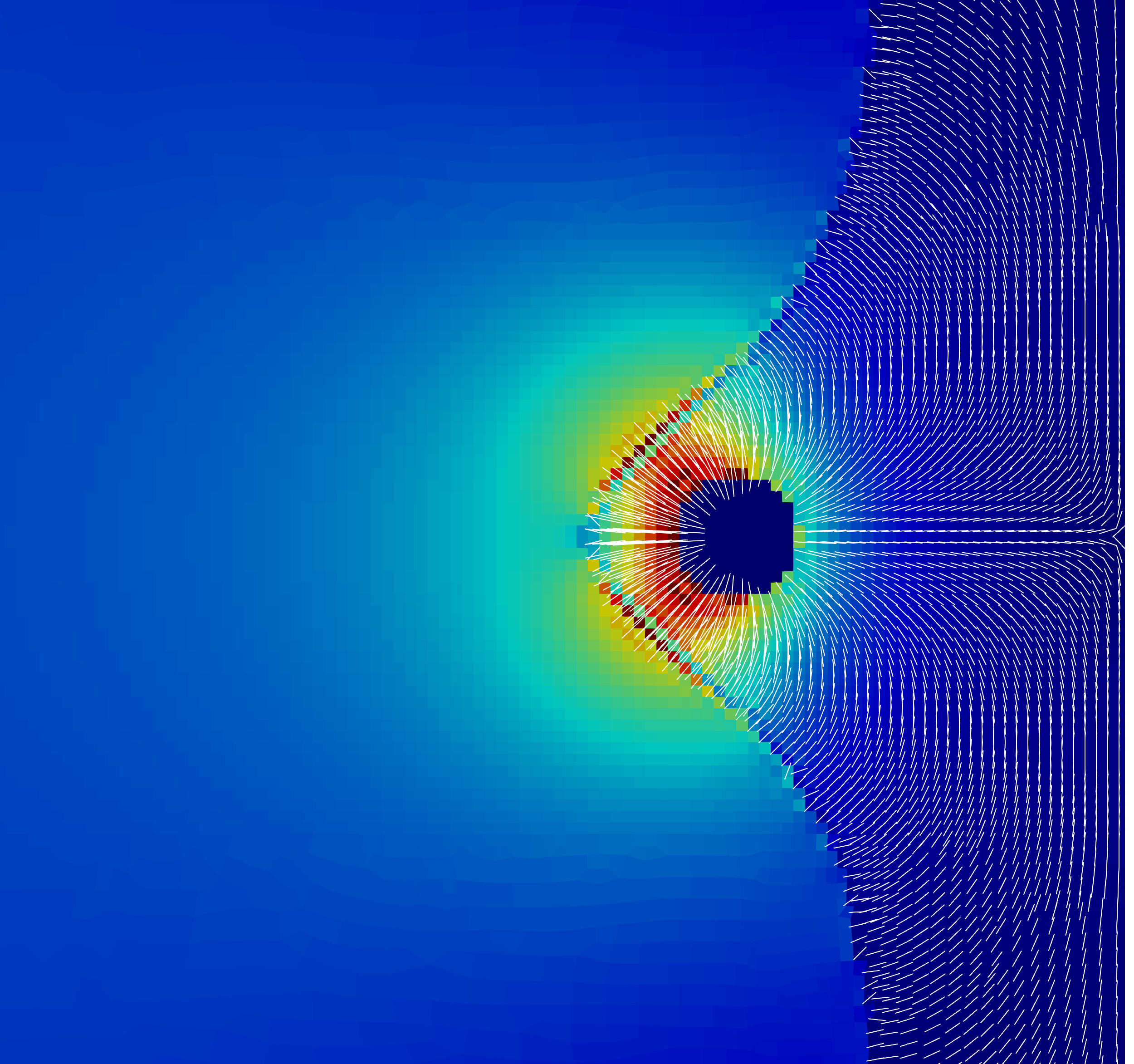}
  \end{subfigure}
  
  \vspace{1mm}
  
  \begin{subfigure}{.34\textwidth} 
      \centering 
      \includegraphics[width=\textwidth]{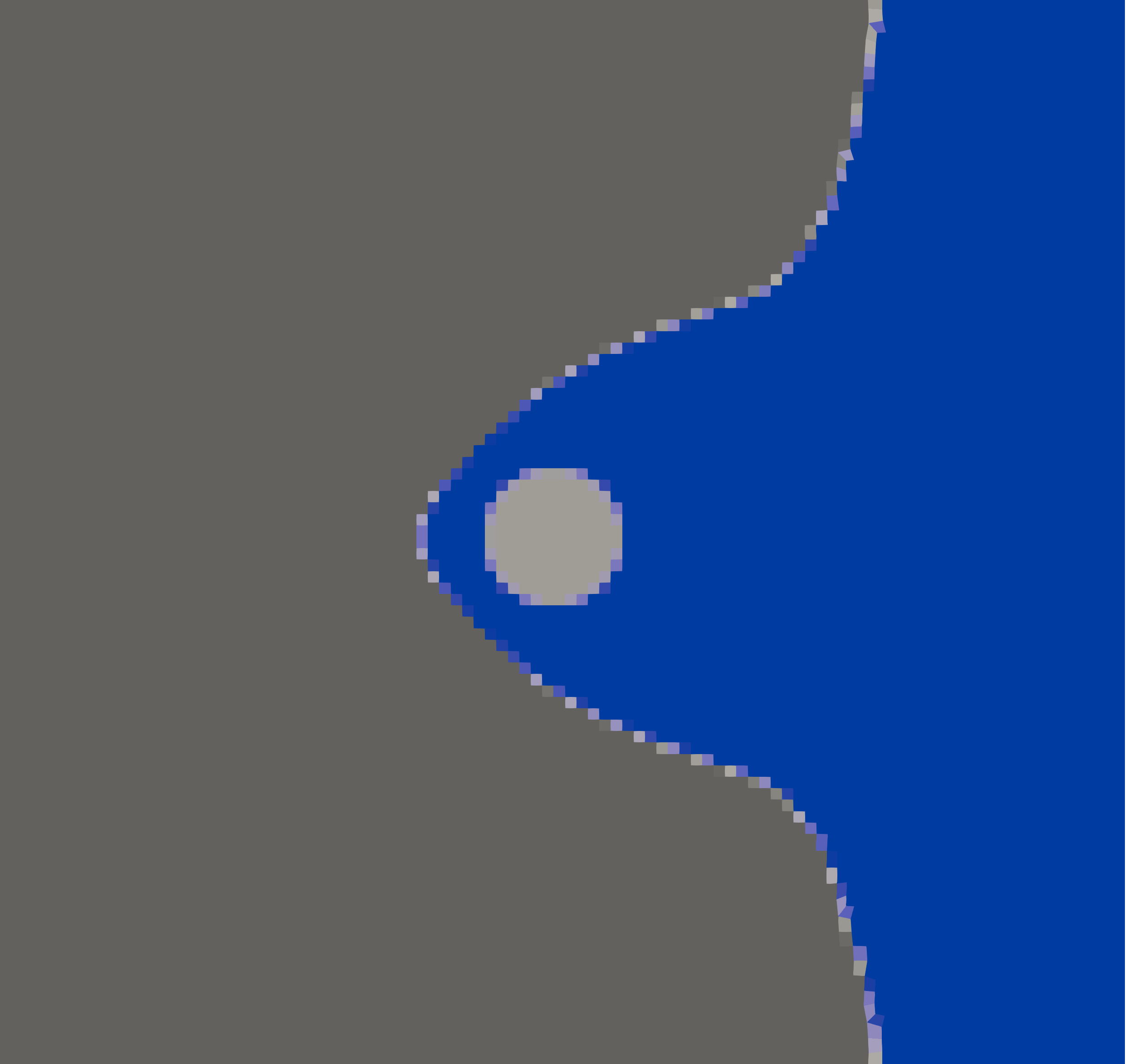}
  \end{subfigure}
  \begin{subfigure}{.34\textwidth} 
      \centering 
      \includegraphics[width=\textwidth]{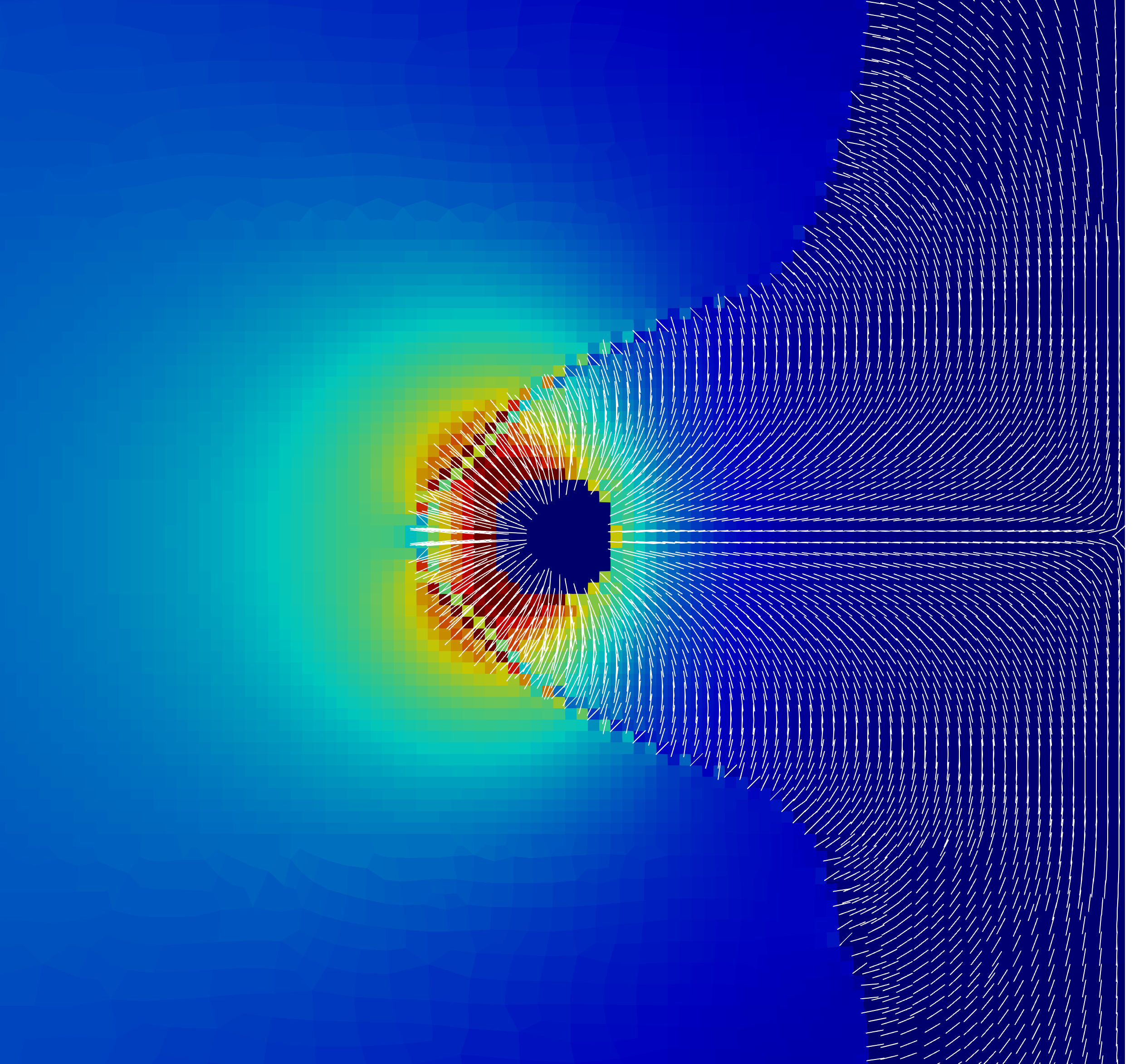}
  \end{subfigure}
  
  \vspace{3mm}

  \begin{subfigure}{.34\textwidth} 
      \centering 
      \begin{tikzpicture} 
        \node[inner sep=0pt] (pic) at (0,0) {\includegraphics[height=5mm, width=30mm]
        {03_Contour/00_Color_Maps/01_dsln_hor.png}};
        \node[inner sep=0pt] (0)   at ($(pic.south)+(-1.80, 0.26)$)  {$0$};
        \node[inner sep=0pt] (1)   at ($(pic.south)+( 1.80, 0.26)$)  {$1$};
        \node[inner sep=0pt] (d)   at ($(pic.south)+( 0.00, 0.85)$)  {$d~\si{[-]}$};
      \end{tikzpicture} 
  \end{subfigure}
  \begin{subfigure}{.34\textwidth} 
      \centering 
      \begin{tikzpicture} 
        \node[inner sep=0pt] (pic) at (0,0) {\includegraphics[height=5mm, width=30mm]
        {03_Contour/00_Color_Maps/05_jet_hor.png}};
        \node[inner sep=0pt] (0)   at ($(pic.south)+(-1.80, 0.26)$)  {$0$};
        \node[inner sep=0pt] (8e5) at ($(pic.south)+( 2.15, 0.26)$)  {$\geq \hspace{-0.8mm} 1$e$6$};
        \node[inner sep=0pt] (j)   at ($(pic.south)+( 0.00, 0.85)$)  {$\|\j\|~\si{[\ampere\per\square\meter]}$};
      \end{tikzpicture} 
  \end{subfigure}  
  \caption{Evolution of the dissolution level $d$ and the norm of the electric current density $\|\j\|$, where the white lines show the orientation of the electric field, for the wire tool. Zoom to the right edge of the specimen at time $0~\si{\s}$, $5~\si{\s}$, $15~\si{\s}$ and $25~\si{\s}$.}
  \label{fig:Ex3_evolution_d_j_zoom}     
\end{figure}

Fig.~\ref{fig:Ex3_mesh} shows the mesh employed in the simulation which possesses strong refinement in the region of the kerf. Moreover, the orange box in Fig.~\ref{fig:Ex3_mesh} shows the image section of Fig.~\ref{fig:Ex3_evolution_d_j_zoom} where the initial phase of the dissolution process is studied.
Here, the dissolution level $d$ and the norm of the electric current density $\|\j\|$ are given. Initially, the electric current density exhibits its maximum in horizontal direction, thereby, inducing the highest material removal at this position. Afterwards, the electric current density's scope becomes larger which causes a widening of the kerf. After $25~\si{\s}$, the distribution of the electric current density between wire and work piece remains approximately constant and yields a clean cut.

\begin{figure}[htbp]    
  \centering
  \begin{tikzpicture}
           \node[inner sep=0pt] (pic) at (0,0) {\includegraphics[width=0.75\textwidth]
           {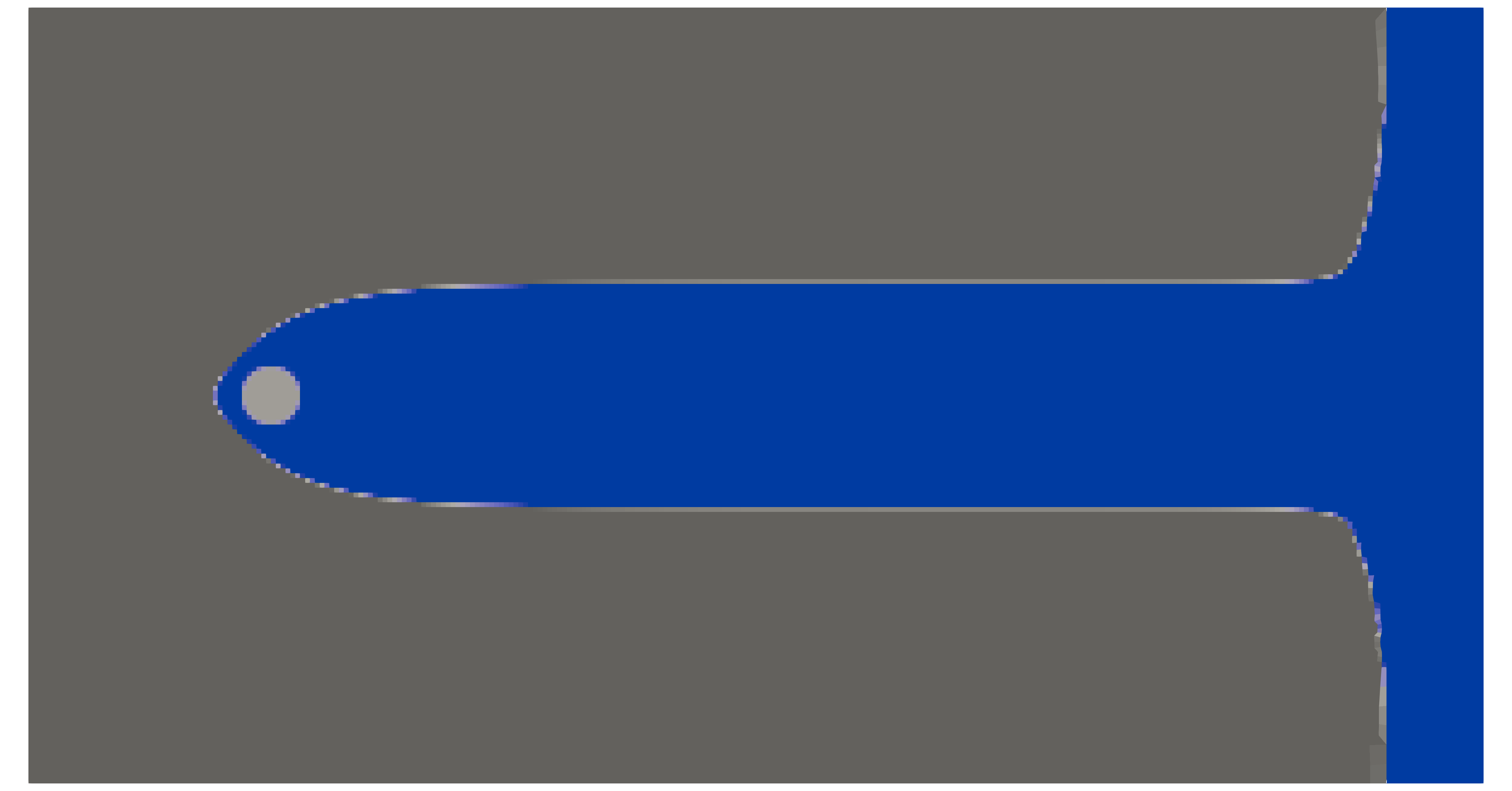}};
  \end{tikzpicture}
  \caption{Final shape of the specimen after machining with the wire tool after a machining time of $150~\si{\s}$.}
  \label{fig:Ex3_final_shape}
\end{figure}

Fig.~\ref{fig:Ex3_final_shape} shows the final shape of the specimen after a machining time of $150~\si{\s}$. Measuring the kerf width at $x = 400~\si{\um}$ yields a width of $118.73~\si{\um}$ which is in the range between $110.08~\si{\um}$ (experimental kerf width) and $124.27~\si{\um}$ (simulated kerf width) of \cite{SharmaSrivastavaEtAl2019}.

The simulation required $10~\si{\hour}~51~\si{\minute}$ and, thus, was faster than the simulation with the fine discretization of \cite{SharmaSrivastavaEtAl2019} whose code was implemented in MATLAB (R2016a) and required $16~\si{\hour}~03~\si{\minute}$. Further speedups are realized by considering an additional moving anode boundary condition. This prescribes $\widetilde{v}_\mathrm{an} = 6~\si{\V}$ in the region $50~\si{\um}$ below the work piece's initial surface and moves horizontally with the cathode's feed rate. Hereby, the computation time is reduced by $59~\si{\%}$ to $4~\si{\hour}~28~\si{\minute}$ and is even faster than the simulation with the coarse discretization of \cite{SharmaSrivastavaEtAl2019} with $6~\si{\hour}~19~\si{\minute}$.


\subsection{Complex cathode - blade machining}
\label{ssec:Ex4}

Inspired by e.g.~\cite{TsuboiYamamoto2009}, \cite{KlockeZeisEtAl2013} or \cite{ZhuZhaoEtAl2017}, we study in this example a blade machining process with a complex cathode geometry, where two tools move towards an enclosed workpiece.

\begin{figure}[htbp] 
     \centering 
     \begin{subfigure}{.45\textwidth} 
         \centering 
         \begin{tikzpicture}
           \node[inner sep=0pt] (pic) at (0,0) {\includegraphics[width=\textwidth]
           {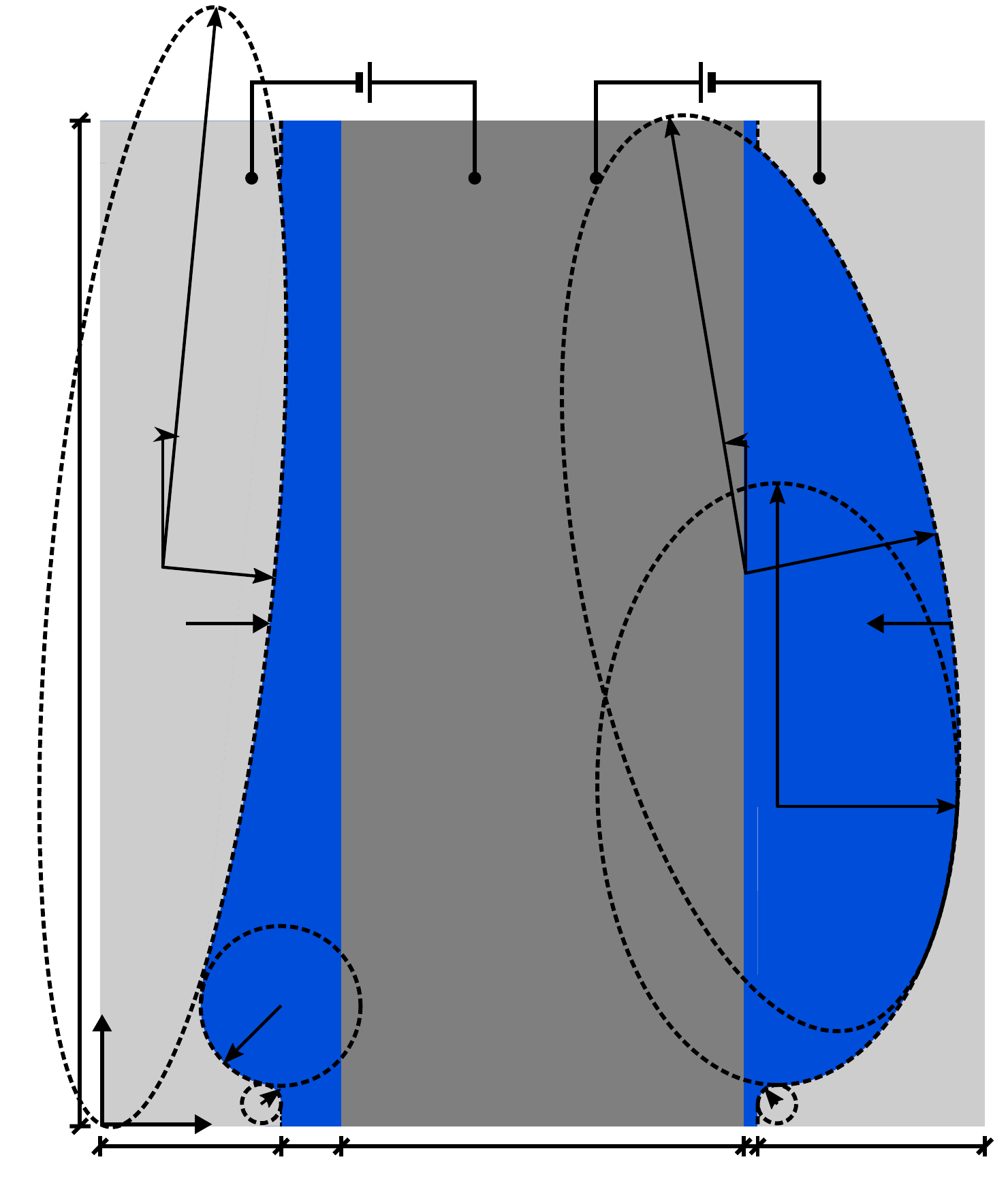}};
           \node[inner sep=0pt] (h1)   at ($(pic.west) +( 0.10,-0.20)$)  {$h$};
           \node[inner sep=0pt] (l1)   at ($(pic.south)+(-2.10, 0.12)$)  {$l_1$};
           \node[inner sep=0pt] (l2)   at ($(pic.south)+( 0.40, 0.12)$)  {$l_2$};
           \node[inner sep=0pt] (l3)   at ($(pic.south)+( 2.60, 0.12)$)  {$l_3$};
           \node[inner sep=0pt] (s1)   at ($(pic.south)+(-1.32, 0.04)$)  {$s_1$};
           \node[inner sep=0pt] (s2)   at ($(pic.south)+( 1.80, 0.04)$)  {$s_2$};
           \node[inner sep=0pt] (x)    at ($(pic.south)+(-2.15, 0.78)$)  {$x$};
           \node[inner sep=0pt] (y)    at ($(x.north)  +(-0.47, 0.42)$)  {$y$};
           \node[inner sep=0pt] (xdc1) at ($(pic.north)+(-2.54,-4.44)$)  {$\dot{x}_\mathrm{ca}$};
           \node[inner sep=0pt] (xdc2) at ($(pic.north)+( 2.30,-4.44)$)  {$\dot{x}_\mathrm{ca}$};
           \node[inner sep=0pt] (dv1)  at ($(pic.north)+(-0.90,-0.25)$)  {$\delv$};
           \node[inner sep=0pt] (dv2)  at ($(pic.north)+( 1.50,-0.25)$)  {$\delv$};
           \node[inner sep=0pt] (r1)   at ($(pic.south)+(-1.34, 0.66)$)  {$r_1$};
           \node[inner sep=0pt] (r2)   at ($(pic.south)+(-1.36, 1.45)$)  {$r_2$};
           \node[inner sep=0pt] (r3)   at ($(pic.south)+(-2.00, 4.65)$)  {$r_3$};
           \node[inner sep=0pt] (r4)   at ($(pic.south)+(-1.96, 6.25)$)  {$r_4$};
           \node[inner sep=0pt] (al1)  at ($(pic.south)+(-2.05, 5.43)$)  {$\alpha_1$};
           \node[inner sep=0pt] (al2)  at ($(pic.south)+( 1.28, 5.43)$)  {$\alpha_2$};
           \node[inner sep=0pt] (r5)   at ($(pic.south)+( 1.22, 6.25)$)  {$r_5$};
           \node[inner sep=0pt] (r6)   at ($(pic.south)+( 2.28, 4.80)$)  {$r_6$};
           \node[inner sep=0pt] (r7)   at ($(pic.south)+( 2.22, 3.45)$)  {$r_7$};
           \node[inner sep=0pt] (r8)   at ($(pic.south)+( 2.55, 3.00)$)  {$r_8$};
           \node[inner sep=0pt] (r9)   at ($(pic.south)+( 2.33, 0.66)$)  {$r_9$};
         \end{tikzpicture} 
         \caption{Geometry}
         \label{fig:Ex4Geom}
     \end{subfigure}
     \qquad
     \begin{subfigure}{.45\textwidth} 
         \centering 
         \begin{tikzpicture}
           \node[inner sep=0pt] (pic) at (0,0) {\includegraphics[width=\textwidth]
           {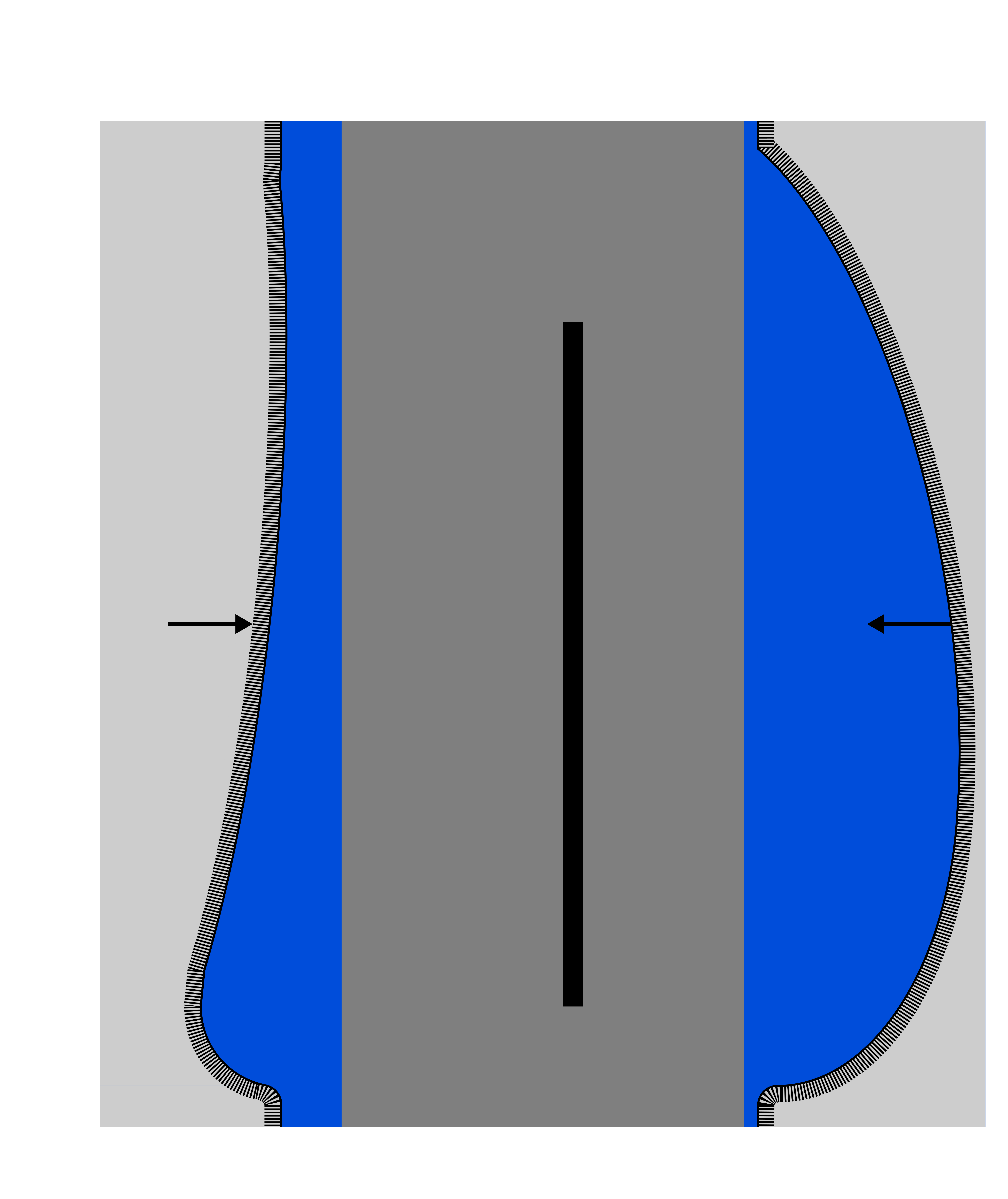}};
           \node[inner sep=0pt] (van)   at ($(pic.south)+( 0.50, 6.70)$)  {$\widetilde{v}_\mathrm{an}$};
           \node[inner sep=0pt] (vca1)  at ($(pic.south)+(-2.15, 6.70)$)  {$\widetilde{v}_\mathrm{ca}$};
           \node[inner sep=0pt] (vca2)  at ($(pic.south)+( 3.00, 6.70)$)  {$\widetilde{v}_\mathrm{ca}$};
         \end{tikzpicture} 
         \caption{MBVP}
         \label{fig:Ex4MBVP}
     \end{subfigure} 
     \caption{Geometry and moving boundary value problem with complex cathode shape.} 
     \label{fig:Ex4}
\end{figure}
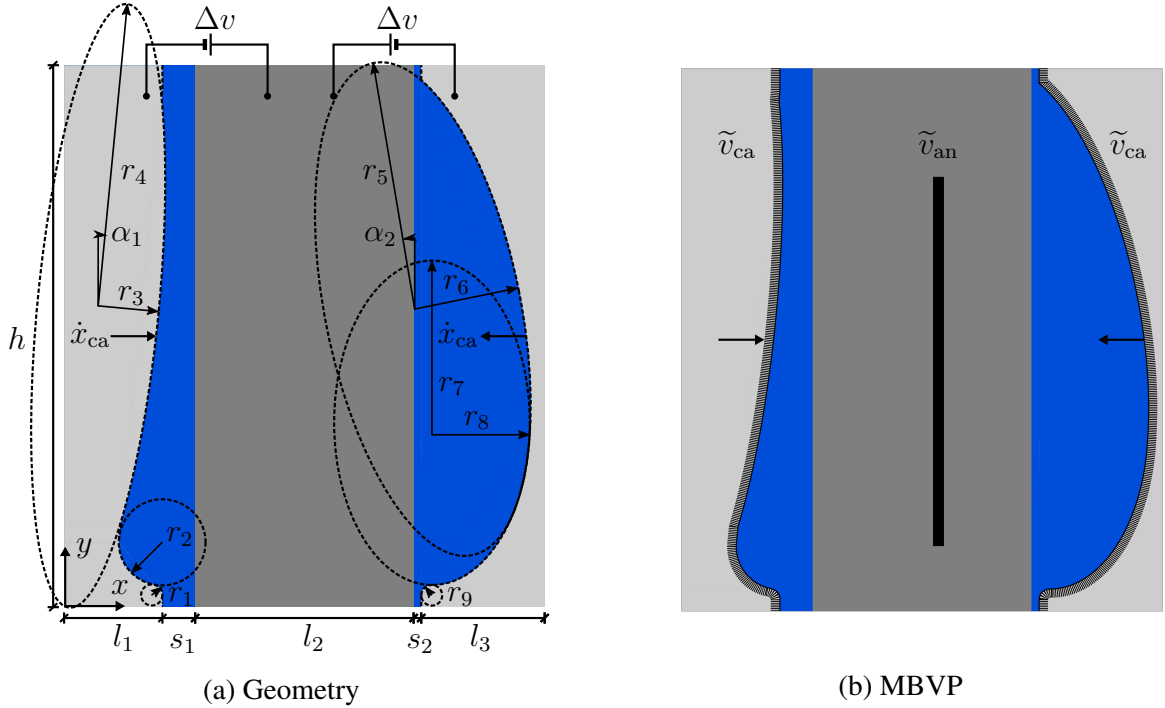

The material parameters stem from Table~\ref{tab:matpar} and the geometric dimensions (in [\si{\mm}]) read $l_1 = 4.5$, $s_1 = 1.5$, $l_2 = 10$, $s_2 = 0.35$, $l_3 = 5.65$, $h = 25$ with a thickness of $0.2$ (Fig.~\ref{fig:Ex4}). Moreover, the following circles define the cathode's surface with $r_1 = 0.5$ (center: $4~|~0.551$), $r_2=2$ (center: $4.5~|~3$), $r_9=0.49$ (center: $16.840~|~0.551$). Additionally, the following ellipses complete the surface with $r_3 = 2.8$, $r_4 = 14$ (center: $1.56~|~13.9$, rotation: $\alpha_1 = 5.44~\si{\degree}$), $r_5 = 11.25$, $r_6 = 4.5$ (center: $16.05~|~13.75$, rotation: $\alpha_2 = 11.5~\si{\degree}$) and $r_7 = 7.5$, $r_8 = 4.5$ (center: $16.35~|~8.5$).

The feed rate is $\dot{x}_\mathrm{ca} = 0.01~\si{\mm\per\s}$ and the potential difference $\delv = 20~\si{\V}$ with $\widetilde{v}_\mathrm{an} = 20~\si{\V}$ which is prescribed in the region $x \in [11.5,12]$ and $y \in [3,20]$. The time increment reads $\dt = 2~\si{\s}$ and method B is utilized.

\begin{figure}[htbp] 
  \centering 
  
  \begin{subfigure}{.34\textwidth} 
      \centering 
      \includegraphics[width=\textwidth]{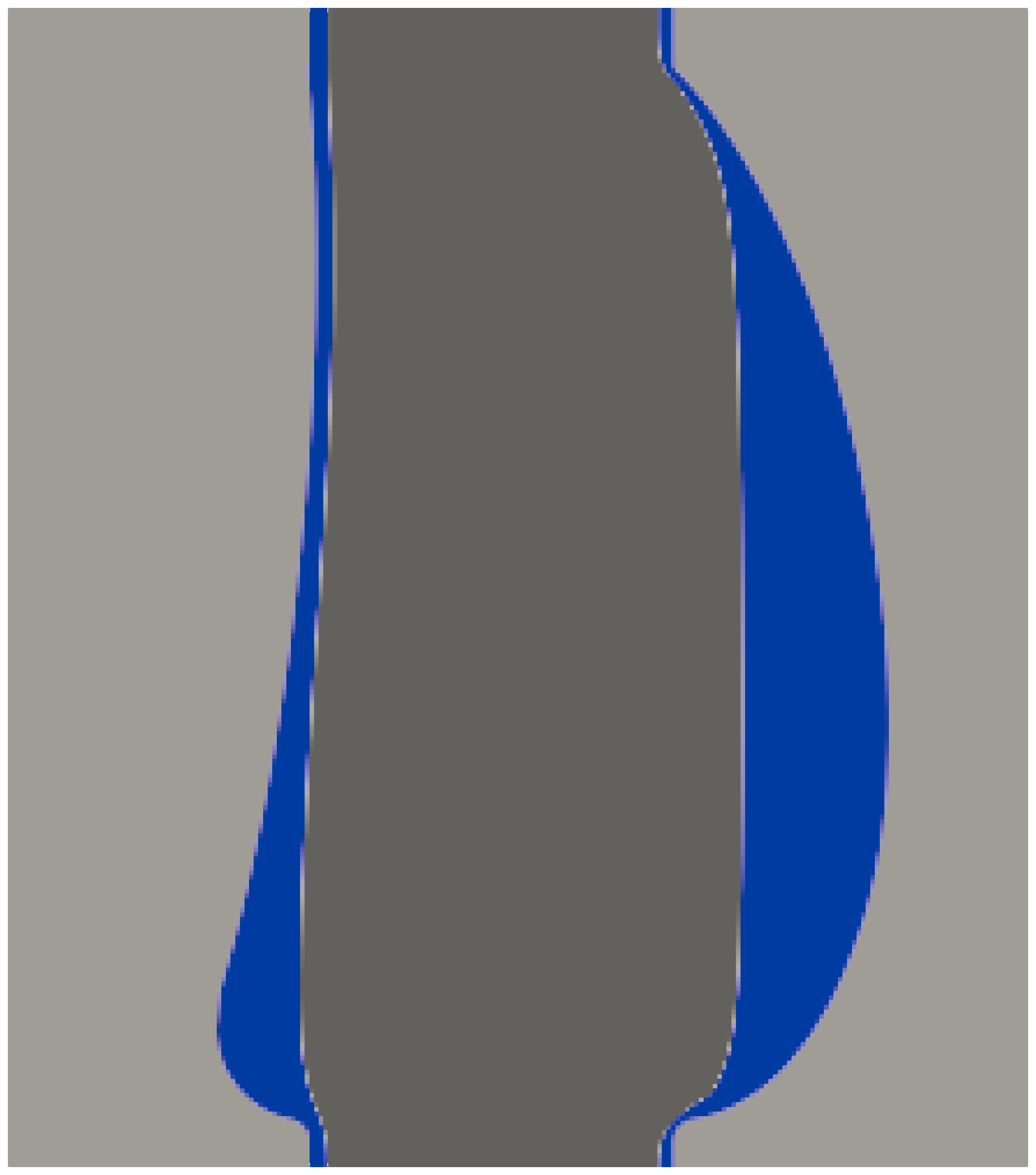}
  \end{subfigure}
  \begin{subfigure}{.34\textwidth} 
      \centering 
      \includegraphics[width=\textwidth]{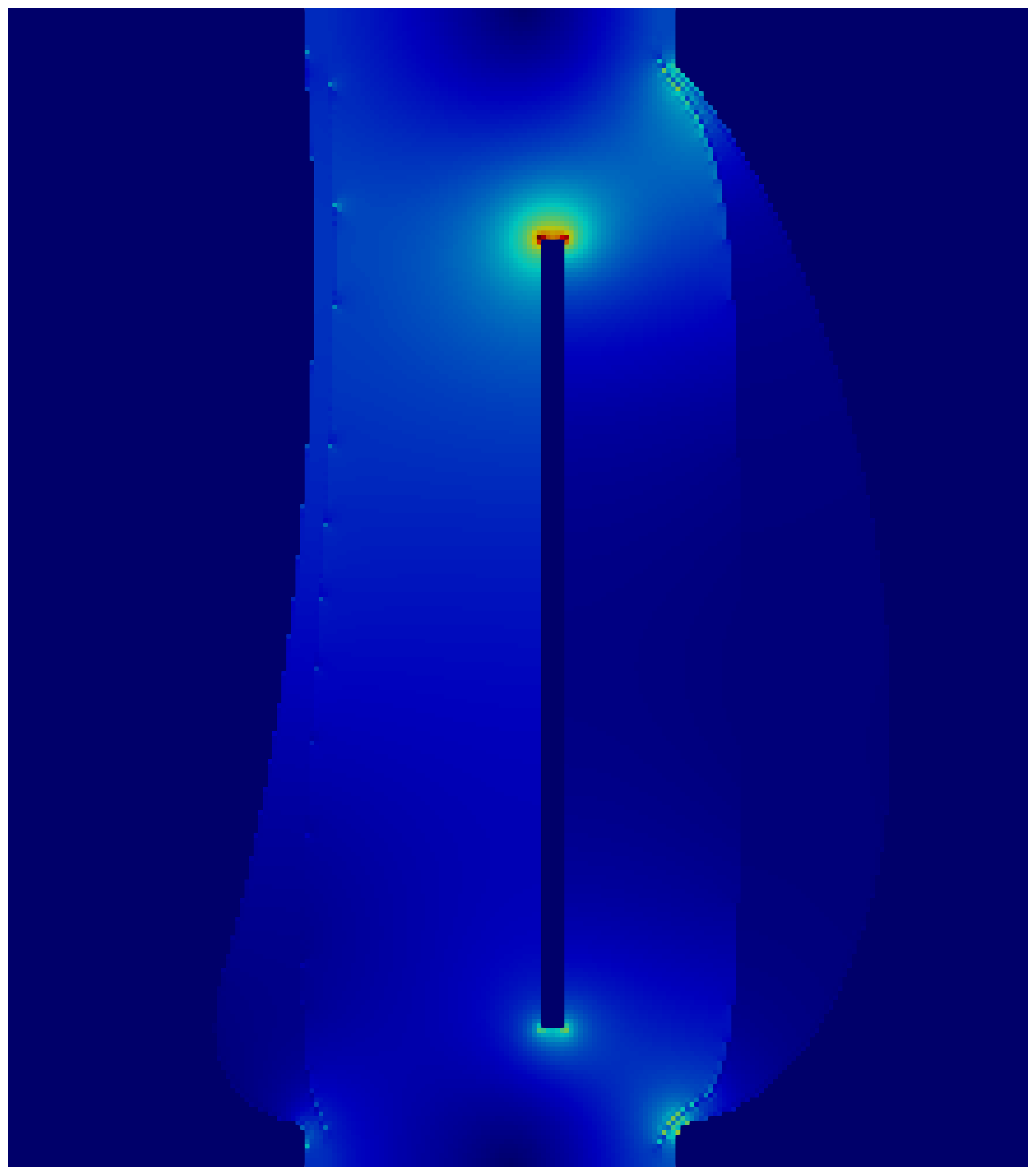}
  \end{subfigure}
  
  \vspace{0.8mm}

  \begin{subfigure}{.34\textwidth} 
      \centering 
      \includegraphics[width=\textwidth]{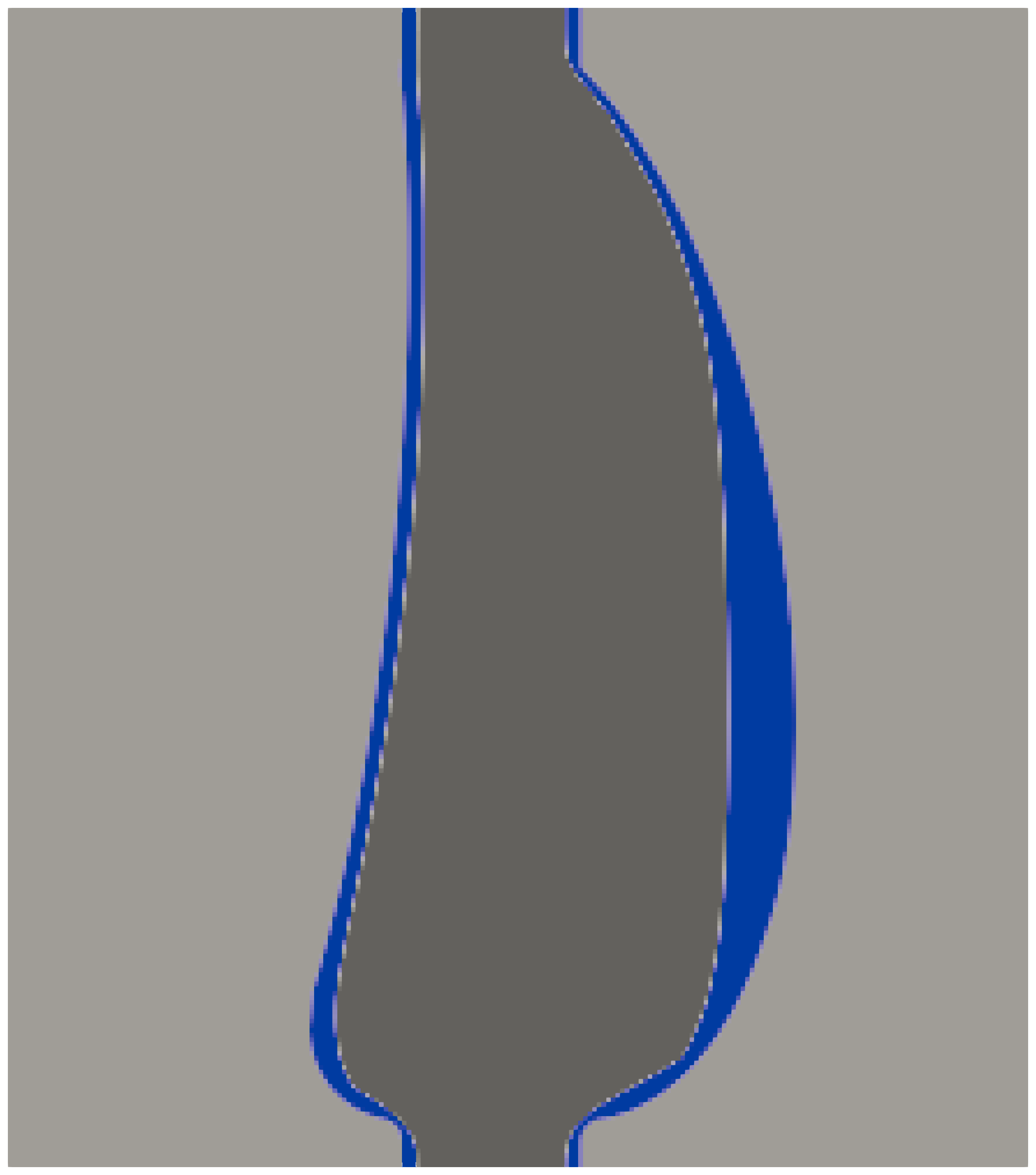}
  \end{subfigure}
  \begin{subfigure}{.34\textwidth} 
      \centering 
      \includegraphics[width=\textwidth]{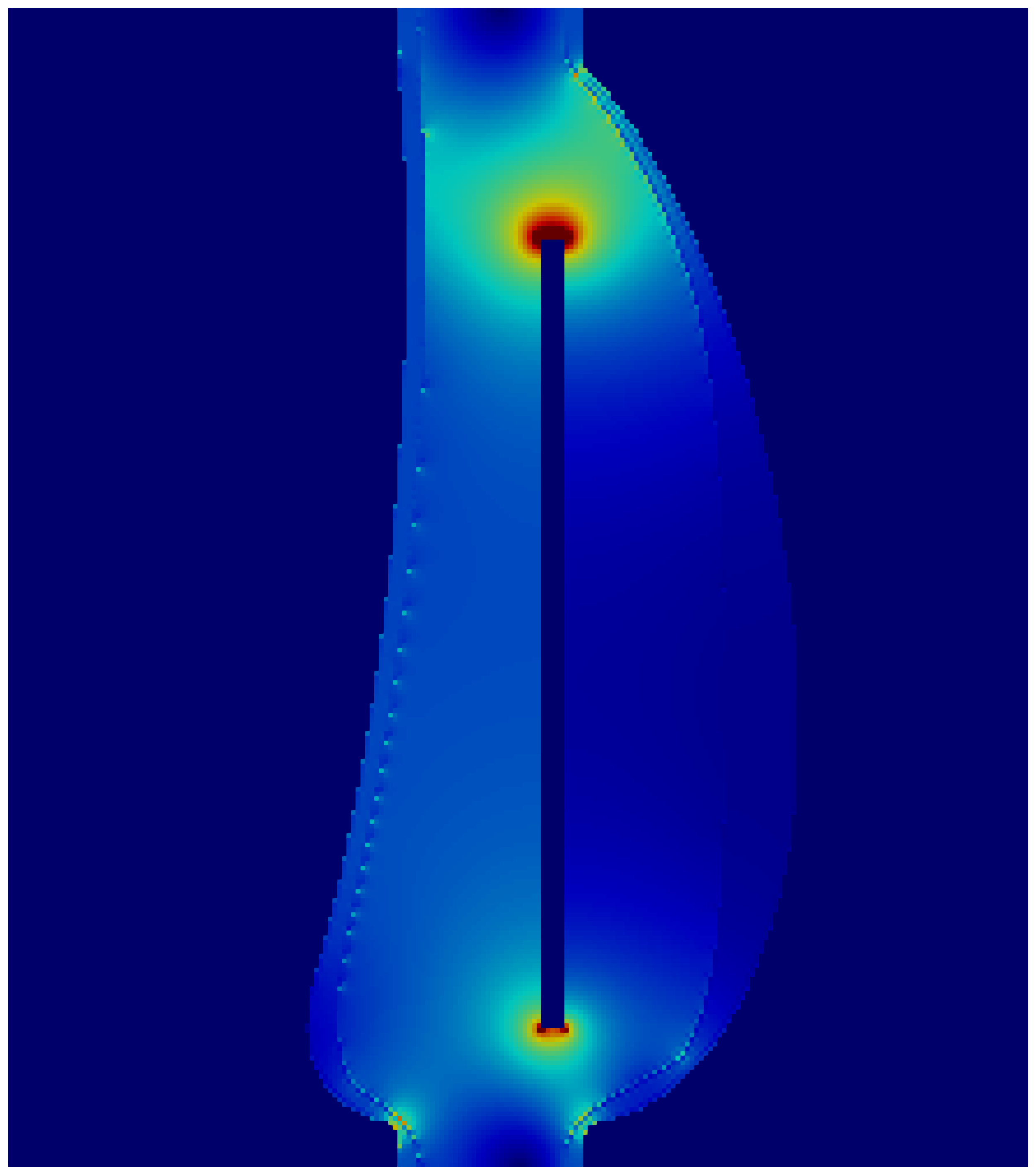}
  \end{subfigure}
  
  \vspace{0.8mm}

  \begin{subfigure}{.34\textwidth} 
      \centering 
      \includegraphics[width=\textwidth]{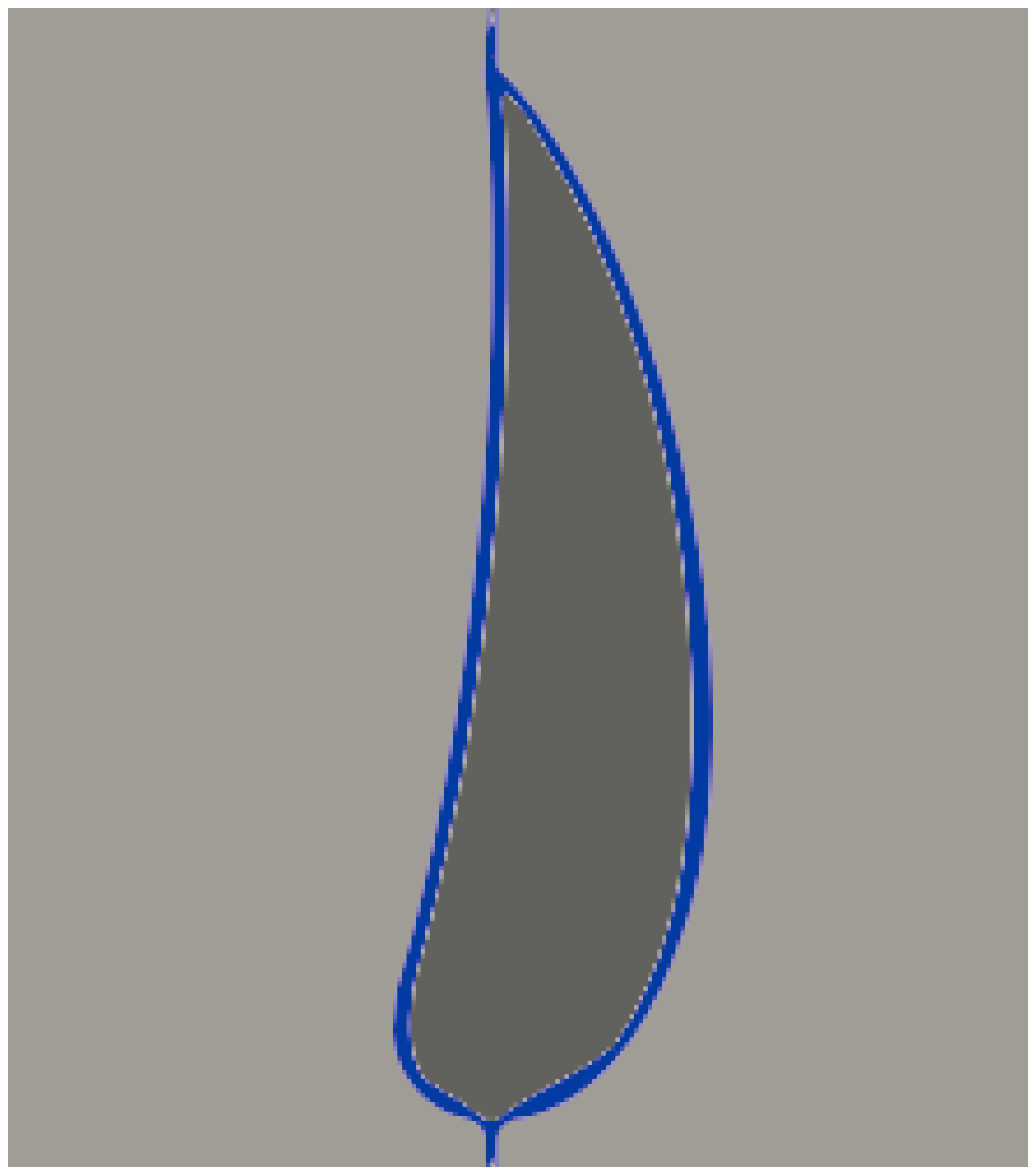}
  \end{subfigure}
  \begin{subfigure}{.34\textwidth} 
      \centering 
      \includegraphics[width=\textwidth]{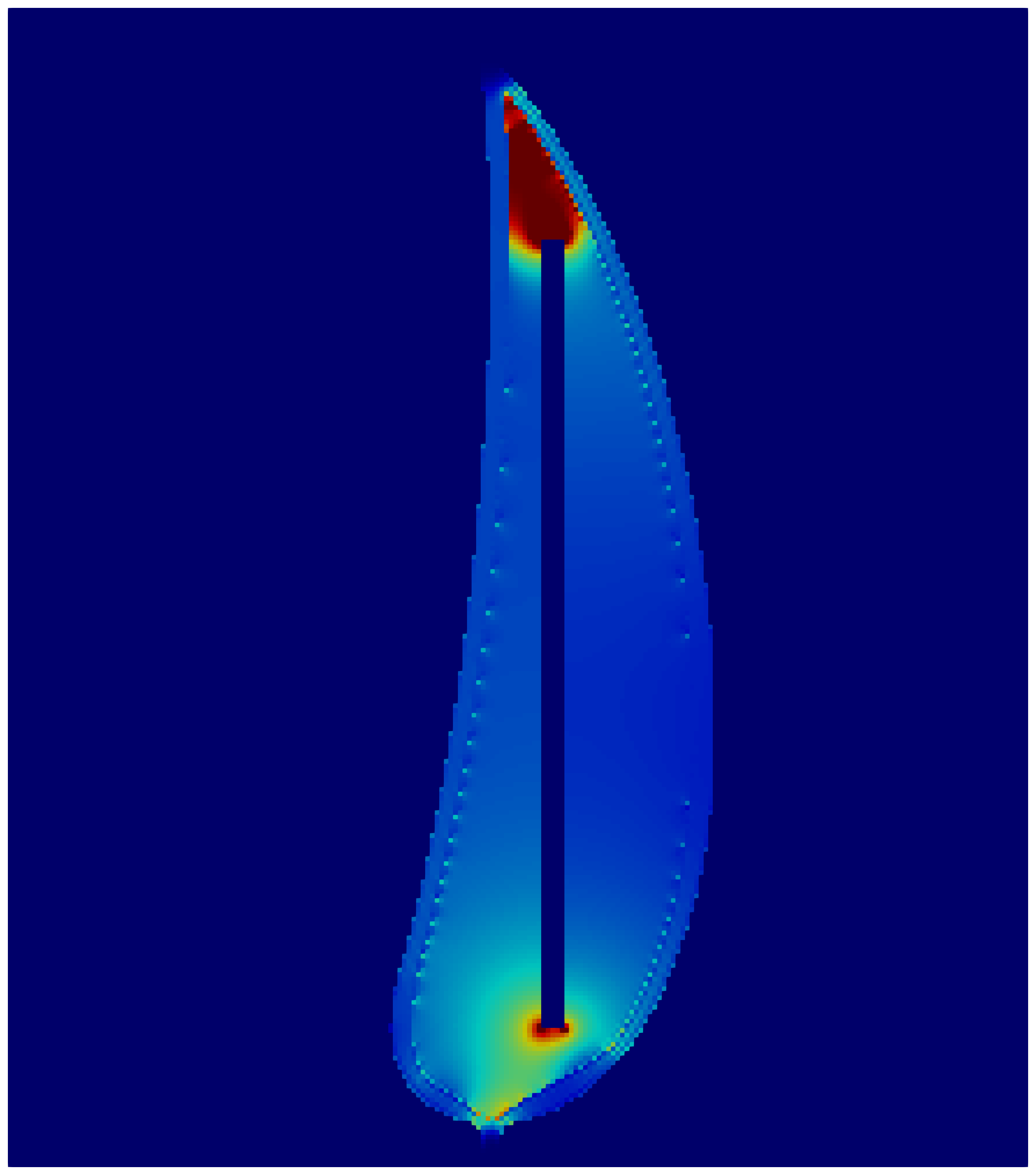}
  \end{subfigure}
  
  \vspace{4mm}
    
  \begin{subfigure}{.34\textwidth} 
      \centering 
      \begin{tikzpicture} 
        \node[inner sep=0pt] (pic) at (0,0) {\includegraphics[height=5mm, width=30mm]
        {03_Contour/00_Color_Maps/01_dsln_hor.png}};
        \node[inner sep=0pt] (0)   at ($(pic.south)+(-1.80, 0.26)$)  {$0$};
        \node[inner sep=0pt] (1)   at ($(pic.south)+( 1.80, 0.26)$)  {$1$};
        \node[inner sep=0pt] (d)   at ($(pic.south)+( 0.00, 0.85)$)  {$d~\si{[-]}$};
      \end{tikzpicture} 
  \end{subfigure}
  \begin{subfigure}{.34\textwidth} 
      \centering 
      \begin{tikzpicture} 
        \node[inner sep=0pt] (pic) at (0,0) {\includegraphics[height=5mm, width=30mm]
        {03_Contour/00_Color_Maps/05_jet_hor.png}};
        \node[inner sep=0pt] (0)   at ($(pic.south)+(-1.80, 0.26)$)  {$0$};
        \node[inner sep=0pt] (8e5) at ($(pic.south)+( 2.15, 0.26)$)  {$\geq \hspace{-0.8mm} 5$e$6$};
        \node[inner sep=0pt] (j)   at ($(pic.south)+( 0.00, 0.85)$)  {$\|\j\|~\si{[\ampere\per\square\meter]}$};
      \end{tikzpicture} 
  \end{subfigure}  
  \caption{Evolution of the dissolution level $d$ and the norm of the electric current density $\|\j\|$ for a complex cathode at times $200~\si{\s}$, $400~\si{\s}$ and $580~\si{\s}$ .}
  \label{fig:Ex4_evolution}     
\end{figure}

In Fig.~\ref{fig:Ex4_evolution}, the evolution of the dissolution level $d$ and the norm of the electric current density $\|\j\|$ is presented for two intermediate and the final configuration. Due to the initially small working gap width at the top and bottom region of the specimen, we observe a pronounced material removal in these sections. With increasing machining time, the workpiece continuously dissolves also in the middle of the specimen (time $400~\si{\s}$) until finally the mapping of the cathode's shape onto the workpiece is completed (time $580~\si{\s}$).

The simulation required a computation time of $2~\si{\hour}~12~\si{\minute}$ and, thus, confirms the method's ability to model complex dissolution processes with good efficiency.


\section{Conclusion}
\label{sec:conclusion}

In this paper, a novel methodology to model the cathode feed in the moving boundary value problem of electrochemical machining was presented. Two approaches are investigated to describe the cathode. In method A, which is characterized by a simple and robust implementation, the electric conductivities of cathode elements are modified. In method B, Dirichlet boundary conditions are applied on all nodes within the cathode and, hence, facilitate considerable speedups. Elements on the cathode's surface are described by effective material parameters. The model's accuracy is confirmed by analytical as well as experimental and numerical reference solutions from the literature. Moreover, low runtimes allow for the efficient investigation of complex shapes.

The presented model formulation is focused on the efficient description of the cathode feed in ECM simulations. Currently, it considers only certain aspects of the multiphysical process. Thus, further work includes the modeling of multiphase materials (cf.~\cite{KozakZybura-Skrabalak2016}, \cite{Harst2019}) and the formation of oxide layers (e.g.~\cite{ZanderSchuppEtAl2021}) to incorporate the polarization voltage. Moreover, a formulation of the dissolution model in the gradient enhanced framework, which was introduced by e.g.~\cite{DimitrijevicHackl2008} and \cite{Forest2009,Forest2016} and applied in e.g.~\cite{BrepolsWulfinghoffEtAl2017,BrepolsWulfinghoffEtAl2020} and \cite{FassinEggersmannEtAl2019b,FassinEggersmannEtAl2019a}, can be investigated. Further, the consideration of mass transfer, hydrogen generation and fluid mechanical effects following \cite{DeconinckVanDamme2012a,DeconinckVanDamme2012b,DeconinckHoogsteen2013} would yield a valuable model extension.

Finally, a transfer of the modeling technique to laserchemical machining (LCM, e.g.~\cite{SchuppPuetzEtAl2021}) is aspired, which requires to adapt the evolution equation of the dissolution level from an electrically to a thermally based formulation (\cite{KlockeVollertsenEtAl2016}).

\subsection*{Acknowledgements}

Funding granted by the German Research Foundation (DFG) for projects number 453715964, 223500200 (M05) and 417002380 (A01) is gratefully acknowledged.

\bibliographystyle{agsm}
\bibliography{literature}

\end{document}